\documentclass[twocolumn,superscriptaddress,prb]{revtex4}

\usepackage{graphicx}
\usepackage{epsfig}
\usepackage{verbatim}
\usepackage{bm}
\usepackage{pgfplots}
\usepackage{tikz}
\usetikzlibrary{arrows, decorations.pathreplacing}

\newcommand{\gammabar}{\ensuremath\gamma\kern-0.53em-}

\usepackage{times}
\usepackage{amsmath}
\usepackage{amsfonts}
\usepackage{amssymb}
\usepackage{graphicx}
\usepackage{pst-all}
\usepackage{leftidx}

\usepackage{tikz}

\newenvironment{changemargin}[2]{%
\begin{list}{}{%
\setlength{\topsep}{0pt}%
\setlength{\leftmargin}{#1}%
\setlength{\rightmargin}{#2}%
\setlength{\listparindent}{\parindent}%
\setlength{\itemindent}{\parindent}%
\setlength{\parsep}{\parskip}%
}%
\item[]}{\end{list}}

\allowdisplaybreaks

\makeatletter
\def\l@subsubsection#1#2{}
\makeatother

\begin{document}

\title{Modular transformations through sequences of topological charge projections}
\author{Maissam Barkeshli}
\affiliation{Station Q, Microsoft Research, Santa Barbara, California 93106-6105, USA}
\affiliation{Department of Physics, Condensed Matter Theory Center, University of Maryland, College Park, Maryland 20742, USA
and Joint Quantum Institute, University of Maryland, College Park, Maryland 20742, USA}
\author{Michael Freedman}
\affiliation{Station Q, Microsoft Research, Santa Barbara, California
  93106-6105, USA}
\affiliation{Department of Mathematics, University of California, Santa Barbara, California 93106 USA}

\begin{abstract}
The ground state subspace of a topological phase of matter forms a representation of the mapping class group
of the space on which the state is defined. We show that elements of the mapping class group
of a surface of genus $g$ can be obtained through a sequence of topological charge projections
along at least three mutually intersecting non-contractible cycles. We demonstrate this both through
the algebraic theory of anyons and also through an analysis of the topology of the space-time manifold. We combine
this result with two observations: (i) that surfaces of genus $g$ can be effectively simulated in planar geometries by
using bilayer, or doubled, versions of the topological phase of interest, and inducing the appropriate types of
gapped boundaries; and (ii) that the required topological charge projections can be implemented as
adiabatic unitary transformations by locally tuning microscopic parameters of the system, such as the energy gap.
These observations suggest a possible path towards effectively implementing modular transformations in physical systems.
In particular, they also show how the Ising $\otimes \overline{\text{Ising}}$ state, in the presence of disconnected gapped boundaries,
can support universal topological quantum computation.
\end{abstract}


\maketitle

Non-Abelian anyons in a topological phase of matter are known to give rise to a topologically protected
space of states, which are indistinguishable by any local operator.\cite{nayak2008} Adiabatically braiding the positions of the
non-Abelian anyons gives rise to a unitary representation of the braid group on this topological subspace, offering
the possibility of implementing non-trivial, topologically protected unitary transformations on the state of the system.

Topological ground state degeneracies can also arise when the system is defined on a space with non-trivial topology,
such as a torus.\cite{witten1989,wen1989,wen1990b} In this case, there is also a group of topologically protected unitary transformations that act
on the ground state subspace, analogous to the case of braiding non-Abelian anyons. These transformations are associated
with adiabatically varying the geometry of the surface $\Sigma$ on which the system is defined.
The analog of the braid group in this case is the mapping class group (MCG) of $\Sigma$, which is the group of self-diffeomorphisms of $\Sigma$, modulo those
which can be continuously connected to the identity. Each element of the MCG is referred
to as a modular transformation. The braid group for $n$ particles can be viewed as a special case of the MCG
of a disk with $n$ punctures.

Given a topological phase of matter, the MCG of a genus $g > 0$ surface
provides access to a much richer group of topologically protected unitary transformations than can be
achieved by braiding the anyons alone. For example, in an Abelian topological phase, the representation of the braid
group for the anyons is always one-dimensional, whereas the representation of the MCG on a higher genus surface is always multi-dimensional.
In some cases, such as for the Ising topological phase, the braid group of non-Abelian anyons is not sufficient
for universal topological quantum computation (TQC), while access to the MCG of the system on a high genus surface can be utilized
to perform universal TQC. \cite{bravyi200universal,freedman2006universal} Moreover, it is conjectured that the modular
transformations on the torus, together with the chiral central charge of the topological phase, can completely characterize
all of the robust universal properties of a topological phase of matter.\cite{wangpc} Thus it is of great interest to perform these
modular transformations both in numerical simulations for the purpose of diagnosing topological order, and also in
real physical systems.

The realization of modular transformations in a topological phase of matter on a torus has been discussed previously
in several ways. One is through adiabatic variations of the geometry of the system, through varying the metric in a
continuum theory\cite{wen1990naberry,read2009,wen2012} or by varying microscopic interactions in a lattice system.\cite{you2015} Modular transformations can also be associated with
basis transformations in the topological ground state subspace\cite{witten1989,moradi2015}; methods for extracting the suitable bases
and thus the basis transformations through entanglement considerations have also been demonstrated, and are useful for numerical
diagnostics of topological order.\cite{zhang2012,zhang2015} Ref. \onlinecite{bonderson2010blueprint} further proposed methods to
effectively generate a specific set of modular transformations for topological phases described by the Ising TQFT,
by depleting the topological phase and performing topological charge measurements of the non-Abelian anyons along
varying cycles of surfaces with nontrivial genus.~\footnote{While the modular transformation ideas of Ref. \onlinecite{bonderson2010blueprint}
apply to any system described by the Ising TQFT, the discussion of physical realization of the ideas in Ref. \onlinecite{bonderson2010blueprint}
focused on utilizing topological superconductors, which are not actually described by the Ising TQFT and so do not
support such methods.~\cite{bonderson2013}}

In this paper, we provide an alternative way of realizing modular transformations on a high genus surface. We show that
any generic modular transformation can be implemented through a series of topological charge projections along at least three mutually intersecting
non-contractible cycles of the system. We further show that each of the required topological charge projections can in principle be realized through
adiabatic unitary evolution, assuming the ability to locally tune certain microscopic parameters, such as the energy gap,
of the system. That the required topological charge projections can be realized as unitary operations is made possible through
the use of an extra handle in the space, which acts as an ancillary set of degrees of freedom. This is related to previous work
demonstrating that measurement-based braiding of non-abelian anyons\cite{bonderson2009} can be achieved
by adiabatically tuning the interactions between them.\cite{bonderson2013braiding}

Finally, we note that genus $g$ surfaces can be realized in planar geometries
by considering doubled, or bilayer, versions of the topological phase of interest, together with either $g+1$ disconnected
gapped boundaries or $2g+2$ genons.\cite{barkeshli2013genon} Altogether, these observations suggest a possible route towards
effectively realizing modular transformations on high genus surfaces in physically realistic systems where both
the genus and the required topological projections can be effectively implemented with experimentally controllable parameters.

This paper is organized as follows. In Section \ref{bilayerSec}, we provide a discussion, most of which is review,
of how high genus surfaces can effectively be realized in planar systems through the use of bilayer, or doubled,
versions of the topological phase of interest, together with either genons or gapped boundaries. This provides
some physical motivation for considering modular transformations and high genus surfaces. In Sec. \ref{gsdRev},
we provide a brief review of topological ground state degeneracies on genus $g$ surfaces, and we establish some
notation that will be used in the subsequent paper. In Section \ref{generalDisc}, we provide a general overview of the relation between
adiabatic variations and topological charge measurements, and between the braid group and the mapping class group.
In Section \ref{tprojSec}, we define the notion of topological charge projections that we use, and discuss how
the ones of interest to us can be implemented through an adiabatic unitary process. In Section \ref{Dehn}, we describe
how modular transformations are realized through topological charge projections, and demonstrate this through an algebraic
calculation in Section \ref{algebraicSec} and also through analysis of the topology of the resulting space-time manifold
in Section \ref{topoSec}. We make a few concluding comments in Section \ref{disc}.

\section{Effectively realizing high genus surfaces through planar geometries}
\label{bilayerSec}

The discussion of topological phases of matter on high genus surfaces seems to be, at first glance, of purely theoretical interest, given that
physical systems are most naturally realized experimentally in planar geometries. However, recent studies of gapped boundaries in topological
phases of matter have demonstrated several ways in which high genus can be effectively simulated in planar systems by using
bilayer, or doubled, versions of the topological phase of interest and inducing the appropriate gapped interfaces or boundaries in
the system. In order to provide some practical motivation for our subsequent considerations, below we briefly review two ways in which a
topological phase of matter, described by a unitary braided tensor category (UBTC) $\mathcal{C}$, can be realized on a genus $g$ surface
in a planar geometry that would be amenable to experimental realization.

\subsection{$\mathcal{C} \otimes \mathcal{C}$ and genons}

\begin{figure}
	\centering
	\includegraphics[width=3.8in]{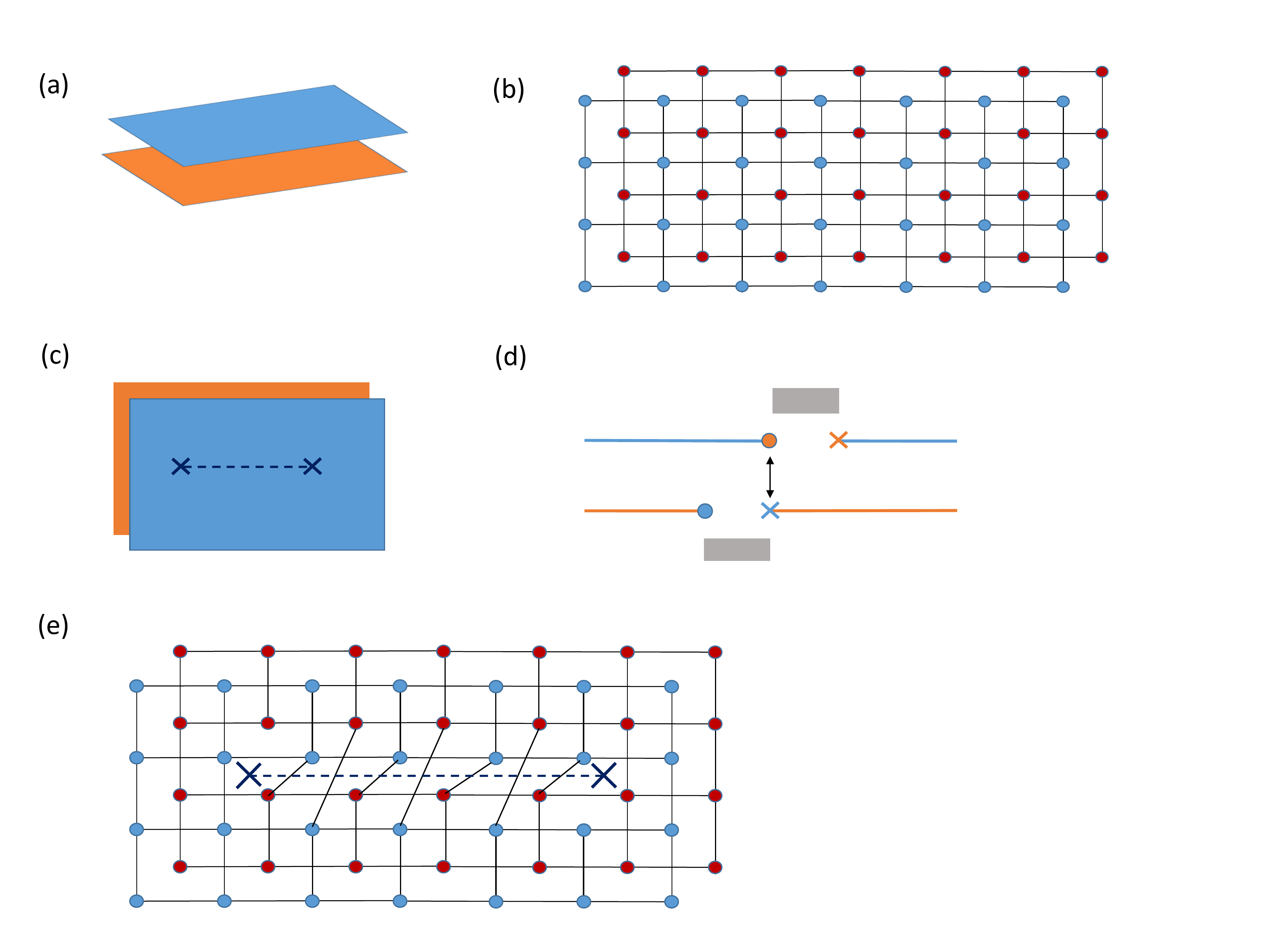}
	\caption{\label{genonsFig} (a) Bilayer system, (b) Two copies of a lattice system, (c) Branch cut that connects the two layers, (d)
Cross-sectional view of a way to realize the branch cut and genons in a bilayer FQH state, proposed in Ref. \onlinecite{barkeshli2014}.
Top and bottom electrical gates (gray) locally deplete the electron fluid, resulting in counterpropagating chiral edge states in each
layer (shown in dots and crosses). Electron tunneling (double arrow) between counterpropagating edge states of different layers can be used
to effectively "glue" the two layers together, realizing half of the branch cut. The other half of the branch cut gluing can be neglected as this merely
leaves behind a puncture which can be projected into the trivial charge sector and subsequently ignored. (Alternatively, second order tunneling proceesses
can also induce the other half of the branch cut.\cite{barkeshli2014})
(e) Lattice defect that effectively couples the two copies of the lattice system.}
\end{figure}

Let us consider a topological phase of the form $\mathcal{C} \otimes \mathcal{C}$, which is topologically equivalent to two
independent copies of a single topological phase $\mathcal{C}$. States of this form have been experimentally
realized in the context of bilayer fractional quantum Hall (FQH) systems.\cite{boebinger1990,suen1991,eisenstein1992} They can also be considered in lattice models, as depicted in
Fig. \ref{genonsFig}. In such a system, one can consider a line segment along which there is a ``branch cut," where the two layers
are connected to each other along the cut. Physically these can be created in bilayer FQH states by using electrical
gates on the top and bottom layers to effectively ``cut" and ``reglue" the FQH state in a twisted manner, as proposed in Ref. \onlinecite{barkeshli2014}.
In lattice models they can be created by changing the connectivity of the lattice, by inserting lattice dislocations into the system.\cite{barkeshli2012a}
(See Ref. \onlinecite{barkeshli2015} for a recent proposed physical realization using Majorana nanowires).

The system with such branch cuts can be thought of as a two-fold branch cover of the disk. The end-point of the
cut, where the branch covering degenerates, is a certain type of topological twist defect referred to as a ``genon."\cite{barkeshli2013genon}
The branch cut itself is not detectable by topological operations such as braiding. The positions of the genons, however,
are well-defined; the genons are non-Abelian twist defects,\cite{bombin2010,kitaev2012,barkeshli2012a,barkeshli2013genon}
whose topological properties are described within the framework of $G$-crossed braided tensor categories,\cite{barkeshli2014SDG}
with $G = Z_2$ corresponding to the permutation of the two copies of $\mathcal{C}$ in $\mathcal{C} \otimes \mathcal{C}$.

The genons effectively change the topology of the manifold: the topological state $\mathcal{C} \otimes \mathcal{C}$
with $n$ pairs of genons is topologically equivalent to a single copy of $\mathcal{C}$, on a genus $g = n-1$ surface.\cite{barkeshli2010,barkeshli2013genon}

\subsection{$\mathcal{C} \otimes \overline{\mathcal{C}}$ and gapped boundaries}

Let us consider a topological phase of matter, $\mathcal{C}$, on a plane. By folding half of the plane over on itself, say along the
line $(x,y=0)$, we obtain a topological phase $\mathcal{C} \otimes \overline{\mathcal{C}}$, together with a boundary to vacuum, where $\overline{C}$
corresponds to the parity-reversed counterpart of $\mathcal{C}$. The process where a quasiparticle
$a \in \mathcal{C}$ propagates across the folding line $(x,y = 0)$ then gets mapped to a quasiparticle  $a \in \mathcal{C}$
propagating to the boundary and reflecting from the boundary as a quasiparticle $\overline{a} \in \overline{\mathcal{C}}$.
Equivalently, this implies that a pair of quasiparticles of the form $(a, \overline{a}) \in \mathcal{C} \otimes \overline{\mathcal{C}}$ can be annihilated
into the vacuum upon approaching the boundary of the system.  This is one particular type of gapped boundary of $\mathcal{C} \otimes \overline{\mathcal{C}}$ with the vacuum.
In general, there can be many other topologically distinct classes of gapped boundaries between $\mathcal{C} \otimes \overline{\mathcal{C}}$ and
vacuum.\cite{kapustin2011,kitaev2012,levin2013,barkeshli2013defect2} The one described above is always one of the possible types of gapped boundaries.

Let us consider now a topological phase $\mathcal{C} \otimes \overline{\mathcal{C}}$ in the presence of $n_b$ disconnected
gapped boundaries, where the boundary condition is as described above. From the above discussion, it is clear that this
situation is equivalent to a single copy of $\mathcal{C}$ on a genus $g = n_b - 1$ surface.

There are several examples which are of particular physical interest. One class of examples corresponds to cases where $\mathcal{C} \otimes \overline{\mathcal{C}}$
is realized as the ground state of a Hamiltonian which is a sum of commuting projectors, as in a Levin-Wen model.\cite{levin2005} Another class of examples that may be
of experimental interest, which we will discuss in some detail below, corresponds to cases where $\mathcal{C}$ describes a fractional quantum Hall
(FQH) state at filling fraction $\nu$.

\begin{figure}
	\centering
	\includegraphics[width=3.6in]{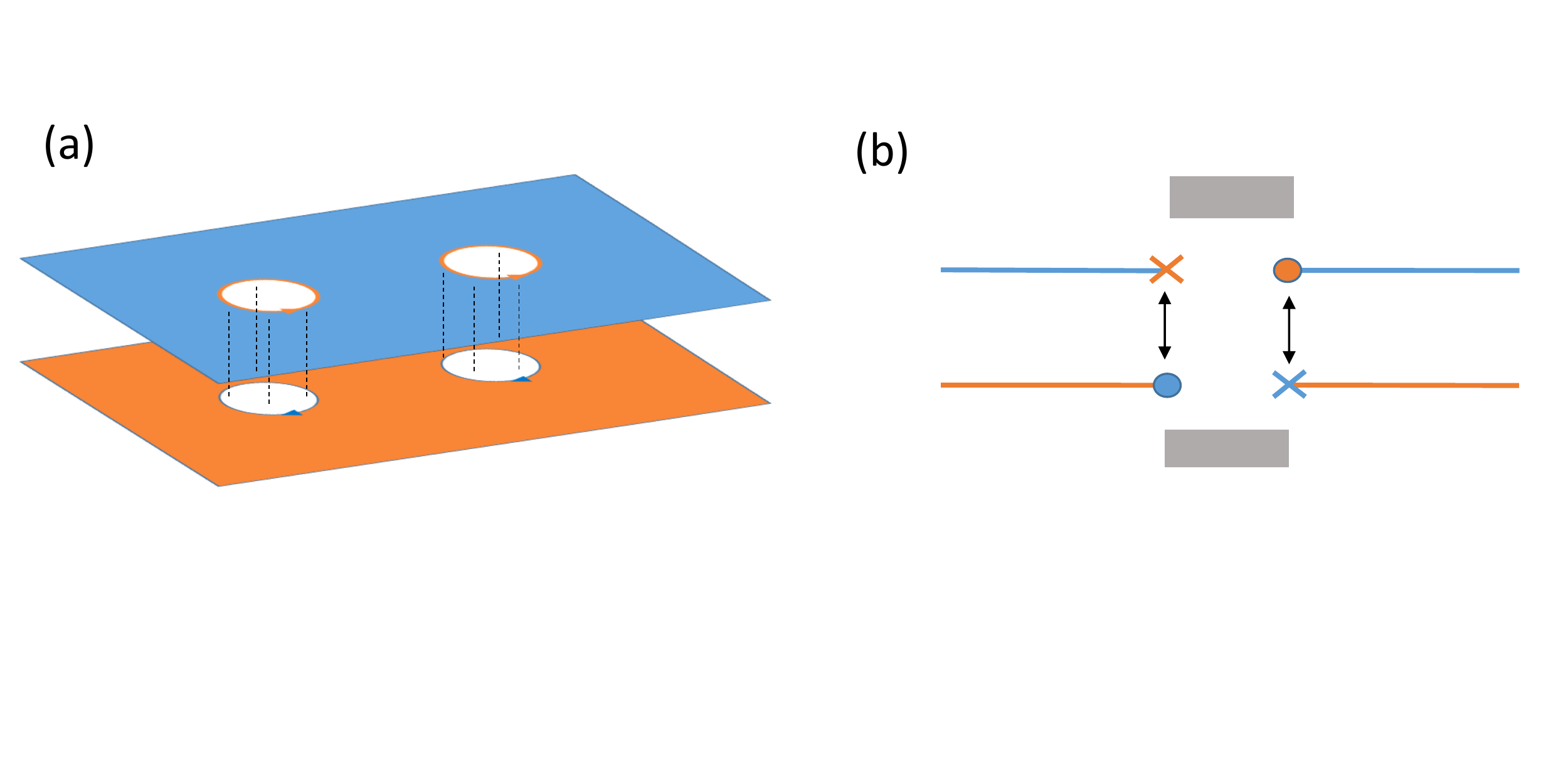}
	\caption{\label{holesFig} (a) Double layer system with boundaries. Dashed lines describe electron tunneling between
counterpropagating chiral edge states.  (b) Cross-sectional view of a way to realize gapped boundaries in electorn-hole bilayer FQH state.
Top and bottom electrical gates (gray) locally deplete the electron/hole fluids in the two layers, resulting in counterpropagating
chiral edge states in each layer (shown in dots and crosses). Electron tunneling (double arrow) between counterpropagating edge states of different layers can be used
to effectively "glue" the two layers together, realizing a gapped boundary where quasiparticles can coherently propagate from one layer to another. .
}
\end{figure}

\subsubsection{Example: Fractional quantum Hall states}

For definiteness let us consider $\mathcal{C}$ to correspond to a $\nu = 1/m$ Laughlin FQH state.
The situation discussed above can be realized by considering a bilayer system, consisting of electrons in one layer and holes
in the second layer. Since the holes have opposite charge relative to the electrons, in a uniform magnetic field the two layers
will form FQH states of opposite chirality, forming $\nu = \pm 1/m$ Laughlin FQH states. At an interface with vacuum,
the two layers will, due to the opposite chiralities, give rise to counterpropagating chiral Luttinger liquid edge modes. These edge modes
are described by the Lagrangian:\cite{wen04}
\begin{align}
\mathcal{L}_0 = \frac{1}{4\pi} (-1)^I m \partial_x \phi_I \partial_t \phi_I - V_{IJ} \partial_x \phi_I \partial_x \phi_J,
\end{align}
where $\phi_I$, for $I = 1,2$ are real scalar fields describing the counterpropagating edge modes in the two layers,
and $V_{IJ}$ is a positive-definite matrix describing the velocities of
and interactions between the edge modes. The electron operator on the two layers is
described by the operator $\Psi_I \sim e^{i m \phi_I}$, while the charge $1/m$ quasiparticle operators in the two layers are given by $e^{i \phi_I}$.
Electron backscattering between the edge modes is therefore described by the term
\begin{align}
\mathcal{L}_t = - t \cos(m (\phi_1 - \phi_2)) .
\end{align}
For large tunneling amplitude $t$, the cosine term can pin its argument: $\langle e^{i (\phi_1 - \phi_2)} \rangle \neq 0$,
thus localizing the edge modes and leading to a gapped boundary. Physically, the fact that $\langle e^{i (\phi_1 - \phi_2)} \rangle \neq 0$
means that quasiparticle-quasihole pairs from the two layers have condensed at the boundary, and thus can disappear into the boundary condensate upon
approaching the boundary from the bulk of the system.

To summarize, an electron-hole bilayer FQH state can be used to simulate a single FQH state on a genus $g$ surface by fabricating
$g + 1$ disconnected boundaries in the system, and causing the counterpropagating edge modes to be localized through electron tunneling (backscattering)
along the boundaries.

\section{Review of Topological Ground State Degeneracy}
\label{gsdRev}

\begin{figure}
	\centering
	\includegraphics[width=3.5in]{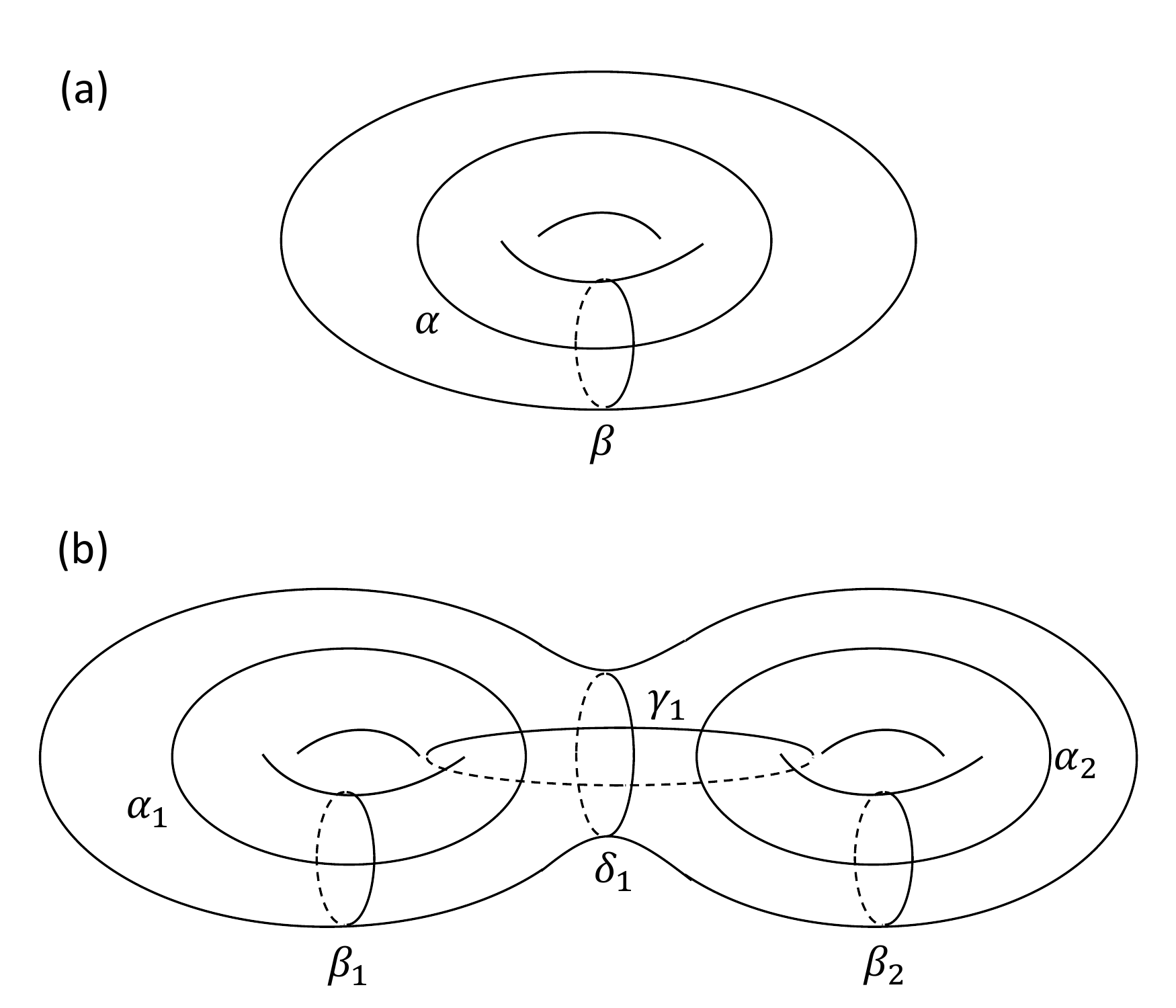}
	\caption{\label{cyclesFig} Labelling of cycles. In the genus 2 case, we will sometimes
drop the subscript in $\gamma_1$ and $\delta_1$.
}
\end{figure}

In this section, we briefly review the understanding of topological ground state degeneracy on genus $g$ surfaces, and
introduce the relevant notation that will be used in subsequent sections.

Let $\mathcal{V}_g$ denote the ground state subspace of a topological phase of matter on a closed genus $g$ surface.
The number of ground states on a genus $g$ surface is given by the dimension of $\mathcal{V}_g$, and is determined by
the Verlinde formula:
\begin{align}
\text{dim } \mathcal{V}_g = \sum_{a\in \mathcal{C}} S_{0a}^{2 -2g},
\end{align}
where $S$ is the modular $S$ matrix of $\mathcal{C}$ and $0 \in \mathcal{C}$ is the identity topological charge.
States in $\mathcal{V}_g$ are labelled by the values of the topological charges along a maximal set of
non-intersecting, non-contractible cycles.

\begin{figure}
	\centering
	\includegraphics[width=3.7in]{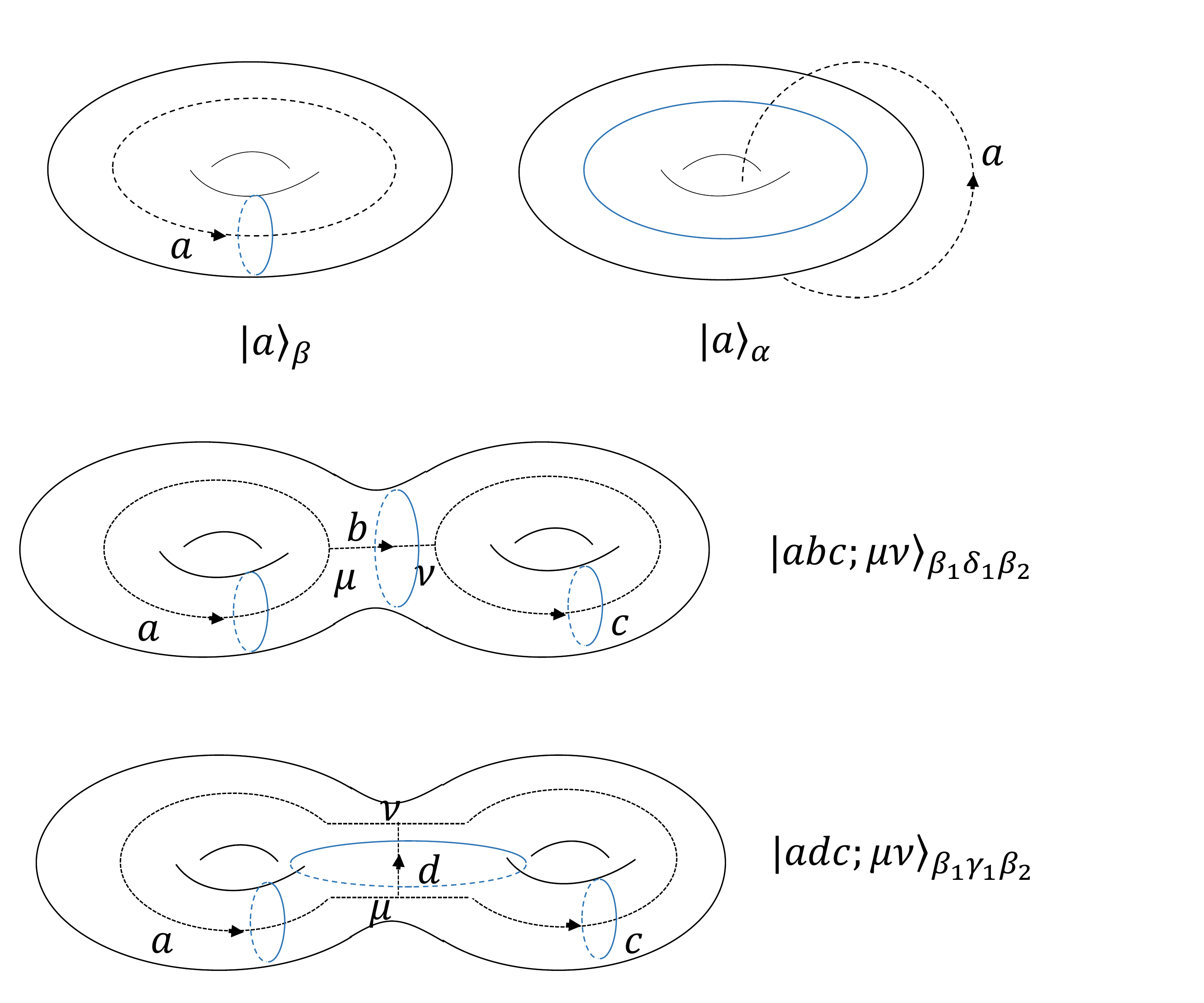}
	\caption{\label{statesFig} Convention for labeling states. Top left: States are labelled by definite topological charge $a \in \mathcal{C}$
along the meridianal (blue) loop, $\beta$. A dual loop is also depicted, as the dashed black loop. The state $|a\rangle_\beta$ can be obtained from
the state $|1\rangle_\beta$ by applying a Wilson loop for the quasiparticle $a$ along a dual loop $\alpha$.
Top right: States labelled by topological charge in the longitudinal (blue)
loop. Bottom two figures: conventions for genus two surfaces, generalizing the notation for the genus one case.
}
\end{figure}

For example, consider the torus, $g = 1$. The ground state degeneracy can be labelled by the
topological charge value $a \in \mathcal{C}$ that would be obtained by measuring the topological charge along
the longitudinal cycle (labelled $\alpha$ in Fig. \ref{cyclesFig}a) of the torus.
Such states will be labelled as $|a \rangle_\alpha$ (see Fig. \ref{statesFig}). Alternatively, the states can be labelled by the topological charge
value $a$ that would be obtained by measuring the topological charge along the meridianal cycle of the torus (labelled $\beta$) in Fig. \ref{statesFig}.
Such states will be labelled as $|a \rangle_\beta$. $\{ |a \rangle_\alpha \}$ and $\{ a \rangle_\beta \}$ each provide a
complete basis for the ground state subspace $\mathcal{V}_1$. They are related to each other by the modular $S$ matrix:
\begin{align}
|a \rangle_\alpha = \sum_{b \in \mathcal{C}} S_{ab} |b \rangle_\beta .
\end{align}

On a genus $g = 2$ surface, we have more choices of cycles. We can for example pick
cycles $\beta_1, \gamma_1, \beta_2$, as shown in Fig. \ref{cyclesFig}(b). The states in $\mathcal{V}_2$
can then be labelled by the values of the topological charges that would be measured along these cycles,
$a, b, c \in \mathcal{C}$, respectively, giving rise to a basis set denoted $|a b c\rangle_{\beta_1\gamma_1\beta_1}$ (see Fig.\ref{statesFig}).
The number of such states is determined by the fusion rules $(N_{a\overline{b}}^c)^2$, as there are two fusion vertices
in this case. Thus, a complete labelling of the basis states is $|a b c;\mu\nu \rangle_{\beta_1,\gamma_1,\beta_1}$,
for $a,b,c \in \mathcal{C}$ and $\mu,\nu = 1, \cdots, N_{a\overline{b}}^c$. Here $\overline{a} \in \mathcal{C}$ refers to the anti-particle (or dual) of
$a \in \mathcal{C}$.

We can fix a different basis by picking a different maximal set of non-intersecting cycles, $\omega_1, \omega_2, \omega_3$,
and labelling the states as $|abc;\mu \nu \rangle_{\omega_1\omega_2\omega_3}$, generalizing the above discussion.
The basis transformations between several different useful bases are listed below:

\begin{align}
|a0c \rangle_{\beta_1\delta_1\alpha_2} = \sum_{c'} S_{cc'} |a0c' \rangle_{\beta_1 \delta_1 \beta_2}
\end{align}
More generally,
\begin{align}
|abc; \mu \nu \rangle_{\beta_1\delta_1\alpha_2} = \sum_{c'} S_{cc'; \mu \nu \mu' \nu'}^{(b)} |abc' ;\mu' \nu' \rangle_{\beta_1 \delta_1 \beta_2},
\end{align}
where $S^{(b)}$ is the ``punctured" $S$ matrix, although we will not explicitly use the definition of $S^{(b)}$ here.

Furthermore, we have:
\begin{align}
|abc;\mu \nu \rangle_{\beta_1 \delta_1 \beta_2} = \sum_{b',\mu',\nu'} [F^{\bar{a} a \bar{c}}_{\bar{c}}]_{(\bar{b}\mu\nu)(b'\mu'\nu')} |ab'c; \mu' \nu'\rangle_{\beta_1 \gamma_1 \beta_2},
\end{align}
where $F^{abc}_d$ are the $F$-symbols of the UBTC $\mathcal{C}$.
Note that
\begin{align}
[F^{\bar{a}a\bar{c}}_{\bar{c}}]_{(000)(b'\mu'\nu')} =
\sqrt{\frac{d_{b'}}{d_a d_c}} \delta_{\mu'\nu'} , \;\;\; \text{ if } N_{a\bar{c}}^{b'} \neq 0,
\end{align}
and is equal to $0$ otherwise, where $d_a = S_{0a}/S_{00}$ is the quantum dimension of $a$.
This implies
\begin{align}
|a0c\rangle_{\beta_1 \delta_1 \beta_2} = \sum_{\{b'| N_{a \bar{c}}^{b'} \neq 0\}} \sum_{\mu' = 1}^{N_{a\bar{c}}^{b'}} \sqrt{\frac{d_{b'}}{d_a d_c}} |ab'c; \mu' \mu' \rangle_{\beta_1 \gamma_1 \beta_2}
\end{align}

\section{General overview}
\label{generalDisc}

In the topological quantum field theory (TQFT) description of topological phases of matter, the worldlines of the anyons must be
given a framing in order to be well-defined; physically, this framing is required to keep track of the topological spin of the
quasiparticles.

\begin{figure}[hbpt]
\begin{tikzpicture}[xscale=0.75]
\draw [dashed] (0,0) to (0,-0.5);
\draw (0,-0.5) to (0,-1) to [out=-90, in=-135] (1,-1.25) to [out=45, in=0] (2,-1)
    to [out=0, in=90] (3,-1.5) to (3,-2.5) to [out=-90, in=0] (2,-3) to [out=180, in=-90] (1,-2.625) to [out=90, in=180] (1.5,-2.25) to [out=0, in=90] (2,-2.5) to (2,-2.875);
\draw (2,-3.1) to (2,-3.15);
\draw (2,-3.35) to (2,-3.5);
\draw [dashed] (2,-3.5) to (2,-3.6);
\draw (2,-3.9) to (2,-4);
\draw [dashed] (0,-4) to (0,-3.5);
\draw (0,-3.5) to (0,-2) to [out=90, in=135] (1,-1.75) to [out=-45, in=180] (1.5,-2)
    to [out=0, in=-90] (2,-1.5) to (2,-1.125);
\draw (2,-0.9) to (2,-0.85); \draw (2,-0.65) to (2,-0.5);
\draw [dashed] (2,-0.5) to (2,-0.4);
\draw (2,-0.1) to (2,0);
\draw [dashed] (3,-0.5) arc (0:73:1 and 0.25);
\draw [dashed] (2,-0.5) ++ (180:1 and 0.25) arc (180:107:1 and 0.25);
\draw (2,-0.5) ++ (65:1 and 0.25) arc (65:71:1 and 0.25);
\draw (2,-0.5) ++ (115:1 and 0.25) arc (115:109:1 and 0.25);
\draw (2,-0.5) ++ (80:1 and 0.25) arc (80:100:1 and 0.25);
\draw (3,-0.5) arc (0:-180:1 and 0.25);
\draw [dashed] (3,-3.5) arc (0:-73:1 and 0.25);
\draw [dashed] (2,-3.5) ++ (180:1 and 0.25) arc (180:253:1 and 0.25);
\draw (2,-3.5) ++ (-65:1 and 0.25) arc (-65:-71:1 and 0.25);
\draw (2,-3.5) ++ (-115:1 and 0.25) arc (-115:-109:1 and 0.25);
\draw (2,-3.5) ++ (-80:1 and 0.25) arc (-80:-100:1 and 0.25);
\draw (3,-3.5) arc (0:180:1 and 0.25);
\draw [->] (4,-1.875) to (3.125,-0.5);
\draw [->] (4,-2.125) to (3.125,-3.5);
\node at (4,-2) [right] {ancilla};
\draw [->] (6.5,-2.5) to (6.5,-1.5);
\node at (6.5,-2) [right] {time};
\draw [->] (-1,-1.875) to (-0.125,-1);
\draw [->] (-1,-2.125) to (-0.125,-2);
\node at (-1,-2) [left] {braid};
\node at (0,-4) [below] {$a_1$};
\node at (1,-4) [below] {$a_2$};
\node at (2,-4) [below] {$a_3$};
\node at (3,-4) [below] {$a_4$};
\end{tikzpicture}
\caption{Braid history of four anyons, labelled $a_1, \cdots, a_4$ left to right, together with three sequential projections to the identity of the fusion channel of two anyons.
These are, $(a_2, a_4)$, followed by $(a_2, a_3)$, followed by $(a_1, a_2)$. This effectively executes a braid between $a_1$ and $a_3$, together with a twist of $a_3$.
Note that the two ellipses shown, which result from the initial and final projections of $(a_2, a_4)$, have only over-crossings with the lines that they cross, which allows
for the product structure discussed in the text.
 }
\label{fig:braid}
\end{figure}
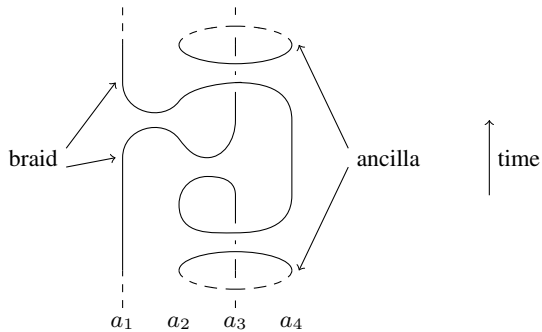

Framed braiding is reversible, hence unitary, whereas measurement is not.  Yet a sequence of collective state projective
measurements (of collective anyon charge) can create a framed braid on one tensor factor, tensor a density
matrix, e.g. $\vert P_1\rangle\langle P_1\vert$, on an ancilla\cite{bonderson2009} (see Fig. \ref{fig:braid}).

Similarly, in this paper we consider a surface $\Sigma$ in a TQFT ground state $\vert \Psi\rangle$ evolving in time.  We define analogs of both braiding and ``measuring collective states,'' which produces a unitary on a tensor factor.  The key common feature on both sides of the analogy, i.e. in all four cases, is that the unitary results by comparing two distinct diffeomorphisms from an initial to a final state.

Some measurement protocols produce exotic unitaries, which cannot possibly be induced by any diffeomorphism, the irrational phase gate of Ref. \onlinecite{levaillant2015} being an example.  This paper explores the simpler case: diffeomorphisms, over subsystems, induced by measurement.

In the case of a braid, the first diffeomorphism from ``bottom'' to ``top'' merely follows the vertical (time) product structure (PS) on $R^2\times R_t$, giving the ``identity map.''  The second follows the braided PS and when compared to the first, is an element of the mapping class group (MCG) of the punctured plane, i.e. the braid group.  This group represents on the TQFT Hilbert space.

In Figure \ref{fig:braid} we see (solid lines) a sequence of four collective state measurements to zero charge which factors as a braid (up to a frame twist) and $\vert P_1\rangle\langle P_1\vert$ on an ancilla.  Again there is the vertical PS which may be compared to a braided PS over a subset $S\subset R^2\times R^1_t$.  Extending to the larger time history (include the dotted lines in Figure \ref{fig:braid}), $S$ can be described as $R^2\times R^1_t\setminus 2$ balls, the balls enclosing the circular loops.  Since in any TQFT the Hilbert space for a sphere $\mathcal{V}\left(S^2\right)\cong\mathbb{C}$ and furthermore \cite{walker1991} $\mathcal{V}\left(S^2\right)$ contains a canonical nonzero element, the Verlinda idempotent, $x$.  This slightly diminished product $S$ still yields a well-defined unitary by filling in $x$ at the puncture, following the braid, and then comparing with the vertical PS.  To summarize, diminishing a PS by removing balls does not affect the linear TQFT map it induces.

Now let us cross to the ``surface side'' of the analogy.  One way to do this is to imagine that Figure \ref{fig:braid} is the time history of $Z_2$-genons and create surfaces interpolated by $3$-manifolds by taking a $2$-fold branched cover.\cite{barkeshli2013genon}  In this case a braid (not drawn) will branch to a product cobordism, whose PS provides an isotopy class which can be compared with the vertical PS (``id'') to produce an element of the MCG.  In the measurement case (drawn as the solid and dashed line in Figure \ref{fig:braid}) the ``braided'' PS is again diminished: it is only defined in the complement of two tubes, each tube is $S^2\times I$, where $S^2\times I\overset{\times 2}{\longrightarrow}D^3$ is the total space of the branched cover of $D^3$ along an unknotted loop.  As before the fact that the ``braided'' PS has ``holes'' with $2$-sphere boundary is no obstacle (again insert copies of $x$) to defining the unitary from initial to final state.

While the genon example is a useful model for creating interesting surfaces $\Sigma^2$ and cobordisms $M^3$ between them---and from these operations on the Hilbert space $\mathcal{V}(\Sigma)$, we focus here on a more general construction of $3$-dimensional cobordisms through two operations:
\begin{enumerate}
    \item\label{op:1} Adiabatic variation of geometry, i.e. a path in Teichm\"{u}ller space $\mathfrak{T}(\Sigma)$ joining a point to its image under a MCG element, thus a loop in moduli space $\mathfrak{M}(\Sigma) = \mathfrak{T}(\Sigma)/\text{MCG}$ (see Refs. \onlinecite{wen1990naberry,you2015}).
    \item\label{op:2} A succession of topological-charge-basis projections on (we will restrict to) constant time Wilson loops of $\Sigma_t$.
\end{enumerate}
While (\ref{op:1}) is associated with a global PS, (\ref{op:2}) is analogous to Figure \ref{fig:braid}: the product PS is partial; in favorable cases it may be arranged that only balls are missing from the PS and unitaries are again then well-defined.

We briefly summarize (\ref{op:1}) before turning in detail to (\ref{op:2}) in the subsequent sections.  The most concrete way to combinatorially mimic an essential loop in $\mathfrak{M}(\Sigma)$ is to consider a lattice Hamiltonian $H$ on $\Sigma$ and a loop $\gamma\subset\Sigma$ transverse to the bonds on $\Sigma$.  The Dehn twist $D_\gamma$ can be implemented on $H$ by gradually rearranging bonds until a full Dehn twist is achieved.
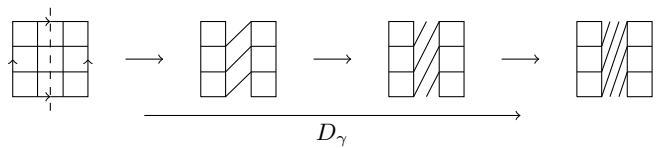
\begin{figure}[hbpt]
\begin{tikzpicture}
\draw [->] (3.75,-0.75) -- (8.75,-0.75);
\node at (6.25,-0.75) [below] {$D_\gamma$};
\draw [->] (2,0) -- (2,0.5) -- (2.5,0.5);
\draw (2.5,0.5) -- (3,0.5) -- (3,0);
\draw [<-] (3,0) -- (3,-0.5) -- (2.5,-0.5);
\draw [<->] (2.5,-0.5) -- (2,-0.5) -- (2,0);
\draw [dashed] (2.5,-0.6875) -- (2.5,0.6875);
\draw (2,0.17) -- (3,0.17);
\draw (2,-0.17) -- (3,-0.17);
\draw (2.33,0.5) -- (2.33,-0.5);
\draw (2.67,0.5) -- (2.67,-0.5);
\draw [->] (3.5,0) -- (4,0);
\draw (4.5,-0.5) -- (4.5,0.5) -- (4.83,0.5) -- (4.83,-0.5) -- (4.5,-0.5);
\draw (4.5,0.17) -- (4.83,0.17); \draw (4.5,-0.17) -- (4.83,-0.17);
\draw (5.17,-0.5) -- (5.17,0.5) -- (5.5,0.5) -- (5.5,-0.5) -- (5.17,-0.5);
\draw (5.17,0.17) -- (5.5,0.17); \draw (5.17,-0.17) -- (5.5,-0.17);
\draw (4.83,0.17) -- (5.17,0.5);
\draw (4.83,-0.17) -- (5.17,0.17);
\draw (4.83,-0.5) -- (5.17,-0.17);
\draw [->] (6,0) -- (6.5,0);
\draw (7,-0.5) -- (7,0.5) -- (7.33,0.5) -- (7.33,-0.5) -- (7,-0.5);
\draw (7,0.17) -- (7.33,0.17); \draw (7,-0.17) -- (7.33,-0.17);
\draw (7.67,-0.5) -- (7.67,0.5) -- (8,0.5) -- (8,-0.5) -- (7.67,-0.5);
\draw (7.67,0.17) -- (8,0.17); \draw (7.67,-0.17) -- (8,-0.17);
\draw (7.33,0.17) -- (7.5,0.5);
\draw (7.33,-0.17) -- (7.67,0.5);
\draw (7.33,-0.5) -- (7.67,0.17);
\draw (7.5,-0.5) -- (7.67,-0.17);
\draw [->] (8.5,0) -- (9,0);
\draw (9.5,-0.5) -- (9.5,0.5) -- (9.83,0.5) -- (9.83,-0.5) -- (9.5,-0.5);
\draw (9.5,0.17) -- (9.83,0.17); \draw (9.5,-0.17) -- (9.83,-0.17);
\draw (10.17,-0.5) -- (10.17,0.5) -- (10.5,0.5) -- (10.5,-0.5) -- (10.17,-0.5);
\draw (10.17,0.17) -- (10.5,0.17); \draw (10.17,-0.17) -- (10.5,-0.17);
\draw (9.83,0.17) -- (9.94,0.5);
\draw (9.83,-0.17) -- (10.06,0.5);
\draw (9.83,-0.5) -- (10.17,0.5);
\draw (9.94,-0.5) -- (10.17,0.17);
\draw (10.06,-0.5) -- (10.17,-0.17);
\end{tikzpicture}
\caption{Adiabatically modifying the interactions in a lattice system can provide a lattice realization of adiabatically varying the geometry of the space
along a non-contractible loop in Teichmuller space. }
\label{fig:twist}
\end{figure}
At this moment $H$ returns to itself and the charge along a fixed loop $\alpha\subset\Sigma$ will now be mapped to charge along $D_\gamma^{-1}(\alpha)$.  In Figure \ref{fig:twist}, $\gamma$ is the vertical dashed arc drawn in the first panel.  Again thinking of $Z_2$-genons makes the analogy to braiding precise.  For example, under branched cover, the conformal moduli spaces $\mathfrak{M}\left(S_4, 4\text{ pts}\right)\cong\mathfrak{M}\left(T^2\right)\cong SL(2,Z)\setminus SL(2,R)/SO(2)$.  For $2g+2$ points, $g>1$, and genus $g$ surfaces, the branched cover arrow, $\longrightarrow$, is onto (only) the hyperelliptic geometries, which is why for $g>2$ the MCG is richer than the corresponding braid group.  However, given a MTC $\mathcal{C}$, and $\mathcal{C}\otimes \mathcal{C}$ represented in the plane with genons, if the rules for measuring charge along simple planar loops permit the breaking of sheet-symmetry, all elements of the projective MCG representation can be realized.

\section{Topological charge projections and adiabatic unitary operations}
\label{tprojSec}

In this paper, we require, as input, the ability to guide the state of the system into subspaces associated with definite
values of topological charge along certain non-contractible cycles of interest. In general, this is implemented
as a projection operator, which corresponds to a measurement of topological charge. As we will describe, the
particular projections of interest to us here can also be implemented through unitary adiabatic evolution.

Let $P_{\omega}^{(a)}$ denote the projection of a state $|\psi\rangle \in \mathcal{V}_g$ into a
subspace corresponding to the topological charge value $a$ associated with the non-contractible cycle $\omega$.
For example, on a torus, we have
\begin{align}
P_{\alpha}^{(a)} | b \rangle_{\alpha} = \delta_{ab} |b \rangle_{\alpha} .
\end{align}
On a genus $2$ surface, we have
\begin{align}
P_{\omega_3}^{(d)} |a b c; \mu \nu \rangle_{\omega_1,\omega_2, \omega_3} = \delta_{dc} |a b c; \mu \nu \rangle_{\omega_1,\omega_2,\omega_3} .
\end{align}
Below, we will describe how the operator $P_{\omega}^{(a)}$ can be implemented through an adiabatic process. This is similar
to the case where the collective measurement of the fusion channel of a pair of non-Abelian anyons can also be obtained
by adiabatically tuning the interactions between them.\cite{bonderson2013braiding}

The ground state degeneracy on a genus $g$ surface will, in finite-size systems, generically acquire exponentially small
splittings. In a finite-size system, there is an effective Hamiltonian $H_{\text{eff}}$ that acts on the ground state
subspace $\mathcal{V}_g$. In general, $H_{\text{eff}}$ will consist of all possible instanton processes, corresponding to virtual tunneling of
quasiparticles along various closed paths (see, e.g., Ref. \onlinecite{wen1990b}). The amplitude for each such process is
exponentially small in the ratio of the length of the path to the correlation length of the system.

As an example, let us first consider the case of a topological phase of matter on a torus, with
an effective Hamiltonian that acts on the ground state subspace $\mathcal{V}_1$ given by
\begin{align}
H_{\text{eff}} =  \sum_{a \in \mathcal{C}} [t^a_\alpha W_a(\alpha) + t^a_\beta W_a(\beta) + H.c.],
\end{align}
where $\alpha$ and $\beta$ are the longitudinal and meridianal cycles of the torus.
The operators $W_a(\omega)$ correspond to the process where a quasiparticle / anti-particle pair of type $a, \overline{a}$ are created out of the vacuum,
one of them tunnels virtually along the cycle $\omega$, and they reannihilate. Here we neglect the Wilson loop operators along
the diagonal loop $W_a(\alpha + \beta)$, as the corresponding amplitudes $t^a_{\alpha+\beta}$ are typically exponentially suppressed
compared to $t^a_\alpha$, $t^a_\beta$ due to the length of the loops. This situation could, however, be changed if desired by locally tuning the
energy gap along $\alpha + \beta$.

The effect of $W_a(\omega)$ on the ground state subspace $\mathcal{V}_1$ can be determined from the modular $S$ matrix. In general,
\begin{align}
W_a(\alpha) |b \rangle_\alpha &= \frac{S_{ab}}{S_{b0}} |b \rangle_\alpha,
\nonumber \\
W_a(\alpha) |b \rangle_\beta &= \sum_c N_{ab}^c |c \rangle_\beta
\end{align}

The tunneling amplitudes $t^a_\alpha$ are non-universal quantities and depend on the microscopic details of the system.
For a given path of length $L$, the corresponding amplitude for a quasiparticle to tunnel along the path is proportional to
$e^{-L / \xi_a}$, where $\xi_a$ is a finite correlation length that can depend on the quasiparticle type $a$.
The tunneling amplitude $t^a_\alpha$ is the sum over the amplitudes for quasiparticles of type $a$ to tunnel along
all possible paths that are topologically equivalent to the cycle $\alpha$.

Let us suppose that the tunneling amplitudes $t^a_\alpha$ can be adiabatically varied. This could be done, for example, by tuning
microscopic parameters along the loop $\alpha$ in such a way as to locally tune the energy gap along $\alpha$.
We then consider a time-dependent Hamiltonian
\begin{align}
H_{\text{eff}}(\tau) = &(1-\tau) \sum_a [t^a_\alpha W_a(\alpha) + H.c.]
\nonumber \\
& + \tau \sum_a [t^a_\beta W_a(\beta) + H.c.]
\end{align}
Let us denote the ground state of $H_{\text{eff}}(\tau)$ as $|\Psi(\tau) \rangle$. It is clear that
\begin{align}
|\Psi(0) \rangle = |b_0 \rangle_\alpha,
\end{align}
where $b_0$ is such that the ground state energy $E = \sum_a t^a_\alpha \frac{S_{ab_0}}{S_{0b_0}} + c.c.$ is minimized.
If there are multiple such $b_0$, then the ground state is degenerate. However generically there should be a unique such $b_0$
unless there is fine-tuning of the parameters $t^a_\alpha$, due to braiding non-degeneracy (which implies unitarity of the $S$ matrix).
Similarly,
\begin{align}
|\Psi(1) \rangle = | b_1 \rangle_\beta,
\end{align}
where $b_1$ is such that $E = \sum_a t^a_\beta \frac{S_{ab_1}}{S_{0b_1}} + c.c.$ is minimized. Again, the choice $b_1$ will generically be unique
unless there is fine-tuning of the tunneling amplitudes $t^a_\beta$.

In fact, aside from accidental degeneracies which can be removed by perturbing the Hamiltonian,
$H_{\text{eff}}$ will generically have a unique ground state for all $\tau$. Therefore, we see that
tuning the tunneling amplitudes can cause the system to be tuned adiabatically between
states of definite topological charge along any desired non-contractible cycle. For
$S_{b_1 b_0} \neq 0$, the above adiabatic process can be expressed in terms of the projector:
\begin{align}
|\Psi(1) \rangle = e^{i\phi} \frac{1}{S_{b_1 b_0}} P_\beta^{(b_1)} | \Psi(0) \rangle ,
\end{align}
up to a non-universal overall phase $e^{i\phi}$.

For more details about implementing the above adiabatic evolution in a concrete system in the context of electron-hole bilayer FQH states,
we refer to Ref. \onlinecite{barkeshli2016bilayersc}, which utilizes the ideas presented above.

Applying the above idea to genus $g$ surfaces, we see that it is possible to implement the projections $P_\omega^{(a)}$ by adiabatically
tuning the exponentially small amplitudes for quasiparticles to tunnel along the various non-contractible cycles.

Diabatic errors associated with such protocols can be managed, as discussed recently in the context of adiabatic braiding operations of non-Abelian
anyons in Ref. \onlinecite{knapp2016}.

\section{Implementing Dehn twists through topological charge projections}
\label{Dehn}

Let us consider starting with a generic state $|\Psi \rangle \in \mathcal{V}_1$:
\begin{align}
|\Psi \rangle = \sum_{a \in \mathcal{C}} \psi_a |a \rangle_\beta
\end{align}
The effect of a Dehn twist $D_\beta$ around $\beta$ is:
\begin{align}
D_\beta |\Psi \rangle = \sum_{a \in \mathcal{C}} e^{i \theta_a} \psi_a |a \rangle_\beta,
\end{align}
where $\theta_a$ is the topological spin of $a$. We wish to demonstrate a protocol for implementing
the Dehn twist $D_\beta$ through a sequence of topological charge projections.

In order to perform $D_\beta$ as described above, we first note that we can embed
$|\Psi\rangle$ into a larger space, on a genus 2 surface, $\mathcal{V}_2$ :
\begin{align}
\label{psieqn}
|\tilde{\Psi} \rangle = \sum_{a \in \mathcal{C}} \psi_a |a 0 0\rangle_{\beta_1 \delta_1 \beta_2},
\end{align}
where we refer to Fig. \ref{statesFig} for the notation.
Formally, there is an inclusion $\mathcal{V}_g \otimes \mathcal{V}_1 \hookrightarrow \mathcal{V}_{g+1}$, such that
$|\Psi\rangle \otimes |0\rangle_\beta \hookrightarrow |\tilde{\Psi} \rangle$.
The additional genus in this case can be thought of as giving rise to an ``ancilla" degree of freedom,
which is useful for performing the operation. In this case, the Dehn twists along $\beta_1$ and $\gamma$
are equivalent:
\begin{align}
\label{DbetaDgamma}
D_{\beta_1} |\tilde{\Psi}\rangle = D_{\gamma} |\tilde{\Psi}\rangle = \sum_a e^{i \theta_a} \psi_a |a00\rangle_{\beta_1 \delta_1 \beta_2}.
\end{align}

In what follows, we will show that $D_{\gamma}$ can be obtained up to overall phase by a series of projections:
\begin{align}
\label{result}
D^\dagger_\gamma |\tilde{\Psi}\rangle &= \mathcal{D}^3 e^{-2\pi i c/8} P_{\beta_2} P_{\gamma + \alpha_2} P_{\alpha_2} |\tilde{\Psi} \rangle,
\nonumber \\
D_\gamma |\tilde{\Psi}\rangle &= \mathcal{D}^3 e^{2\pi i c/8} P_{\beta_2} P_{\alpha_2} P_{\gamma + \alpha_2} |\tilde{\Psi} \rangle,
\end{align}
where $P_{\omega} \equiv P_{\omega}^{(0)}$ is the projection onto the identity topological charge, $c$ is the chiral
central charge of the topological phase, and $\mathcal{D} = \sqrt{\sum_a d_a^2}$ is the total quantum dimension.

The above statements can be immediately generalized: a Dehn twist $D_\omega$ along an arbitrary loop $\omega$ acting on a state $|\Psi\rangle \in \mathcal{V}_g$
on a genus $g$ surface can be implemented as follows. We first embed $|\Psi\rangle$ into a larger space $\mathcal{V}_{g+1}$, on a genus $g+1$ surface by
considering the state $|\tilde{\Psi}\rangle$, which satisfies
\begin{align}
P_{\beta_{g+1}} |\tilde{\Psi}\rangle = |\tilde{\Psi}\rangle.
\end{align}
Then,
\begin{align}
D_\omega^\dagger | \tilde{\Psi}\rangle &= \mathcal{D}^3 e^{-2\pi i c/8} P_{\beta_{g+1}} P_{\alpha_{g+1} + \omega} P_{\alpha_{g+1}} |\tilde{\Psi}\rangle,
\nonumber \\
D_\omega| \tilde{\Psi}\rangle &= \mathcal{D}^3 e^{2\pi i c/8} P_{\beta_{g+1}} P_{\alpha_{g+1}} P_{\alpha_{g+1}+\omega} |\tilde{\Psi}\rangle.
\end{align}

To phrase the above result differently, consider three cycles $\omega_1, \omega_2, \omega_3$, which have the property that
$\omega_i$ and $\omega_j$ intersect exactly once (for $i \neq j$). If we then start with a state $|\tilde{\Psi} \rangle $ which satisfies
$P_{\omega_1} |\tilde{\Psi}\rangle = |\tilde{\Psi}\rangle$, we have
\begin{align}
D_{\omega_3 - \omega_2}^\dagger |\tilde{\Psi} \rangle &= \mathcal{D}^3 e^{-2\pi i c/8} P_{\omega_1} P_{\omega_2} P_{\omega_3} |\tilde{\Psi}\rangle,
\nonumber \\
D_{\omega_3 - \omega_2} |\tilde{\Psi} \rangle &= \mathcal{D}^3 e^{2\pi i c/8} P_{\omega_1} P_{\omega_3} P_{\omega_2} |\tilde{\Psi}\rangle.
\end{align}

Below, we will establish the above result in two ways, first through a calculation using the algebraic theory of anyons,
and then by studying in detail the topology of the space-time history of the surface.

\section{Algebraic calculation}
\label{algebraicSec}

In order to understand the effect of the projections of interest, note that
$P_{\alpha_2} | a b c \rangle_{\beta_1 \delta_1 \alpha_2} = \delta_{b0} \delta_{c0} |a 0 0 \rangle_{\beta_1 \delta_1 \alpha_2}$.
This implies
\begin{align}
P_{\alpha_2} | a  b  c \rangle_{\beta_1 \delta_1 \beta_2}.
= \delta_{b0} \frac{d_c}{\mathcal{D}^2} \sum_{c' \in \mathcal{C}} d_{c'} |a 0 c'\rangle_{\beta_1 \delta_1 \beta_2} .
\end{align}

Furthermore, we also have the following relation,
\begin{align}
P_{\gamma + \alpha_2} = D_{\gamma}^\dagger P_{\alpha_2} D_{\gamma} .
\end{align}


We will use the following identity (Gauss-Milgram sum):
\begin{align}
\label{GMsum}
\frac{1}{\mathcal{D}} \sum_{a \in \mathcal{C}} d_a^2 e^{i \theta_a} = e^{2\pi i c/8}
\end{align}

\subsection{Abelian Phases}

At this stage, it is simpler to first consider the case where the topological phase is Abelian. In this case,
\begin{align}
D_{\gamma} |a 0 c\rangle_{\beta_1 \delta_1 \beta_2} = e^{i\theta_{a - c}} |a0c\rangle_{\beta_1 \delta_1 \beta_2},
\end{align}
and $d_a = 1$ for all $a \in \mathcal{C}$. Thus, we find for the Abelian case:
\begin{align}
P_{\beta_2} P_{\gamma + \alpha_2} &P_{\alpha_2} |a00 \rangle_{\beta_1 \delta_1 \beta_2} = P_{\beta_2} D_{\gamma}^\dagger P_{\alpha_2} D_{\gamma}
\sum_{c'} \frac{1}{\mathcal{D}^2} |a0c' \rangle_{\beta_1 \delta_1 \beta_2}
\nonumber \\
&= P_{\beta_2} D_{\gamma}^\dagger P_{\alpha_2} \sum_{c'} e^{i\theta_{a-c'}} \frac{1}{\mathcal{D}^2} |a 0 c'\rangle_{\beta_1 \delta_1 \beta_2}
\nonumber \\
&= P_{\beta_2} D_{\gamma}^\dagger \sum_{c',d'} e^{i\theta_{a-c'}} \frac{1}{\mathcal{D}^4} |a 0 d' \rangle_{\beta_1 \delta_1 \beta_2}
\nonumber \\
&= P_{\beta_2} \sum_{c',d'} e^{i\theta_{a-c'}} e^{-i\theta_{a-d'}} \frac{1}{\mathcal{D}^4} |a 0 d' \rangle_{\beta_1 \delta_1 \beta_2}
\nonumber \\
&= e^{-i\theta_{a}} \frac{1}{\mathcal{D}^4} \sum_{c'} e^{i\theta_{a-c'}}  |a 00 \rangle_{\beta_1 \delta_1 \beta_2}
\nonumber \\
&= \frac{1}{\mathcal{D}^3} e^{-i\theta_a} e^{2\pi i c/8} |a 0 0 \rangle_{\beta_1 \delta_1 \beta_2}
\end{align}
where to obtain the last line we have used Eq. (\ref{GMsum}). This then implies the result, Eq. (\ref{result}).

\subsection{Non-Abelian Phases}

Let us now consider the more complicated case where the topological phase is non-Abelian. Note that
\begin{align}
D_{\gamma} |ab'c;\mu\nu \rangle_{\beta_1 \gamma_1 \beta_2} = e^{i\theta_{b'}} |a b'c;\mu \nu \rangle_{\beta_1 \gamma_1 \beta_2}
\end{align}
\begin{widetext}
Thus,
\begin{align}
D_{\gamma}^\dagger |abc;\mu \nu \rangle_{\beta_1 \delta_1 \beta_2}
&= \sum_{b',\mu'\nu'} e^{-i\theta_{b'}} [F^{\bar{a} a \bar{c}}_{\bar{c}}]_{(\bar{b}\mu\nu)(b'\mu'\nu')} |ab'c;\mu'\nu'\rangle_{\beta_1 \gamma_1 \beta_2}
\nonumber \\
&= \sum_{b',\mu',\nu', d,\alpha,\beta} e^{-i\theta_{b'}} [F^{\bar{a} a \bar{c}}_{\bar{c}}]_{(\bar{b}\mu\nu)(b'\mu'\nu')} [F^{\bar{a} a \bar{c}}_{\bar{c}}]^\dagger_{(b'\mu'\nu')(d\alpha\beta)} |adc;\alpha\beta\rangle_{\beta_1 \delta_1 \beta_2}
\end{align}
In particular,
\begin{align}
D_{\gamma}^\dagger |a0c \rangle_{\beta_1 \delta_1 \beta_2}
&= \sum_{\{b'| N_{a \bar{c}}^{b'}\neq 0\}} \sum_{\mu', d,\alpha,\beta} e^{-i\theta_{b'}} \sqrt{\frac{d_{b'}}{d_a d_c}} [F^{\bar{a} a \bar{c}}_{\bar{c}}]^\dagger_{(b'\mu'\mu')(d\alpha\beta)} |adc;\alpha\beta\rangle_{\beta_1 \delta_1 \beta_2}
\end{align}
Thus,
\begin{align}
P_{\beta_2} P_{\gamma + \alpha_2} P_{\alpha_2} |a00 \rangle_{\beta_1 \delta_1 \beta_2} &= P_{\beta_2} D_{\gamma}^\dagger P_{\alpha_2} D_{\gamma}
\sum_{c'} \frac{d_{c'}}{\mathcal{D}^2} |a0c' \rangle_{\beta_1 \delta_1 \beta_2}
\nonumber \\
&= P_{\beta_2} D_{\gamma}^\dagger P_{\alpha_2} \sum_{c',d} \sum_{\{b'|N_{a\bar{c'}}^{b'} \neq 0\}} \sum_{\mu',\alpha,\beta} \frac{d_{c'}}{\mathcal{D}^2} e^{i \theta_{b'}} \sqrt{\frac{d_{b'}}{d_ad_{c'}}} [F^{\bar{a}a\bar{c'}}_{\bar{c'}}]^{\dagger}_{(b'\mu'\mu')(d\alpha\beta)} |a d c';\alpha\beta \rangle_{\beta_1 \delta_1 \beta_2}
\nonumber \\
&= P_{\beta_2} D_{\gamma}^\dagger \sum_{c',e} \sum_{\{b'|N_{a\bar{c'}}^{b'} \neq 0\}} \sum_{\mu'} \frac{d_e}{\mathcal{D}} \frac{d_{c'}^2}{\mathcal{D}^3} e^{i \theta_{b'}} \sqrt{\frac{d_{b'}}{d_ad_{c'}}} [F^{\bar{a}a\bar{c'}}_{\bar{c'}}]^{\dagger}_{(b'\mu'\mu')(000)} |a 0 e \rangle_{\beta_1 \delta_1 \beta_2}
\nonumber \\
&= P_{\beta_2} D_{\gamma}^\dagger \sum_{c',e} \sum_{\{b'|N_{a\bar{c'}}^{b'} \neq 0\}} \sum_{\mu'} \frac{d_e}{\mathcal{D}} \frac{d_{c'}}{\mathcal{D}^3} e^{i \theta_{b'}} \frac{d_{b'}}{d_a} |a 0 e \rangle_{\beta_1 \delta_1 \beta_2}
\nonumber \\
&= P_{\beta_2}  \sum_{c',e} \sum_{f',g,\kappa,\delta,\sigma} \sum_{\{b'|N_{a\bar{c'}}^{b'} \neq 0\}} \sum_{\mu'} \frac{d_e}{\mathcal{D}} \frac{d_{c'}}{\mathcal{D}^3}
e^{i \theta_{b'}} \frac{d_{b'}}{d_a}
 e^{-i\theta_{f'}} \sqrt{\frac{d_{f'}}{d_a d_e}} [F^{\bar{a} a \bar{e}}_{\bar{e}}]^\dagger_{(f'\kappa\kappa)(g\delta\sigma)}|a g e;\delta \sigma
\rangle_{\beta_1 \delta_1 \beta_2}
\nonumber \\
&=  \sum_{c'} \sum_{\{b'|N_{a\bar{c'}}^{b'}\neq 0\}} \sum_{\mu'}  \frac{d_{c'}}{\mathcal{D}^4} e^{i \theta_{b'}} \frac{d_{b'}}{d_a}
e^{-i\theta_{a}} [F^{\bar{a} a 0}_{0}]^\dagger_{(a00)(000)}|a 0 0 \rangle_{\beta_1 \delta_1 \beta_2}
\\
\label{na0}
&=  \sum_{c'} \sum_{\{b'|N_{a\bar{c'}}^{b'} \neq 0\}}  \sum_{\mu'} \frac{1}{\mathcal{D}^4} e^{i \theta_{b'}} \frac{d_{c'} d_{b'}}{d_a}
 e^{-i\theta_{a}} |a 0 0\rangle_{\beta_1 \delta_1 \beta_2}
\\
\label{na1}
&=   \sum_{c'} \sum_{b'}  \frac{1}{\mathcal{D}^4} e^{i \theta_{b'}} \frac{N_{a \bar{c'}}^{b'} d_{c'} d_{b'}}{d_a}
 e^{-i\theta_{a}} |a 0 0  \rangle_{\beta_1 \delta_1 \beta_2}
 \\
 \label{na2}
&=  \sum_{c'} \sum_{b'}  \frac{1}{\mathcal{D}^4} e^{i \theta_{b'}} \frac{N_{a \bar{b'}}^{c'} d_{c'} d_{b'}}{d_a}
 e^{-i\theta_{a}} |a 0 0  \rangle_{\beta_1 \delta_1 \beta_2}
 \\
 \label{na3}
&=  \sum_{b'}  \frac{1}{\mathcal{D}^4} e^{i \theta_{b'}} d_{b'}^2
e^{-i\theta_{a}} |a 0 0  \rangle_{\beta_1 \delta_1 \beta_2}
  \\
  \label{na4}
&=   \frac{1}{\mathcal{D}^3} e^{2\pi i c/8} e^{-i\theta_{a}} |a 0 0 \rangle_{\beta_1 \delta_1 \beta_2},
\end{align}
\end{widetext}
which establishes the result, Eq. (\ref{result}). The first several equalities are obtained by straightforward applications of the
projectors $P_{\alpha_2}$ and $P_{\beta_2}$, and the Dehn twists $D_{\gamma}$, as described in the previous sections.
To get Eq. (\ref{na0}) from the preceding equation, we used the fact that $[F^{\bar{a} a 0}_{0}]^\dagger_{(a00)(000)} = 1$.
Eq. (\ref{na1}) is obtained from Eq. (\ref{na0}) by replacing the restricted sum over $b$ and the fusion
channel index $\mu$, $\sum_{\{b'|N_{a\bar{c'}}^{b'} \neq 0\}}  \sum_{\mu'}$, by the unrestricted sum $\sum_{b'} N_{a \bar{c'}}^{b'}$. Eq. (\ref{na2})
is obtained from Eq. (\ref{na1}) using the identity $N_{a \bar{c'}}^{b'} = N_{a \bar{b'}}^{c'}$. Eq. (\ref{na3}) is obtained from Eq. (\ref{na2})
using the identity $\sum_{c'} N_{a \bar{b'}}^{c'} d_{c'} = d_a d_{\bar{b'}}$, together with the fact that $d_b = d_{\bar{b}}$. Finally, Eq. (\ref{na4})
is obtained from Eq. (\ref{na3}) by using the Gauss-Milgram sum, Eq. (\ref{GMsum}).

\subsection{Projections onto other topological charge sectors}
\label{algebraicSecOther}

Let us consider now a slightly more general setup. We consider the projectors $P_{\alpha}^{(b)}$, which project onto the topological charge $b$. The previous considerations
are associated with taking $b = 0$.

Let us consider starting with a state
\begin{align}
|a 0 b_0\rangle_{\beta_1 \delta_1 \beta_2}.
\end{align}
Note that
\begin{align}
P_{\alpha_2}^{(b)} | a 0 c \rangle_{\beta_1 \delta_1 \alpha_2} = \delta_{cb} |a 0 b \rangle_{\beta_1 \delta_1 \alpha_2}
\end{align}
This implies
\begin{align}
P_{\alpha_2}^{(b)} | a 0 c \rangle_{\beta_1 \delta_1 \beta_2}
&= S_{cb} \sum_{c'} S^{\dagger}_{bc'} |a 0 c'\rangle_{\beta_1 \delta_1 \beta_2}
\end{align}


Now, it useful to note that
\begin{align}
P_{\gamma + \alpha_2}^{(b)} = D_{\gamma}^\dagger P_{\alpha_2}^{(b)} D_{\gamma}
\end{align}

Therefore, our task is to compute
\begin{align}
P_{\beta_2}^{(b_3)} D_\gamma^\dagger P_{\alpha_2}^{(b_2)} D_{\gamma}  S_{b_0b_1} \sum_{c'}
S^\dagger_{b_1c'} |a0c' \rangle_{\beta_1 \delta_1 \beta_2}
\end{align}

\subsubsection{Abelian Phases}

Let us now specialize to the Abelian case. We use the following identities:
\begin{align}
S_{ab} = \frac{1}{\mathcal{D}} e^{i \theta_{a,b}},
\end{align}
where $\theta_{a,b}$ is the mutual statistics between $a$ and $b$, and
\begin{align}
\theta_{a+b} = \theta_a + \theta_b + \theta_{a,b} .
\end{align}

Following the preceding calculation together with repeated application of the above identities, we obtain
\begin{widetext}
\begin{align}
P_{\beta_2}^{(b_3)} D_\gamma^\dagger P_{\alpha_2}^{(b_2)} & D_{\gamma}  S_{b_0b_1} \sum_{c'}
S^\dagger_{b_1,c'} |a0c' \rangle_{\beta_1 \delta_1 \beta_2}
\nonumber \\
&= P_{\beta_2}^{(b_3)} D_\gamma^\dagger P_{\alpha_2}^{(b_2)} S_{b_0b_1} \sum_{c'}
S^\dagger_{b_1c'} e^{i \theta_{a-c'}} |a0c' \rangle_{\beta_1 \delta_1 \beta_2}
\nonumber \\
&=  P_{\beta_2}^{(b_3)} D_\gamma^\dagger
S_{b_0b_1} \sum_{c'} S^\dagger_{b_1c'} e^{i \theta_{a-c'}} S_{c'b_2} \sum_{d'} S^\dagger_{b_2 d'} |a 0 d'\rangle_{\beta_1 \delta_1 \beta_2}
\nonumber \\
&=  P_{\beta_2}^{(b_3)}
S_{b_0b_1} \sum_{c'} S^\dagger_{b_1 c'} e^{i \theta_{a-c'}} S_{c' b_2} \sum_{d'} S^\dagger_{b_2 d'} e^{-i \theta_{a-d'}} |a 0 d'\rangle_{\beta_1 \delta_1 \beta_2}
\nonumber \\
&= S_{b_0 b_1} \sum_{c'} S^\dagger_{b_1 c'} e^{i \theta_{a-c'}} S_{c' b_2} S^\dagger_{b_2 b_3} e^{-i \theta_{a-b_3}} |a 0 b_3\rangle_{\beta_1 \delta_1 \beta_2}
\nonumber \\
&= \frac{1}{\mathcal{D}^4} e^{i\theta_{b_0,b_1} - i \theta_{b_2,b_3}} \sum_{c'} e^{-i\theta_{b,c'} + i \theta_{a-c'} + i \theta_{c',b_2}} e^{-i\theta_{a-b_3}}
|a 0 b_3\rangle_{\beta_1 \delta_1 \beta_2}
\nonumber \\
&= \frac{1}{\mathcal{D}^4} e^{i\theta_{b_0,b_1} - i \theta_{b_2,b_3}} \sum_{c'} e^{i\theta_{c',b_2-b_1-a} + i \theta_{c'}} e^{-i\theta_{b_3} + i\theta_{a,b_3}}
|a 0 b_3\rangle_{\beta_1 \delta_1 \beta_2}
\nonumber \\
&= \frac{1}{\mathcal{D}^4} e^{i\theta_{b_0,b_1} - i \theta_{b_2,b_3}-i\theta_{b_3}} e^{i\theta_{a,b_3}} \sum_{c'} e^{i\theta_{c'+b_2-b_1-a} - i \theta_{b_2-b_1-a}}
|a 0 b_3\rangle_{\beta_1 \delta_1 \beta_2}
\nonumber \\
&= \frac{1}{\mathcal{D}^3} e^{2\pi i c/ 8} e^{i\theta_{b_0,b_1} - i \theta_{b_2,b_3}-i\theta_{b_3}} e^{i\theta_{a,b_3}} e^{- i \theta_{b_2-b_1-a}}
|a 0 b_3\rangle_{\beta_1 \delta_1 \beta_2}
\nonumber \\
&= \frac{1}{\mathcal{D}^3} e^{2\pi i c/ 8} e^{i\theta_{b_0,b_1} - i \theta_{b_2,b_3}-i\theta_{b_3} - i \theta_{b_2-b_1}} e^{-i\theta_a} e^{i\theta_{a,b_3+b_2-b_1}}
|a 0 b_3\rangle_{\beta_1 \delta_1 \beta_2}
\nonumber \\
&= \frac{1}{\mathcal{D}^3} e^{2\pi i c/ 8} e^{i\theta_{b_0,b_1} - i \theta_{b_2,b_3}-i\theta_{b_3} - i \theta_{b_2-b_1} + i \theta_{b_1-b_2-b_3}} e^{-i\theta_{a+b_1-b_2-b_3}}
|a 0 b_3\rangle_{\beta_1 \delta_1 \beta_2}
\nonumber \\
&= \frac{1}{\mathcal{D}^3} e^{2\pi i c/ 8} e^{i\theta_{b_0-b_3,b_1}} e^{-i\theta_{a+b_1-b_2-b_3}}
|a 0 b_3\rangle_{\beta_1 \delta_1 \beta_2}
\end{align}

The final result can be rewritten as:
\begin{align}
W^\dagger_{b_1 - b_2 - b_3}(\alpha_1) D_\gamma^\dagger W_{b_1 - b_2 - b_3}(\alpha_1) |a 0 b_3\rangle_{\beta_1 \delta_1 \beta_2}
= \mathcal{D}^3 e^{-2\pi i c/8} e^{ i\theta_{b_3-b_0,b_1} }
P_{\beta_2}^{(b_3)} P_{\gamma + \alpha_2}^{(b_2)} P_{\alpha_2}^{(b_1)} |a 0 b_3\rangle_{\beta_1 \delta_1 \beta_2}
\end{align}
\end{widetext}

The analogous calculation of $P_{\beta_2}^{(b_3)} P_{\gamma + \alpha_2}^{(b_2)} P_{\alpha_2}^{(b_1)} |a0b_3\rangle $ can now also be performed for
a general non-Abelian topological phase as well. However the computation quickly becomes quite complex. Below, we introduce a different way to evaluate
the result of the projections, by studying the topology of the space-time history of the surface. In addition to providing a new persepective, this
method allows one to express the results in a simpler fashion.

\section{Topology of space-time history}
\label{topoSec}


We have shown that a succession of measurements along time constant Wilson loops $\{\omega_i\}\subset\Sigma$ can be used to realize a Dehn twist.
As explained in Refs. \onlinecite{bonderson2013twisted, bonderson2016twisted} this can be given a $3$-dimensional interpretation as framed surgeries $\mathfrak{S}_i$ on $\Sigma\times[0,1]_\text{time}$ (and further $4$-dimensional interpretation which we will soon come to).  The 3D interpretation is that around each $\omega_i$ a solid torus $S_i\subset\Sigma\times[0,1]$ is deleted and another solid torus $S_i^\prime$ is glued back.  The gluing back is done so that the meridianal curve on $S_i^\prime$ (i.e. the isotopy class of curve that bounds a disk in $S_i^\prime$) glues to an in-surface parallel $\omega_i^\prime$ to $\omega_i$ where $\omega_i^\prime$, $\omega_i\in\Sigma\times t_i$.  This is the interpretation when the \emph{trivial} charge is measured along $\omega_i$---meaning that when $\Sigma\times t_i$ is cut along $\omega_i$ the trivial charge appears on the two new boundary circles, $\omega_i^\prime$ and $\omega_i^{\prime\prime}$.  It is not difficult to understand why this is so: trivial charge on $\omega_i^\prime$ is equivalent to the topological ground state being extendable over a disk bounding $\omega_i^\prime$.  The solid torus $S_i^\prime$ is nothing more than a circle's worth, the circle being the normal linking circle to $\omega_i\subset \Sigma\times[0,1]$, of disks $D_\theta$ in the TQFT ground state.  Similarly,
if instead of the trivial charge being measured along $\omega_i$, the topological charge $a_i$ is measured, then the glued back solid torus $S_i^\prime$ will have a charge $a_i$ Wilson loop at its core corresponding to the fact that each $D_\theta$ will now have an $a_i$-anyon at its center.  We denote the manifold after the regluings on $\{\omega_i\}$, the surgered manifold, by $\mathfrak{S}(\Sigma\times I)$.  However, in what follows it will not contain $a_i$-Wilson loops.

The 3D picture contains the requisite pair of PSs.  The first is easy: the vertical time lines, they identify $\Sigma_0$ with $\Sigma_1$.  To get a unitary we need to locate a second PS in the complement of some balls inside $\mathfrak{S}(\Sigma\times I)$.  We need to find an embedding $i$ where $f$ can be an interesting map, for example a Dehn twist $D_{\beta_1}$, for $\beta_1\subset\Sigma$, the case we treat in detail.
\begin{figure}[hbpt]
\begin{tikzpicture}
\node (1) at (0,0) {$\Sigma\times 1$};
\node (2) at (0,-1) {$\Sigma\times I\setminus\text{balls}$};
\node (3) at (0,-2) {$\Sigma\times 0$};
\node (4) at (3,0) {$\Sigma\times 1$};
\node (5) at (3,-1) {$\mathfrak{S}(\Sigma\times I)$};
\node (6) at (3,-2) {$\Sigma\times 0$};
\path [right hook->] (1) edge (2);
\path [right hook->] (3) edge (2);
\path [->] (1) edge node [above] {$f$} (4);
\path [right hook->] (2) edge node [above] {$i$} (5);
\path [<->] (3) edge node [above] {$id$} (6);
\path [right hook->] (4) edge (5);
\path [right hook->] (6) edge (5);
\end{tikzpicture}
\caption{}
\label{fig:embedding}
\end{figure}

Figures \ref{fig:neighborhood1} and \ref{fig:neighborhood2} show an annulary neighborhood $\mathfrak{N}\left(\beta_1\right)$ of any $\beta_1\subset\Sigma$, expanded by creating a small nearby handle with meridian $\beta_2$ carrying trivial charge.
\begin{changemargin}{2cm}{2cm}
\begin{figure*}
\begin{tikzpicture}[scale=2]
\draw (0.2,0) arc (0:180:0.2 and 0.1);
\draw (0.1,0) arc (0:-180:0.1 and 0.05);
\draw (-1,0.6) to [out=10, in=170] (0,0.75) to [out=0, in=90] (1,0)
    to [out=-90, in=0] (0,-0.75) to [out=170, in=10] (-1,-0.6);
\draw (-1,0.4) ellipse (0.15 and 0.2);
\draw (-1,-0.4) ellipse (0.15 and 0.2);
\draw (-1,0.2) arc (90:-90:0.15 and 0.2);
\draw (-0.85,0.4) to [out=0, in=0] (-0.85,-0.4);
\draw [dashed] (-0.2,0) arc (0:180:0.325 and 0.108);
\draw (-0.2,0) arc (0:-180:0.325 and 0.108);
\draw [dashed] (0,0.125) to [out=180, in=90] (-0.85,0);
\draw (-0.85,0) to [out=-90, in=-90] (0.75,0) to [out=90, in=0] (0.375,0.375) to
    [out=180, in=0] (0,0.125);
\draw (0,0) ellipse (0.5625 and 0.3125);
\draw [dashed] (1,0) arc (0:180:0.375 and 0.125);
\draw (1,0) arc (0:-180:0.375 and 0.125);
\node at (-2,0) {Notation:};
\node at (1,0) [right] {$\beta_2$};
\node at (0,-0.58) {$\alpha_2+\gamma$};
\node at (0,0.3125) [above] {$\alpha_2$};
\node at (-0.65,0.3125) [above] {$\delta$};
\node at (-0.2,-0.17) {$\gamma$};
\node [align=left] at (2.5,0) {whenever $\delta$ slides over\\$x$ using ``part of
    $y$,''\\we darken the part of\\$y$ involved};
\node at (0,0.9) [above] {2D-handle slide perspective};
\node at (0,-0.9) [below] {$D_\gamma\vert \tilde{\Psi} \rangle = P_{\beta_2} P_{\alpha_2}  P_{\alpha_2 + \gamma} \vert\tilde{\Psi} \rangle$ (up to phase and normalization)};
\end{tikzpicture}\vspace{.1in}
\begin{tikzpicture}[scale=2]
\draw (0.2,0) arc (0:180:0.2 and 0.1);
\draw (0.1,0) arc (0:-180:0.1 and 0.05);
\draw (-1,0.6) to [out=10, in=170] (0,0.75) to [out=0, in=90] (1,0)
    to [out=-90, in=0] (0,-0.75) to [out=170, in=10] (-1,-0.6);
\draw (-1,0.4) ellipse (0.15 and 0.2);
\draw (-1,-0.4) ellipse (0.15 and 0.2);
\draw (-1,0.2) arc (90:-90:0.15 and 0.2);
\draw (-0.85,0.4) to [out=0, in=0] (-0.85,-0.4);
\draw [dashed] (0,0.125) to [out=180, in=90] (-0.85,0);
\draw (-0.85,0) to [out=-90, in=150] (-0.69,-0.31);
\draw [line width=2] (-0.69,-0.31) to [out=-30, in=-110] (0.73,-0.12);
\draw (0.73,-0.12) [out=-110, in=-90] to (0.75,0) to [out=90, in=0] (0.375,0.375) to
    [out=180, in=0] (0,0.125);
\draw (0,0) ellipse (0.5625 and 0.3125);
\draw [dashed] (1,0) arc (0:180:0.375 and 0.125);
\draw (1,0) arc (0:-180:0.375 and 0.125);
\node at (-0.45,0.25) [above] {$\delta_\text{initial}$};
\node at (0,-0.9) [below] {(a)};
\draw [->] (1.25,0) -- (2.25,0);
\node [align=left] at (1.75,0) {slide $\delta$ over\\$\beta_2$ using\\
    part of\\$\alpha_2+\gamma$};
\end{tikzpicture}
\begin{tikzpicture}[scale=2]
\draw (0.2,0) arc (0:180:0.2 and 0.1);
\draw (0.1,0) arc (0:-180:0.1 and 0.05);
\draw (-1,0.6) to [out=10, in=170] (0,0.75) to [out=0, in=90] (1,0)
    to [out=-90, in=0] (0,-0.75) to [out=170, in=10] (-1,-0.6);
\draw (-1,0.4) ellipse (0.15 and 0.2);
\draw (-1,-0.4) ellipse (0.15 and 0.2);
\draw (-1,0.2) arc (90:-90:0.15 and 0.2);
\draw (-0.85,0.4) to [out=0, in=105] (-0.65,0.2);
\draw (-0.85,-0.4) to [out=0, in=-135] (-0.75,-0.37);
\draw (-0.65,0.2) to [out=-120, in=165] (-0.25,-0.3) to [out=-15, in=-135]
    (0.625,-0.125);
\draw (-0.75,-0.37) to [out=-30, in=-105] (0.82,-0.11);
\draw [dashed] (0,0.125) to [out=180, in=90] (-0.85,0);
\draw (-0.85,0) to [out=-90, in=-90] (0.75,0) to [out=90, in=0] (0.375,0.375) to
    [out=180, in=0] (0,0.125);
\draw (0,0) ellipse (0.4 and 0.25);
\draw [line width=2] (0.4,0) arc (0:65:0.4 and 0.25);
\draw [line width=2] (0.4,0) arc (0:-22:0.4 and 0.25);
\draw [dashed] (1,0) arc (0:180:0.375 and 0.125);
\draw (1,0) arc (0:-60:0.375 and 0.125);
\draw (0.25,0) arc (-180:-90:0.375 and 0.125);
\node at (0,-0.9) [below] {(b)};
\draw [->] (1.25,0) -- (2.25,0);
\node [align=left] at (1.75,0) {slide $\delta$ over\\$\alpha_2+\gamma$\\
    using part\\of $\alpha_2$};
\end{tikzpicture}\vspace{.1in}
\begin{tikzpicture}[scale=2]
\draw (0.2,0) arc (0:180:0.2 and 0.1);
\draw (0.1,0) arc (0:-180:0.1 and 0.05);
\draw (-1,0.6) to [out=10, in=170] (0,0.75) to [out=0, in=90] (1,0)
    to [out=-90, in=0] (0,-0.75) to [out=170, in=10] (-1,-0.6);
\draw (-1,0.4) ellipse (0.15 and 0.2);
\draw (-1,-0.4) ellipse (0.15 and 0.2);
\draw (-1,0.2) arc (90:-90:0.15 and 0.2);
\draw (-0.85,0.4) to [out=0, in=105] (-0.65,0.2);
\draw (-0.85,-0.4) to [out=0, in=-135] (-0.75,-0.37);
\draw (-0.65,0.2) to [out=-120, in=165] (-0.25,-0.3) to [out=-15, in=-135]
    (0.625,-0.125);
\draw (-0.75,-0.37) to [out=-30, in=-105] (0.82,-0.11);
\draw (0.44,-0.11) to [out=45, in=-180] (0.375,0.375);
\draw [dashed] (0,0.125) to [out=180, in=90] (-0.85,0);
\draw (-0.85,0) to [out=-90, in=-90] (0.75,0) to [out=90, in=0] (0.375,0.375);
\draw (0,0.125) to [out=0, in=180] (0.14,0.1875) to [out=0, in=90] (0.2,0);
\draw (0,0) ellipse (0.4 and 0.25);
\draw [dashed] (1,0) arc (0:180:0.375 and 0.125);
\draw (1,0) arc (0:-60:0.375 and 0.125);
\draw (0.625,-0.125) arc (-90:-120:0.375 and 0.125);
\node at (0,-0.9) [below] {(c)};
\draw [->] (1.25,0) -- (2.25,0);
\node at (1.75,0) [above] {redraw};
\end{tikzpicture}
\begin{tikzpicture}[scale=2]
\draw (0.2,0) arc (0:180:0.2 and 0.1);
\draw (0.1,0) arc (0:-180:0.1 and 0.05);
\draw (-1,0.6) to [out=10, in=170] (0,0.75) to [out=0, in=90] (1,0)
    to [out=-90, in=0] (0,-0.75) to [out=170, in=10] (-1,-0.6);
\draw (-1,0.4) ellipse (0.15 and 0.2);
\draw (-1,-0.4) ellipse (0.15 and 0.2);
\draw (-1,0.2) arc (90:-90:0.15 and 0.2);
\draw (-0.85,0.4) to [out=0, in=90] (-0.77,0.35) to [out=-90,in=57] (-0.89,0.125);
\draw (-0.85,-0.4) to [out=-15, in=-180] (0,-0.6) to [out=0, in=-30] (0.85,0.35);
\draw [dashed] (0.85,0.35) to [out=150, in=0] (0,0.55) to [out=-180, in=45] (-0.89,0.125);
\draw (0,0) ellipse (0.4 and 0.25);
\draw [dashed] (1,0) arc (0:180:0.375 and 0.125);
\draw (1,0) arc (0:-180:0.375 and 0.125);
\draw [line width=2] (0.625,0) ++ (-47.5:0.375 and 0.125) arc (-47.5:-132.5:0.375 and 0.125);
\node at (0,-0.9) [below] {(d)};
\draw [->] (1.25,0) -- (2.25,0);
\node [align=left] at (1.75,0.115) {slide $\delta$ over\\$\alpha_2$ using\\
    part of $\beta_2$};
\end{tikzpicture}\vspace{.1in}
\begin{tikzpicture}[scale=2]
\draw (0.2,0) arc (0:180:0.2 and 0.1);
\draw (0.1,0) arc (0:-180:0.1 and 0.05);
\draw (-1,0.6) to [out=10, in=170] (0,0.75) to [out=0, in=90] (1,0)
    to [out=-90, in=0] (0,-0.75) to [out=170, in=10] (-1,-0.6);
\draw (-1,0.4) ellipse (0.15 and 0.2);
\draw (-1,-0.4) ellipse (0.15 and 0.2);
\draw (-1,0.2) arc (90:-90:0.15 and 0.2);
\draw (-0.85,0.4) to [out=0, in=90] (-0.77,0.35) to [out=-90, in=57] (-0.89,0.125);
\draw (0,0.25) to [out=0, in=-30] (0.85,0.35);
\draw (0,-0.25) to [out=0, in=-165] (0.625,-0.09) to [out=15, in=45] (0.875,-0.1) to
    [out=-135, in=0] (0,-0.6) to [out=-180, in=0] (-0.85,-0.4);
\draw [dashed] (0.85,0.35) to [out=150, in=0] (0,0.55) to [out=-180, in=45] (-0.89,0.125);
\draw (0,0.25) arc (90:270:0.4 and 0.25);
\node at (0,-0.9) [below] {(e)};
\draw [->] (1.25,0) -- (2.25,0);
\node at (1.75,0.115) [above] {redraw};
\end{tikzpicture}
\begin{tikzpicture}[scale=2]
\draw (0.2,0) arc (0:180:0.2 and 0.1);
\draw (0.1,0) arc (0:-180:0.1 and 0.05);
\draw (-1,0.6) to [out=10, in=170] (0,0.75) to [out=0, in=90] (1,0)
    to [out=-90, in=0] (0,-0.75) to [out=170, in=10] (-1,-0.6);
\draw (-1,0.4) ellipse (0.15 and 0.2);
\draw (-1,-0.4) ellipse (0.15 and 0.2);
\draw (-1,0.2) arc (90:-90:0.15 and 0.2);
\draw (-0.85,0.4) to [out=0, in=90] (-0.7,0.2625) to [out=-90, in=0] (-0.89,0.125);
\draw [dashed] (-0.89, 0.125) to [out=60, in=-165] (-0.3,0.75);
\draw (-0.3,0.75) to [out=-15, in=90] (-0.85,-0.4);
\node at (0,-0.9) [below] {(f)};
\path (1.25,0) -- (2.25,0);
\end{tikzpicture}
\caption{The initial state $|\tilde{\Psi}\rangle$ satisfies $P_{\beta_2} |\tilde{\Psi} \rangle = |\tilde{\Psi} \rangle$ (see Sec. \ref{Dehn}), which justifies the first slide
from $(a)$ to $(b)$. The effect of the final projection $P_{\beta_2}$ is not shown explicitly; it effectively removes the handle associated with $\beta_2$. }
\label{fig:neighborhood1}
\end{figure*}
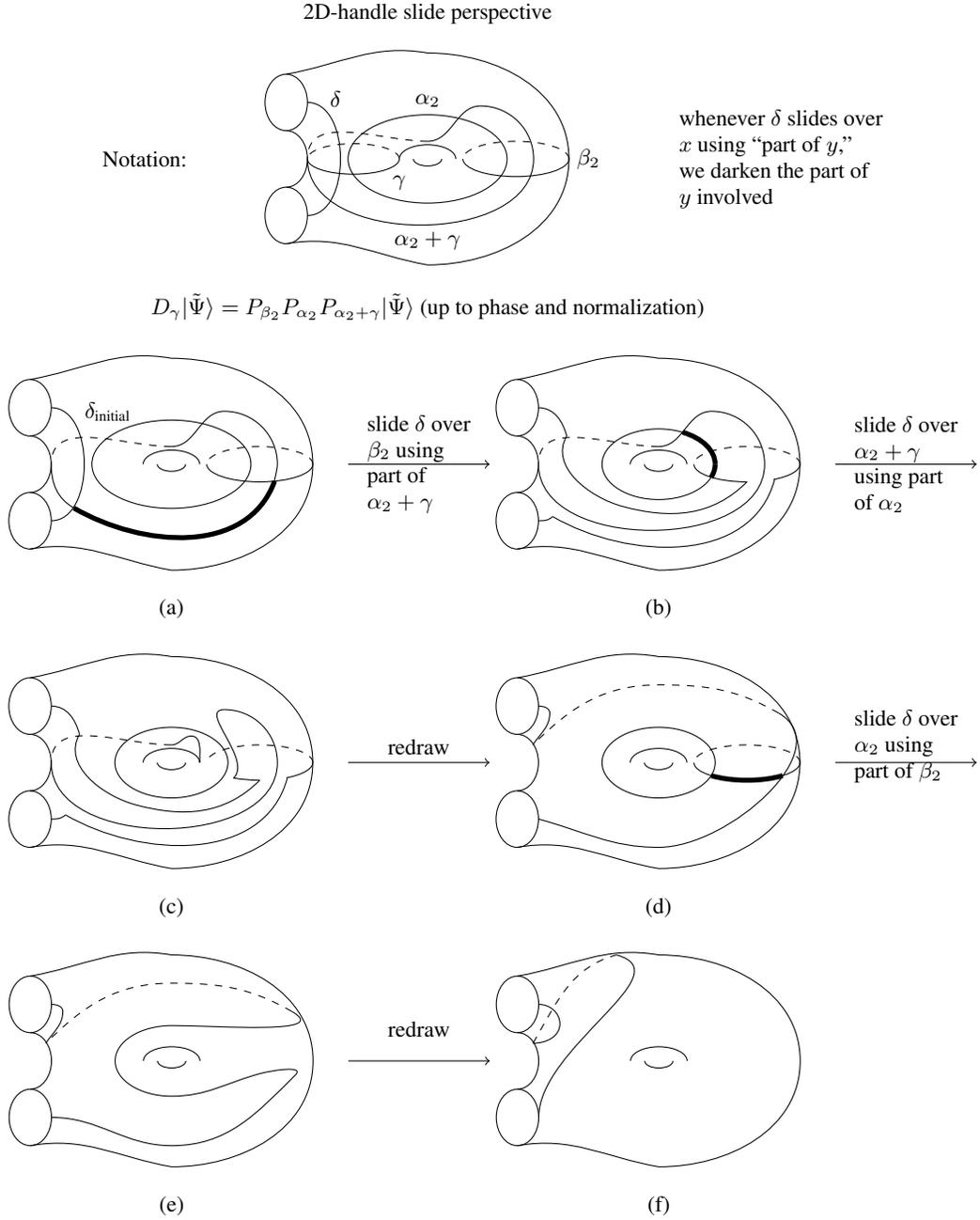

\begin{figure*}[hbpt]
\[D_\gamma^{-1}\vert \tilde{\Psi} \rangle = P_{\beta_2} P_{\alpha_2 + \gamma} P_{\alpha_2} \vert \tilde{\Psi} \rangle\text{ (up to phase and normalization)}\]
\begin{tikzpicture}[scale=2]
\draw (0.2,0) arc (0:180:0.2 and 0.1);
\draw (0.1,0) arc (0:-180:0.1 and 0.05);
\draw (-1,0.6) to [out=10, in=170] (0,0.75) to [out=0, in=90] (1,0)
    to [out=-90, in=0] (0,-0.75) to [out=170, in=10] (-1,-0.6);
\draw (-1,0.4) ellipse (0.15 and 0.2);
\draw (-1,-0.4) ellipse (0.15 and 0.2);
\draw (-1,0.2) arc (90:-90:0.15 and 0.2);
\draw (-0.85,0.4) to [out=0, in=0] (-0.85,-0.4);
\draw [dashed] (0,0.125) to [out=180, in=90] (-0.85,0);
\draw (-0.85,0) to [out=-90, in=150] (-0.69,-0.31);
\draw [line width=2] (-0.69,-0.31) to [out=-30, in=-90] (0.75,0) to [out=90, in=0]
    (0.375,0.375) to [out=180, in=45] (0.23,0.27);
\draw (0.23,0.27) to [out=45, in=0] (0,0.125);
\draw (0,0) ellipse (0.5625 and 0.3125);
\draw [dashed] (1,0) arc (0:180:0.375 and 0.125);
\draw (1,0) arc (0:-180:0.375 and 0.125);
\node at (-0.45,0.25) [above] {$\delta_\text{initial}$};
\node at (0,-0.9) [below] {(a)};
\draw [->] (1.25,0) -- (2.25,0);
\node [align=left] at (1.75,0) {slide $\delta$ over\\$\alpha_2$ along\\
    part of\\$\alpha_2+\gamma$};
\end{tikzpicture}
\begin{tikzpicture}[scale=2]
\draw (0.2,0) arc (0:180:0.2 and 0.1);
\draw (0.1,0) arc (0:-180:0.1 and 0.05);
\draw (-1,0.6) to [out=10, in=170] (0,0.75) to [out=0, in=90] (1,0)
    to [out=-90, in=0] (0,-0.75) to [out=170, in=10] (-1,-0.6);
\draw (-1,0.4) ellipse (0.15 and 0.2);
\draw (-1,-0.4) ellipse (0.15 and 0.2);
\draw (-1,0.2) arc (90:-90:0.15 and 0.2);
\draw [dashed] (0,0.125) to [out=180, in=90] (-0.85,0);
\draw (-0.85,0) to [out=-90, in=-90] (0.75,0) to [out=90, in=0] (0.375,0.375) to
    [out=180, in=0] (0,0.125);
\draw (0,0.25) arc(90:390:0.45 and 0.25);
\draw (0.39,0.125) to [out=120, in=180] (0.45,0.2) to [out=0, in=120] (0.55,0.125) to
    [out=-60, in=0] (0,-0.325) to [out=180, in=-75] (-0.66,0) to [out=105, in=-30] (-0.85,0.4);
\draw (0,0.25) to [out=0, in=180] (0.375,0.475) to [out=0, in=90] (0.85,0) to
    [out=-90, in=-30] (-0.75,-0.38) to [out=-165, in=-15] (-0.85,-0.4);
\node at (0,-0.9) [below] {(b)};
\draw [->] (1.25,0) -- (2.25,0);
\node at (1.75,0) [above] {redraw};
\end{tikzpicture}\vspace{.1in}
\begin{tikzpicture}[scale=2]
\draw (0.2,0) arc (0:180:0.2 and 0.1);
\draw (0.1,0) arc (0:-180:0.1 and 0.05);
\draw (-1,0.6) to [out=10, in=170] (0,0.75) to [out=0, in=90] (1,0)
    to [out=-90, in=0] (0,-0.75) to [out=170, in=10] (-1,-0.6);
\draw (-1,0.4) ellipse (0.15 and 0.2);
\draw (-1,-0.4) ellipse (0.15 and 0.2);
\draw (-1,0.2) arc (90:-90:0.15 and 0.2);
\draw (-0.85,0.4) to [out=0, in=90] (0.85,0) to [out=-90, in=-15] (-0.85,-0.4);
\draw [dashed] (0,0.125) to [out=180, in=90] (-0.85,0);
\draw (-0.85,0) to [out=-90, in=-90] (0.75,0) to [out=90, in=0] (0.375,0.375) to
    [out=180, in=0] (0,0.125);
\draw [dashed] (1,0) arc (0:180:0.375 and 0.125);
\draw (1,0) arc (0:-180:0.375 and 0.125);
\draw [line width=2] (0.85,-0.1) to [out=-160, in=10] (0.725,-0.125);
\node at (0,-0.9) [below] {(c)};
\draw [->] (1.25,0) -- (2.25,0);
\node [align=left] at (1.75,0) {slide $\delta$ over\\$\alpha_2 + \gamma$\\
    along part\\of $\beta_2$};
\end{tikzpicture}
\begin{tikzpicture}[scale=2]
\draw (0.2,0) arc (0:180:0.2 and 0.1);
\draw (0.1,0) arc (0:-180:0.1 and 0.05);
\draw (-1,0.6) to [out=10, in=170] (0,0.75) to [out=0, in=90] (1,0)
    to [out=-90, in=0] (0,-0.75) to [out=170, in=10] (-1,-0.6);
\draw (-1,0.4) ellipse (0.15 and 0.2);
\draw (-1,-0.4) ellipse (0.15 and 0.2);
\draw (-1,0.2) arc (90:-90:0.15 and 0.2);
\draw (-0.85,-0.4) to [out=-15, in=-15] (0,-0.3) to [out=165, in=-90] (-0.85,0);
\draw (-0.85,0.4) to [out=15, in=120] (0.8,0.25) -- (0.55,0.25) to
    [out=120, in=0] (0.375,0.375) to [out=180, in=0] (0,0.125);
\draw [dashed] (0,0.125) to [out=180, in=90] (-0.85,0);
\draw (0.2,0.1) ellipse (0.5 and 0.375);
\draw [line width=2] (0.7,0.1) arc (0:25:0.5 and 0.375);
\draw [line width=2] (0.7,0.1) arc (0:-38:0.5 and 0.375);
\draw [dashed] (1,0) arc (0:180:0.375 and 0.125);
\draw (1,0) arc (0:-180:0.375 and 0.125);
\node at (-0.3,0.1) [below left] {$\alpha_2$};
\node at (0,-0.9) [below] {(d)};
\draw [->] (1.25,0) -- (2.25,0);
\node [align=left] at (1.75,0.115) {slide $\delta$ over\\$\beta_2$ along\\
    part of $\alpha_2$};
\end{tikzpicture}\vspace{.1in}
\begin{tikzpicture}[scale=2]
\draw (0.2,0) arc (0:180:0.2 and 0.1);
\draw (0.1,0) arc (0:-180:0.1 and 0.05);
\draw (-1,0.6) to [out=10, in=170] (0,0.75) to [out=0, in=90] (1,0)
    to [out=-90, in=0] (0,-0.75) to [out=170, in=10] (-1,-0.6);
\draw (-1,0.4) ellipse (0.15 and 0.2);
\draw (-1,-0.4) ellipse (0.15 and 0.2);
\draw (-1,0.2) arc (90:-90:0.15 and 0.2);
\draw (-0.85,-0.4) to [out=-15, in=-90] (-0.6,-0.3) to [out=90, in=-90] (-0.85,0);
\draw (-0.85,0.4) to [out=15, in=120] (0.8,0.25) to [out=-60, in=75] (0.85,-0.1) to
    [out=-75, in=-90] (1,0);
\draw [dashed] (0,0.125) to [out=180, in=90] (-0.85,0);
\draw (0,0.125) to [out=0, in=180] (0.375,0.375) to [out=0, in=90] (0.7,0.1) to
    [out=-90, in=-90] (0.25,0);
\draw [dashed] (1,0) arc (0:180:0.375 and 0.125);
\node at (0,-0.9) [below] {(e)};
\draw [->] (1.25,0) -- (2.25,0);
\node at (1.75,0) [above] {redraw};
\end{tikzpicture}
\begin{tikzpicture}[scale=2]
\draw (0.2,0) arc (0:180:0.2 and 0.1);
\draw (0.1,0) arc (0:-180:0.1 and 0.05);
\draw (-1,0.6) to [out=10, in=170] (0,0.75) to [out=0, in=90] (1,0)
    to [out=-90, in=0] (0,-0.75) to [out=170, in=10] (-1,-0.6);
\draw (-1,0.4) ellipse (0.15 and 0.2);
\draw (-1,-0.4) ellipse (0.15 and 0.2);
\draw (-1,0.2) arc (90:-90:0.15 and 0.2);
\draw (-0.85,-0.4) to [out=-15, in=-90] (-0.6,-0.3) to [out=90, in=-90] (-0.85,0);
\draw [dashed] (-0.85,0) to [out=90, in=180] (-0.3,0.75);
\draw (-0.3,0.75) to [out=0, in=0] (-0.85,0.4);
\node at (0,-0.9) [below] {(f)};
\path (1.25,0) -- (2.25,0);
\end{tikzpicture}
\caption{}
\label{fig:neighborhood2}
\end{figure*}
\end{changemargin}

Then from this starting point, a succession of $3$ measurements of trivial charge produce $\mathfrak{S}(\Sigma\times I)$, which admits an inclusion $i$, as in Figure \ref{fig:embedding}, inducing $f=D_{\beta_1}$ or $D_{\beta_1}^{-1}$ in Figures \ref{fig:neighborhood1} and \ref{fig:neighborhood2}, respectively.  We will give successively 2D, 3D, and 4D accounts of the effect of measurement.  In the 2D account, the $\pm$ twist is verified by tracking a typical fiber $\delta$ in the normal annular collar about $\beta_1\subset\Sigma$ under the ``handle slides'' shown.  Handle slides of $\delta$ are merely isotopies over the ``new material'' provided by giving in $\{S_i^\prime\}\subset\mathfrak{S}(\Sigma\times I)$ (see Section \ref{sec:2d}). Note that when acting on states of the form $|\tilde{\Psi}\rangle$ as described in Sec. \ref{Dehn}, $D_{\beta_1} = D_{\gamma}$ (see Eq. (\ref{DbetaDgamma})).

The embedding $i$ (from Figure \ref{fig:embedding}) is the identity outside $\mathfrak{N}\left(\beta_1\right)\times I$.  On $\mathfrak{N}\left(\beta_1\right)\times I$ it is defined (with compatible boundary conditions except on a single $3$-ball). One may visualize $\operatorname{domain}(i)$ as swept out in two parameters by the arc $\delta$ in Figure \ref{fig:neighborhood1} (or \ref{fig:neighborhood2}).  One parameter is the progression through the subfigures (a)$\longrightarrow$(f).  The other parameter is obtained by pushing $\delta$ as far as possible around the annulus $\mathfrak{N}\left(\beta_1\right)$.  Before the creation and after the destruction of the additional genus, $\delta$ can be slid $2\pi$ around $\beta_1$ staying normal to $\beta_1$.  While the additional genus is present, there is an obstacle that $\delta$ cannot cross, a ``missing'' $2$-disk.  These missing $2$-disks sweep out a $3$-ball in time; this is the $3$-ball missing from the PS of $\operatorname{domain}(i)$.  Section \ref{sec:3d} gives additional details on the 3D interpretation.

\subsection{2D Interpretation}\label{sec:2d}

We start with the surface $\Sigma$ and state $\vert\Psi\rangle\in \mathcal{V}(\Sigma)$.  All operations take place near $\beta_1$ so we restrict attention to the annular neighborhood $\mathfrak{N}\left(\beta_1\right)=Y$ with boundary components $\{\beta_1^+,\beta_1^-\}$.  The most naive, but still accurate, point of view is that any simple loop $\omega$ which is definitely in a trivial charge state behaves as if it bounds a disk $\Delta$ of material in the TQFT ground state.  Thus, for all physical purposes, there is \emph{no} distinction between an arc passing along a segment $\omega^\prime\subset \omega$ and its complementary segment $\omega^{\prime\prime} = \omega\setminus \omega^\prime$.  If $\Delta$ were physically present, replacing $\omega^\prime$ by $\omega^{\prime\prime}$ would be an ``isotopy across $\Delta$''.  In Section \ref{sec:3d} we will explain why this isotopy is called a \emph{handle slide}.

In $Y$ we see a product structure (PS) of arcs joining $\beta_1^+$ to $\beta_1^-$, one of those arcs we call $\delta$.  The first operation is to expand $Y$ to $Y^+$ by adding one to the genus; $Y^+$ is the surface studied in Figure \ref{fig:neighborhood1}.  The three arrows labeled by ``slide $\delta$ over...'' indicate isotoping $\delta$ until it runs parallel to a tiny segment $\omega^\prime$ of the curve $\omega$ which it is to be slid over, and then replacing the bit parallel to $\omega^\prime$ with a bit parallel to $\omega^{\prime\prime}$.  The measurements of zero charge successively along $\beta_2$, $\alpha_2+\gamma$, and $\alpha_2$ each \emph{temporarily} secure the condition needed for this handle slide---that at that moment, the charge on $\omega$ ($=\beta_2$, then $\alpha_2+\gamma$, then $\alpha_2$) is trivial. The composition of these slides allows us to follow the progress of one fiber $\delta$ of the 2D PS of $Y$ under these three measurements.  After the measurements the extra genus is removed by a final projection $P_{\beta_2}$.  In this time history, contained in the righthand column of Figure \ref{fig:embedding}, we see the punctured product structure embedded by the map $i$ as described above.  The central conclusion is that $\delta$ as drawn in \ref{fig:neighborhood1}(f) has picked up a $2\pi$-twist relative to its position in \ref{fig:neighborhood1}(a).  The bridge between measurement and Dehn twist is already evident in this 2D analysis, and clarified further within the 3D interpretation; however, the extraction of the precise overall phase factor must wait for the 4D interpretation.

\subsection{3D Interpretation}\label{sec:3d}

Here we use Morse theory to build the 3D bordism $\mathfrak{S}(\Sigma\times I)$ (see Figure \ref{fig:embedding}) as a time history.  We discuss this first from the perspective of the surface $\Sigma$ evolving in time and then in the language of 3D surgery.  The later perspective belongs to the next section since 3D surgeries are accomplished by passing across 4D cobordisms.

In the first perspective (evolving surface) there are eight events, each of which is an ($i-1$)-surgery on the surface, or equivalently from the perspective of the space-time the attachment of a 3D, index $=i$-handle $h_i$, for $i=1$ or $2$.  The eight values for $i$, in order, are $(1, 2, 1, 2, 1, 2, 1, 2)$.  Let us explain this language.  Since we will shortly turn to 4D, we give the definitions in all dimensions.

A $d$-dimensional $p$-handle is a $d$-ball which contains a distinguished subset called the attaching region on its boundary:
\begin{align}
d\text{-dimensional $p$-handle $h$: }\left(D^p\times D^q, \partial D^p\times D^q\right),
\end{align}
where $d=p+q$.
The ``belt'' or non-attaching region is $D^p\times\partial D^q$.  Given a $d$-manifold with boundary $(M,\partial M)$, we \emph{attach} $h$ by embedding $\partial D^p\times D^q\overset{e}{\hookrightarrow}\partial M$.  The result is shown in Fig. \ref{fig:attachment3d}.

\begin{figure}[hbpt]
\begin{tikzpicture}[scale=1.25]
\node at (0.375,0) {$M\cup_e h=$};
\begin{scope}[decoration={random steps,segment length=3pt,amplitude=0.5pt},decorate]
\draw[decorate] (3,-0.5) ellipse (1.25 and 0.625) node {$M$};
\draw (3.5,-0.1875) ++ (-2:0.625 and 1) arc (-2:162:0.625 and 1);
\draw (3.5,-0.015625) ++ (2:0.25 and 0.5) arc (2:166:0.25 and 0.5);
\node at (3.5,0.625) {$h$};
\end{scope}
\end{tikzpicture}
\caption{}
\label{fig:attachment3d}
\end{figure}

The effect on the boundary is $\partial M\rightarrow \left(\partial M\setminus e\left(\partial D^p\times D^q\right)\right)\cup\left(D^p\times \partial D^q\right)$; this is called a ($p-1$)-surgery.

So for cobording a surface to make a $3$-manifold, we have two operations:

\begin{figure*}[hbpt]
\begin{tikzpicture}
\draw (0,0.125) -- (0,-0.375) -- (2,-0.375) -- (2,0.125) -- (0,0.125) -- (0.5,1.375) --
    (2.5,1.375) -- (2,0.125)
    (2.5,1.375) -- (2.5,0.875) -- (2,-0.375);
\node at (1,-0.125) {$\Sigma_0$};
\draw [->] (3,0.5) -- (4,0.5);
\begin{scope}[shift={(4.5,0)}]
\draw (0,0.125) -- (0,-0.375) -- (2,-0.375) -- (2,0.125) -- (0,0.125) -- (0.5,1.375) --
    (2.5,1.375) -- (2,0.125)
    (2.5,1.375) -- (2.5,0.875) -- (2,-0.375);
\node at (1,-0.125) {$\Sigma_0$};
\draw (0.625,0.75) ellipse (0.125 and 0.125);
\draw (1.875,0.75) ellipse (0.125 and 0.125);
\draw (0.5,0.75) arc (180:0:0.75 and 1.25);
\draw (0.75,0.75) arc (180:0:0.5 and 1);
\node at (1.25,0.75) {$\Sigma_1$};
\end{scope}
\node at (3.5,-1) {(a) attach 3D $1$-handle, accomplish $0$-surgery};
\begin{scope}[shift={(0,-3.625)}]
\draw (0,0.125) -- (0,-0.375) -- (2,-0.375) -- (2,0.125) -- (0,0.125) -- (0.5,1.375) --
    (2.5,1.375) -- (2,0.125)
    (2.5,1.375) -- (2.5,0.875) -- (2,-0.375);
\draw [->] (3,0.5) -- (4,0.5);
\draw (0.75,0.75) -- (1,0.5) -- (1.5,0.5) -- (1.75,0.75) -- (1.5,1) -- (1,1) -- (0.75,0.75);
\draw [dashed] (0.75,0.75) -- (0.75,0.25) -- (1,0) -- (1.5,0) -- (1.75,0.25) -- (1.75,0.75)
    (1,0) -- (1,1)
    (1.5,0) -- (1.5,1);
\node at (0.5,-0.125) {$\Sigma_0$};
\begin{scope}[shift={(4.5,0)}]
\draw (0,0.125) -- (0,-0.375) -- (2,-0.375) -- (2,0.125) -- (0,0.125) -- (0.5,1.375) --
    (2.5,1.375) -- (2,0.125)
    (2.5,1.375) -- (2.5,0.875) -- (2,-0.375);
\draw (0.75,0.75) -- (1,0.5) -- (1.5,0.5) -- (1.75,0.75) -- (1.5,1) -- (1,1) -- (0.75,0.75);
\draw [dashed] (0.75,0.75) -- (0.75,0.25) -- (1,0) -- (1.5,0) -- (1.75,0.25) -- (1.75,0.75)
    (1,0) -- (1,1)
    (1.5,0) -- (1.5,1);
\draw (0.5,0.75) arc (180:0:0.75 and 1.25);
\draw (0.625,0.75) arc (180:0:0.625 and 1.125);
\draw (1.25,0.75) ellipse (0.75 and 0.5);
\draw (1.25,0.75) ellipse (0.625 and 0.375);
\end{scope}
\node at (3.5,-1) {(b) attach 3D $2$-handle, accomplish $1$-surgery};
\end{scope}
\end{tikzpicture}
\caption{}
\label{fig:cobordops}
\end{figure*}

The first event is type (a) (attach $1$-handle, do $0$-surgery); it adds genus (presuming the 1-handle joins already connected regions of the surface).  The last event is type (b) (attach $2$-handle, do $1$-surgery); it removes genus (or disconnects the surface).  It is important to note \emph{how} the genus is removed; it is removed by attaching to $\beta_2$; that is, the genus is removed by surgery on the meridian created by the first $0$-surgery. (This gives a very different 3D trace than would removing the genus by surgery on that dual curve $\alpha_2$.)  The remaining six events are in three pairs.  Measuring the trivial charge (along first $\beta_2$, then $\alpha_2+\gamma$, then $\alpha_2$) attaches a $2$-handle and effects a $1$-surgery (first on $\beta_2$, then $\alpha_2+\gamma$, then $\alpha_2$) but immediately after each measurement, since we want our surface back as it was (not with reduced genus), we should immediately reverse the $2$-handle attachment (in all three cases) by attaching a $1$-handle to the ``belt region''.  This way, the 3D cobordism records the charge measurement but does not ultimately alter the topology of the surface.  This pair of 3D handle attachments may seem ad hoc from the perspective of the evolving space time, but we will next see that it is quite essential from the perspective of 3D surgery.

In the 3D surgery perspective, we take as the starting point the 3D cobordism (from $\Sigma_\text{int}$ to $\Sigma_\text{final}$) consisting of only the first and last events---which actually alter topology.  From this starting point, we interpret the three measurements of trivial charge as three 3D $1$-surgeries.

\subsection{4D Interpretation}\label{sec:4d}

The 4D interpretation of 3D surgery is 4D handle attachment.  The 4D cobordism provides a geometric way of understanding the overall phases intrinsic to Chern-Simons theory and in particular the relevant central extensions of the MCG (see Ref. \onlinecite{walker1991} Chapters 16 and 17, Shale-Weil cocycle).  A 4D handle is merely a ball with the following portion of its boundary specified as a subspace, and called the ``attaching region'':
\[\left(D^2\times D^2, \partial D^2\times D^2\right) = \left(D^2\times D^2, S^1\times D^2\right).\]
Figure \ref{fig:handle4d} attempts to capture this 4D $2$-handle in pictures by cutting dimensions in half.  Figure \ref{fig:attachment} illustrates this handle being attached to $M^3\times[0,1]$.

\begin{figure}[hbpt]
\begin{tikzpicture}[scale=1.5]
\draw [line width=2] (0,0) -- (0,1);
\draw (0,1) -- (1,1);
\draw [line width=2] (1,1) -- (1,0);
\draw (1,0) -- (0,0);
\draw [->] (-0.25,0.25) -- (-0.25,0.75);
\node at (-0.25,0.5) [left] {$D^2_\text{second}$};
\draw [->] (0.25,-0.25) -- (0.75,-0.25);
\node at (0.5,-0.25) [below] {$D^2_\text{first}$};
\node at (1.25,0.5) [right] {``attaching region'' in bold $=\left(S^1\times D^2\right)$};
\end{tikzpicture}
\caption{}
\label{fig:handle4d}
\end{figure}

\begin{figure}[hbpt]
\begin{tikzpicture}[scale=1.5]
\draw [line width=2] (-1,0) -- (-0.5,0);
\draw [line width=2] (0.5,0) -- (1,0);
\draw (-0.5,0) to [out=90, in=180] (0,0.75) to [out=0, in=90] (0.5,0);
\draw (-1,0) to [out=90, in=180] (0,1.25) to [out=0, in=90] (1,0);
\draw (-2,0) -- (2,0) -- (2,-0.75) -- (-2,-0.75) -- (-2,0);
\node (1) at (0,-0.5) {$S^1\times D^2$ glues to $S_i$};
\path [->] (1) edge (-0.75,-0.125);
\path [->] (1) edge (0.75,-0.125);
\node (2) at (1.5,1) {$S_i^\prime$};
\path [->] (2) edge (0.25,0.75);
\path [->] (2) edge (0.875,0.75);
\end{tikzpicture}
\caption{}
\label{fig:attachment}
\end{figure}

Note that $S_i$ and $S_i^\prime$ appearing disconnected is merely an artifact of the low dimensionality of the picture.  In Figures \ref{fig:handle4d} and \ref{fig:attachment}, subspaces diffeomorphic to $S^1\times D^2$ are drawn as $S^0\times D^1$.  $S^0$ is by definition the boundary of the unit ball in $R^1$, that is $\{-1, +1\}$, two points.

Also note that gluing $S^1\times D^2$ to $S_i$ requires choosing a normal framing to the core circle of $S_i$.  Up to isotopy, the framing lies in a $Z$-torsor.  In $3$-space or $S^3$, the $Z$-torsor is based ($0$ is the linking number $=$ zero push-off) so the data for handle attachment (and surgery), a framed link, can also be thought of as a link with integral labels in $S^3$.  A literal picture of 3D $2$-handles was drawn in Figure \ref{fig:cobordops}(b). In general handles of the same index---in this paper it is always $2$---may be ``slid over each other.''  This process isotopes the attaching region of one $k$-handle in the ``upper'' boundary which results from the attachment of the other $k$-handles.  Slides may occur in sequences of arbitrary length; it is only necessary at each step to choose which $k$-handle is mobile and fix the rest.  Importantly, handle sliding does not change the topology (or smooth structure) of the manifold; it merely changes its combinatorial description, or handle decomposition.  Our 4D handle body pictures begin with a 4-ball ($0$-handle) and the entire diagram should be seen as lying on its boundary $\partial D^4=S^3$.

We will exploit a useful 3D notation: An unknotted circle with a dot on it \tikz[scale=0.5, baseline=-0.5ex]{\draw (0,0) ellipse (0.5 and 0.5); \node at (0,0.5) {$\bullet$};} means delete a $2$-handle from the $0$-handle, $D^4$.  As far as the boundary is concerned, this gives the same result as attaching a $0$-framed $2$-handle to the unknot $\tikz[baseline=-0.5ex]{\draw (0,0) ellipse (0.25 and 0.25);}^0$.  But the bulks are different.  \tikz[scale=0.5, baseline=-0.5ex]{\draw (0,0) ellipse (0.5 and 0.5); \node at (0,0.5) {$\bullet$};} yields $S^1\times D^3$ and so represents (dually) the attachment of a 4D $1$-handle, whereas $\tikz[baseline=-0.5ex]{\draw (0,0) ellipse (0.25 and 0.25);}^0$ yields $S^2\times D^2$ as one would expect from attaching a trivial $2$-handle.  Thus, instead of attaching a 4D $1$-handle to a $4$-ball $D^4$, the same $4$-manifold results from deleting from $D^4$ a properly embedded disk whose boundary is the dotted circle.  In one dimension lower this is the duality; ``a topologist cannot tell a bridge from a (perpendicular) tunnel.''  For more details on handles and handle sliding, see Ref. \onlinecite{gompf1999}.

\begin{figure*}[hbpt]
\begin{tikzpicture}
\draw (0,0) -- (0,0.5) -- (3,0.5);
\draw (0,0.5) -- (0.5,2);
\draw [dashed] (1.5,1.125) arc (0:180:0.5 and 0.4);
\draw (1.5,1.125) arc (0:65:0.5 and 0.4);
\draw (0.5,1.125) arc (180:115:0.5 and 0.4);
\draw (1.5,1.125) arc (0:-180:0.5 and 0.4);
\draw (1,1.125) ellipse (0.25 and 0.2);
\draw (1.5,1.125) arc (0:180:0.5 and 1);
\draw (1.25,1.125) arc (0:180:0.25 and 0.75);
\draw [dashed] (2.875,1.125) arc (0:180:0.5 and 0.4);
\draw (2.875,1.125) arc (0:65:0.5 and 0.4);
\draw (1.875,1.125) arc (180:115:0.5 and 0.4);
\draw (2.875,1.125) arc (0:-180:0.5 and 0.4);
\draw (2.375,1.125) ellipse (0.25 and 0.2);
\draw (2.875,1.125) arc (0:180:0.5 and 1);
\draw (2.625,1.125) arc (0:180:0.25 and 0.75);
\draw [->] (3.25,0.625) -- (3.75,0.625);
\draw (4,0) -- (4,0.5) -- (7,0.5);
\draw (4,0.5) -- (4.5,2);
\draw [dashed] (6.38,1.66) to [out=0, in=120] (6.6,1.49);
\draw [dashed] (5.25,1.47) to [out=-30, in=195] (5.95,1.6);
\draw [dashed] (4.5,1.125) arc (180:60:0.5 and 0.4);
\draw (4.5,1.125) arc (180:115:0.5 and 0.4);
\draw (4.5,1.125) arc (-180:-60:0.5 and 0.4);
\draw (5.25,0.78) to [out=30, in=180] (6.025,0.85) to [out=0, in=-120] (6.5,2.1);
\draw [dashed] (6.5,2.1) to [out=-75, in=90] (6.6,1.485);
\draw (6.5,2.1) to [out=105, in=0] (6.28,2.3) to [out=180, in=-60] (5.25,2);
\draw (4.5,1.125) arc (180:60:0.5 and 1);
\draw [dashed] (6.38,1.55) to [out=-15, in=120] (6.47,1.51);
\draw (5.125,1.3) to [out=-30, in=195] (5.9,1.4);
\draw (4.75,1.125) arc (180:60:0.25 and 0.2);
\draw (4.75,1.125) arc (-180:-60:0.25 and 0.2);
\draw (5.125,0.95) to [out=30, in=180] (5.9,1.05) to [out=0, in=-120] (6.28,2.1);
\draw [dashed] (6.43, 1.92) to [out=-75, in=90] (6.47,1.51);
\draw (6.28, 2.1) to [out=180, in=0] (5.4,1.6) to [out=180, in=-60] (5.125,1.77);
\draw (4.75,1.125) arc (180:60:0.25 and 0.75);
\draw (6.875,1.125) arc (0:-135:0.5 and 0.4);
\draw (5.875,1.125) arc (-180:-170:0.5 and 0.4);
\draw (5.875,1.125) arc (180:115:0.5 and 0.4);
\draw [dashed] (6.875,1.125) arc (0:180:0.5 and 0.4);
\draw (6.875,1.125) arc (0:65:0.5 and 0.4);
\draw (6.625,1.125) arc (0:92:0.25 and 0.2);
\draw (6.625,1.125) arc (0:-118:0.25 and 0.2);
\draw (6.875,1.125) arc (0:75:0.5 and 1);
\draw (5.875,1.125) arc (180:103:0.5 and 1);
\draw (6.625,1.125) arc (0:80:0.25 and 0.75);
\draw [->] (7.25,0.625) -- (7.75,0.625);
\draw (8,0) -- (8,0.5) -- (11.5,0.5);
\draw (8,0.5) -- (8.5,2);
\draw [dashed] (11.075,1.125) arc (0:150:0.7 and 0.56);
\draw (9.84,1.48) to [out=210, in=-30] (9.25,1.47);
\draw (8.5,1.125) arc (180:60:0.5 and 0.4);
\draw (8.5,1.125) arc (-180:-60:0.5 and 0.4);
\draw (9.25,0.78) to [out=30, in=150] (10.03,0.64);
\draw (11.075,1.125) arc (0:-120:0.7 and 0.56);
\draw (11.075,1.125) arc (0:120:0.7 and 1.2);
\draw (9.25,2) to [out=-60, in=-135] (10.025,2.16);
\draw (8.5,1.125) arc (180:60:0.5 and 1);
\draw [dashed] (10.975,1.125) arc (0:150:0.6 and 0.48);
\draw (9.9,1.42) to [out=210, in=-30] (9.2,1.4);
\draw (8.6,1.125) arc (180:60:0.4 and 0.32);
\draw (8.6,1.125) arc (-180:-60:0.4 and 0.32);
\draw (9.2,0.85) to [out=30, in=150] (10.07,0.71);
\draw (10.975,1.125) arc (0:-120:0.6 and 0.48);
\draw (10.975,1.125) arc (0:120:0.6 and 1.1);
\draw (9.2,1.91) to [out=-60, in=-135] (10.075,2.08);
\draw (8.6,1.125) arc (180:60:0.4 and 0.9);
\draw [dashed] (10.875,1.125) arc (0:180:0.5 and 0.4);
\draw (10.875,1.125) arc (0:65:0.5 and 0.4);
\draw (9.875,1.125) arc (180:115:0.5 and 0.4);
\draw (10.875,1.125) arc (0:-180:0.5 and 0.4);
\draw (10.375,1.125) ellipse (0.25 and 0.2);
\draw (10.875,1.125) arc (0:180:0.5 and 1);
\draw (10.625,1.125) arc (0:180:0.25 and 0.75);
\end{tikzpicture}
\[\text{(a) slide of 3D $2$-handles}\]
\begin{tikzpicture}
\draw (0,0) -- (0,0.5) -- (3,0.5);
\draw (0.25,0.5) to [out=90, in=180] (0.6875,1.125) to [out=0, in=90] (1.125,0.5);
\draw (0.4375,0.5) to [out=90, in=180] (0.6875,0.9375) to [out=0, in=90] (0.9375,0.5);
\draw (1.875,0.5) to [out=90, in=180] (2.3125,1.125) to [out=0, in=90] (2.75,0.5);
\draw (2.0625,0.5) to [out=90, in=180] (2.3125,0.9375) to [out=0, in=90] (2.5625,0.5);
\draw [->] (3.25,0.625) -- (3.75,0.625);
\draw (4,0) -- (4,0.5) -- (7,0.5);
\draw (4.25,0.5) to [out=90, in=180] (4.6875,1.125) to [out=0, in=100] (5.0625,0.875)
    to [out=-80, in=-90] (5.5,0.875) to [out=90, in=180] (6.125,1.5) to [out=0, in=90] (6.375,1.125);
\draw (4.4375,0.5) to [out=90, in=180] (4.6875,0.9375) to [out=0, in=120] (4.9375,0.6875)
    to [out=-60, in=180] (5.375,0.5625) to [out=0, in=-90] (5.6,0.6875) to [out=90, in=180] (6.125,1.3125) to [out=0, in=90] (6.22,1.125);
\draw (5.875,0.5) to [out=90, in=180] (6.3125,1.125) to [out=0, in=90] (6.75,0.5);
\draw (6.0625,0.5) to [out=90, in=180] (6.3125,0.9375) to [out=0, in=90] (6.5625,0.5);
\draw [->] (7.25,0.625) -- (7.75,0.625);
\draw (8,0) -- (8,0.5) -- (11.5,0.5);
\draw (8.25,0.5) to [out=90, in=180] (8.6875,1.125) to [out=0, in=100] (9.0625,0.875)
    to [out=-80, in=-90] (9.5,0.875) to [out=90, in=180] (10.3125, 1.625) to [out=0, in=90] (11.25,0.5);
\draw (8.4375,0.5) to [out=90, in=180] (8.6875,0.9375) to [out=0, in=120] (8.9375,0.6875)
    to [out=-60, in=180] (9.375,0.5625) to [out=0, in=-90] (9.6,0.6875) to [out=90, in=180] (10.3125,1.4375) to [out=0, in=90] (11.0625,0.5);
\draw (9.875,0.5) to [out=90, in=180] (10.3125,1.125) to [out=0, in=90] (10.75,0.5);
\draw (10.0625,0.5) to [out=90, in=180] (10.3125,0.9375) to [out=0, in=90] (10.5625,0.5);
\end{tikzpicture}
\[\text{(b) slide of 2D $1$-handles, and schematic for slide of 4D $2$-handles}\]
\caption{}
\label{fig:sliding}
\end{figure*}

Recall the notation in Figure \ref{fig:neighborhood1}.  According to Ref. \onlinecite{walker1991}, to keep track of phase information one must work with \emph{extended surfaces}, i.e. a surface $\Sigma$ with a maximal isotropic subspace $L$ of $H_1(\Sigma; R)$ specified.  In our case, all the action is confined to the subsurface $\mathfrak{N}\left(\beta_1\right) = Y\subset\Sigma$.  After adding one genus to $Y$, we obtain $Y^+$:
\[\tikz{
\draw (0.2,0) arc (0:180:0.2 and 0.1);
\draw (0.1,0) arc (0:-180:0.1 and 0.05);
\draw [dashed] (1,0) arc (0:180:0.375 and 0.125);
\draw (1,0) arc (0:-180:0.375 and 0.125);
\draw (-1,0.6) to [out=10, in=170] (0,0.75) to [out=0, in=90] (1,0)
    to [out=-90, in=0] (0,-0.75) to [out=170, in=10] (-1,-0.6);
\draw (-1,0.4) ellipse (0.15 and 0.2);
\draw (-1,-0.4) ellipse (0.15 and 0.2);
\draw (-1,0.2) arc (90:-90:0.15 and 0.2);
\node at (-2.25,0) {$Y^+=$};
\node at (-1,0.4) [above left] {$\beta_1^+$};
\node at (-1,-0.4) [below left] {$\beta_1^-$};
\node at (1,0) [right] {$\beta_2$};
}\]

\begin{figure*}[hbpt]
\begin{tikzpicture}
\fill (0,0) ellipse (0.25 and 0.5);
\draw [dashed] (1.5,0.75) arc (90:270:0.25 and 0.5);
\draw (1.5,0.75) arc (90:-90:0.25 and 0.5);
\fill (3,0.5) ellipse (0.25 and 0.5);
\draw (0,0.5) -- (3,1);
\draw (0,-0.5) -- (3,0);
\node at (-0.25,0) [left] {cap};
\node at (1.5,0.75) [above] {$S^2_\text{top}$};
\node at (1.75,0.25) [right] {$\beta_1$};
\node at (3.25,0.5) [right] {cap};
\fill (0,-2) ellipse (0.25 and 0.5);
\draw [dashed] (0.75,-1.375) arc (90:270:0.25 and 0.5);
\draw (0.75,-1.375) arc (90:-90:0.25 and 0.5);
\draw [dashed] (1.5,-1.25) arc (90:270:0.075 and 0.15);
\draw (1.5,-1.25) arc (90:-90:0.075 and 0.15);
\draw (1.125,-1.75) arc (180:0:0.375 and 0.1875);
\draw (1.25,-1.75) arc (-180:0:0.25 and 0.125);
\fill (3,-1.5) ellipse (0.25 and 0.5);
\draw (0,-1.5) -- (3,-1);
\draw (0,-2.5) -- (3,-2);
\node at (-0.25,-2) [left] {cap};
\node at (1.25,-2.0625) {$\beta_1$};
\draw [->] (1.875,-1.4) -- (1.67,-1.4);
\node at (2.125,-1.4) {$\beta_2$};
\node at (3.25,-1.5) [right] {cap};
\fill (0,-4) ellipse (0.25 and 0.5);
\draw [dashed] (1.5,-3.25) arc (90:270:0.25 and 0.5);
\draw (1.5,-3.25) arc (90:-90:0.25 and 0.5);
\fill (3,-3.5) ellipse (0.25 and 0.5);
\draw (0,-3.5) -- (3,-3);
\draw (0,-4.5) -- (3,-4);
\node at (-0.25,-4) [above left] {cap};
\draw [->] (-0.5,-4.17) -- (-1.25,-4.3);
\node at (-1.25,-4.6) {rest of $\Sigma$};
\node at (1.75,-3.75) [right] {$\beta_1$};
\node at (1.5,-4.25) [below] {$S^2_\text{bot}$};
\node at (3.25,-3.5) [below right] {cap};
\draw [->] (3.5,-3.33) -- (4.25,-3.2);
\node at (4.25,-2.9) {rest of $\Sigma$};
\node at (8.5,-1.75) {$\cong \left(S^1\times S^2\setminus (D^3 \amalg D^3), S^2_\text{bot} \cup S^2_\text{top}\right)$};
\draw [decorate,decoration={brace,amplitude=10pt}] (11.75,-2) -- (5.75,-2);
\node [rotate=-90] at (8.75,-2.75) {$\cong$};
\node at (8.75,-3.375) {$\left(S^2\times I\sharp S^1\times S^2, S^2\times 0\cup S^2\times 1\right)$};
\draw [->] (8.75,-3.825) -- (4,-4.825);
\node [rotate=10] at (6.375,-4.325) [above] {picture};
\end{tikzpicture}
\begin{tikzpicture} [scale=1.25]
\draw (0,0) ellipse (1 and 0.5);
\draw [dashed] (-1,-4) arc (180:0:1 and 0.5);
\draw (-1,-4) arc (-180:0:1 and 0.5);
\draw (-1,0) -- (-1,-1.25);
\draw (-1,-2.75) -- (-1,-4);
\draw (1,0) -- (1,-4);
\draw [dashed] (-1,-1.25) arc (90:270:0.375 and 0.75);
\draw (-1,-1.25) arc (90:-90:0.375 and 0.75);
\draw (-4,-2) arc (180:44:1.25 and 1);
\draw (-4,-2) arc (180:316:1.25 and 1);
\draw (-1.85,-1.305) to [out=-45, in=180] (-1,-1.25);
\draw (-1.85,-2.695) to [out=45, in=180] (-1,-2.75);
\draw (-3.15,-2) arc (180:0:0.4 and 0.2);
\draw (-3,-2) arc (-180:0:0.25 and 0.125);
\node at (-2.5,-2.5) {$S^2\times S^1\setminus B^3$};
\node at (1,0) [right] {$S^2_\text{top}$};
\node at (1,-4) [right] {$S^2_\text{bot}$};
\draw [->] (1.25,-2) -- (4,-2);
\node [align=left] at (2.625,-2) {simplified picture\\for Figure \ref{fig:cobordism4d}};
\path (1.125,-2) -- (4.125,-2);
\end{tikzpicture}
\begin{tikzpicture} [scale=1.25]
\draw (-1,0) -- (-1,-1.25);
\draw (-1,-2.75) -- (-1,-4);
\draw (-4,-2) arc (180:44:1.25 and 1);
\draw (-4,-2) arc (180:316:1.25 and 1);
\draw (-1.85,-1.305) to [out=-45, in=-90] (-1,-1.25);
\draw (-1.85,-2.695) to [out=45, in=90] (-1,-2.75);
\node at (-1,0) {$\bullet$};
\node at (-1,0) [left] {$S^2_\text{top}$};
\node at (-1.3,-1.48) {$\bullet$};
\node at (-1.3,-2.52) {$\bullet$};
\node at (-1,-4) {$\bullet$};
\node at (-1,-4) [left] {$S^2_\text{bot}$};
\path (-1,0.5) -- (-1,-4.5);
\end{tikzpicture}
\caption{3D cobordism. The unlabelled two dots in the bottom right figure denote the 2-sphere along which the connected sum occurs.}
\label{fig:cobordism3d}
\end{figure*}
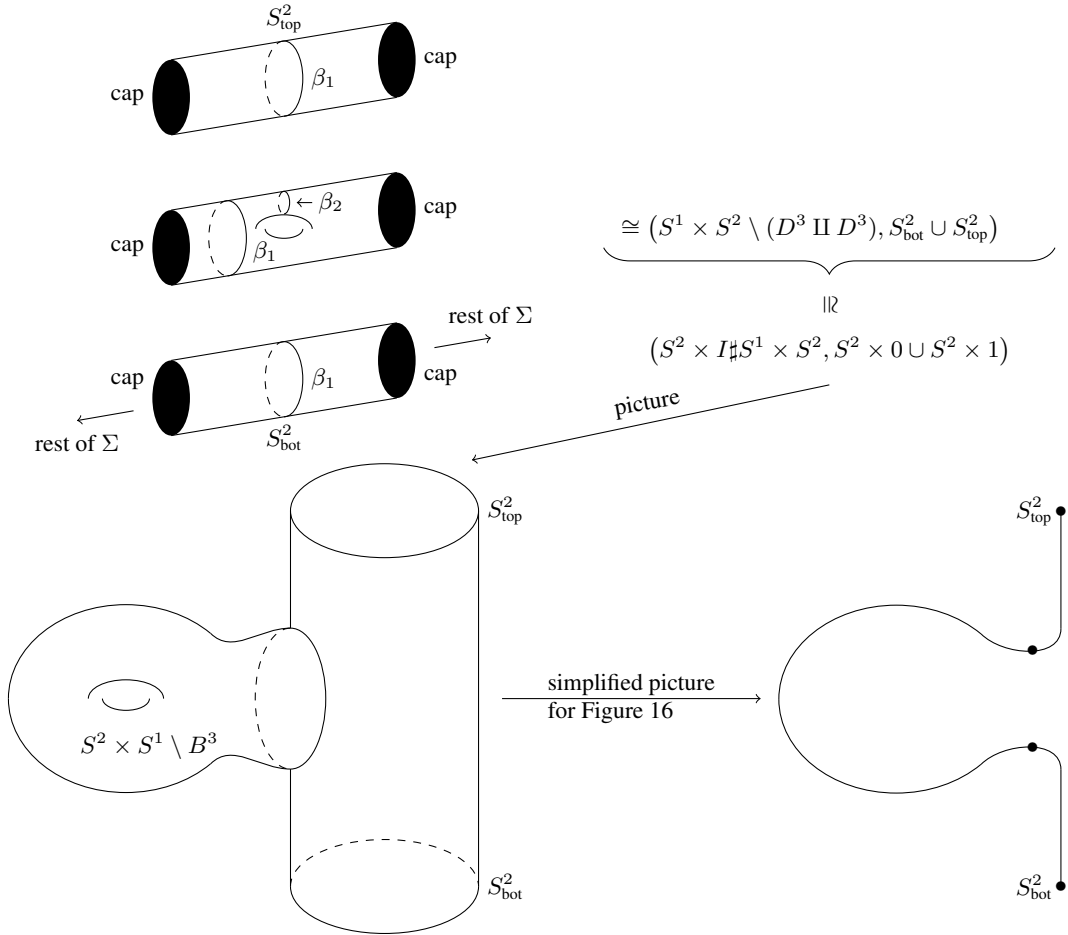

We should choose $L$ to be spanned by $\beta_2$ (since $P_{\beta_2}\vert\psi\rangle=1$) and either $\beta_1^+$ or $\beta_1^-$, let's say $\beta_1^+$.

To match the algebraic calculation (Sec. \ref{algebraicSec}), the surface $\Sigma$ should be extended by choosing $L=\operatorname{span}\{\beta\text{s}\}$.  The $\beta$s include $\beta_1^+$ and the stabilizing $\beta_2$ and the other meridians to $\Sigma$ (if genus $\Sigma > 1$).  Restricting to $Y^+$ this is the $L$ specified above.  Again, following the notation of Ref. \onlinecite{walker1991} we pass to the capped off surface $\hat{Y}^+$, a torus with $L=\operatorname{span}\left(\beta_2\right)$, and compute the signature $\sigma$ of the 4D cobordism associated to the three measurement ($=$ 4D $2$-handle attachments).

From the 4D perspective, the entire measurement protocol for $D_\gamma$ (up to phase) (see Figure \ref{fig:neighborhood1}) has five (not eight) steps: a 3D $1$-handle is attached to add one genus and a 3D $2$-handle is attached to $\beta_2$ to remove the added genus.  This produces a 3D cobordism which we product with $I$ and then attach three 4D $2$-handles.  We may localize to where the action is, $Y$.  $\hat{Y}$ is a $2$-sphere and the (localized) 3D cobordism we construct turns out to be $(S^2\times I\sharp S^1\times S^2, S^2\times 0\cup S^2\times 1)$ as shown in Figure \ref{fig:sliding}(b).

\begin{figure*}[hbpt]
\begin{tikzpicture}
\draw (-0.25,0) -- (0.25,0);
\draw (-0.25,-4) -- (0.25,-4);
\draw (-0.25,0) -- (-0.25,-1) to [out=-90, in=0] (-0.5,-1.25);
\draw (-0.5,-1.25) arc (90:270:1 and 0.75);
\draw (-0.5,-2.75) to [out=0, in=90] (-0.25,-3) -- (-0.25,-4);
\draw (0.25,0) -- (0.25,-1.5) to [out=-90, in=0] (-0.25,-1.75);
\draw (-0.25,-1.75) arc (90:270:0.6 and 0.25);
\draw (-0.5,-2.25) to [out=0, in=90] (0.25,-2.5) -- (0.25,-4);
\draw (0.25,-0.25) arc (90:-90:0.75 and 0.5);
\draw (0.25,-0.45) arc (90:-90:0.6 and 0.3);
\draw (0.25,-1.5) arc (90:-90:0.75 and 0.5);
\draw (0.25,-1.7) arc (90:-90:0.6 and 0.3);
\draw (0.25,-1.7) arc (90:100:0.6 and 0.3);
\draw (0.25,-2.3) arc (-90:-100:0.6 and 0.3);
\draw (0.25,-2.75) arc (90:-90:0.75 and 0.5);
\draw (0.25,-2.95) arc (90:-90:0.6 and 0.3);
\node at (0,-0.75) {$\overset{\times I}{\rightarrow}$};
\node at (0,-3.4325) {$\rightarrow$};
\node at (-0.25,0) {$\bullet$};
\node at (0,0) [above] {$S^2_\text{top}\times I$};
\node at (-0.25,-4) {$\bullet$};
\node at (0,-4) [below] {$S^2_\text{bot}\times I$};
\draw [->] (2.75,0) to [out=-90, in=0] (1.125,-0.75);
\draw [->] (2.75,0) to [out=-90, in=0] (1.125,-2);
\draw [->] (2.75,0) to [out=-90, in=0] (1.125,-3.25);
\node at (2.75,0) [above] {three 4D $2$-handles};
\node at (0,-5) {(a)};
\draw [->] (3,-2) -- (5.5,-2);
\node [align=left] at (3.75,-2) {or even more\\schematically};
\end{tikzpicture}\hspace{-0.9in}
\begin{tikzpicture}
\draw (-0.25,0) -- (0.25,0);
\draw (-0.25,-4) -- (0.25,-4);
\draw (-0.25,0) -- (-0.25,-1) to [out=-90, in=0] (-0.5,-1.25);
\draw (-0.5,-1.25) arc (90:270:1 and 0.75);
\draw (-0.5,-2.75) to [out=0, in=90] (-0.25,-3) -- (-0.25,-4);
\draw (0.25,0) -- (0.25,-1) to [out=-90, in=180] (0.5,-1.25);
\draw (0.5,-1.25) arc (90:-90:1 and 0.75);
\draw (0.5,-2.75) to [out=180, in=90] (0.25,-3) -- (0.25,-4);
\node at (0,0) [above] {$S^2_\text{top}\times I$};
\node [align=left] at (0,-2) {4D bulk with\\signature $\sigma$};
\node at (0,-4) [below] {$S^2_\text{bot}\times I$};
\node at (0,-5) {(b)};
\draw [<->] (-0.75,0) to [out=-90, in=90] (-1.75,-2) to [out=-90, in=90] (-0.75,-4);
\draw [<->] (0.75,0) to [out=-90, in=90] (1.75,-2) to [out=-90, in=90] (0.75,-4);
\node at (-1.25,-3.25) [left] {no Dehn twist};
\node [align=left] at (1.5,-3.25) [right] {Dehn twist $D_\gamma$\\between the caps};
\end{tikzpicture}\\
\begin{tikzpicture}[scale=0.75]
\path [fill=lightgray, draw=black] (0,1.5) ellipse (0.25 and 1);
\path [fill=lightgray, draw=black] (0,-1.5) ellipse (0.25 and 1);
\filldraw [fill=lightgray, draw=black] (5.5,-0.375) ellipse (1.5 and 0.5);
\fill [lightgray] (5.5,0.375) ellipse (1.5 and 0.5);
\draw [dashed] (7,0.375) arc (0:197:1.5 and 0.5);
\begin{scope}
    \clip (5.5,0.375) ellipse (1.5 and 0.5);
    \draw [dashed] (5.5,-0.375) ellipse (1.5 and 0.5);
\end{scope}
\draw (7,-0.375) to [out=-90, in=0] (0,-2.5) (0,-0.5) to [out=0, in=0] (0,0.5) (0,2.5) to
    [out=0, in=90] (7,0.375) arc (0:-163:1.5 and 0.5);
\node at (7,-0.375) [below right] {$S^2_\text{top} = S^2\times 1$};
\begin{scope}
    \clip (4.5,0.5) to [out=135, in=90] (2.5,0.5) -- (1,2) to [out=0, in=135] (5,1);
    \filldraw [white, line width=1.1] (5.5,0.375) ellipse (1.5 and 0.5);
\end{scope}
\begin{scope}
    \clip (4.5,-0.5) to [out=-135, in=-90] (2.5,-0.5) -- (1,-2) to [out=0, in=-135] (5,-1);
    \filldraw [white, line width=1.1] (5.5,-0.375) ellipse (1.5 and 0.5);
\end{scope}
\path [fill=lightgray, draw=black] (1,1.5) ellipse (0.33 and 0.5);
\path [fill=lightgray, draw=black] (1,-1.5) ellipse (0.33 and 0.5);
\filldraw [fill=lightgray, draw=black, rotate around={-45:(4.75,0.75)}]
    (4.75,0.75) ellipse (0.17 and 0.35);
\filldraw [fill=lightgray, draw=black, rotate around={45:(4.75,-0.75)}]
    (4.75,-0.75) ellipse (0.17 and 0.35);
\draw (4.5,0.5) to [out=135, in=90] (2.5,0.5) -- (2.5,-0.5) to
    [out=-90, in=-135] (4.5,-0.5) (5,-1) to [out=-135, in=0] (1,-2) (1,-1) to [out=0, in=0] (1,1) (1,2) to [out=0, in=135] (5,1);
\draw (4.11,0.1875) to [out=135, in=90] (3,0) to [out=-90, in=-135] (4.11,-0.1875);
\draw [rotate around={-30:(4,1.25)}, line width=5, white] (4,1.25) ++ (90:0.33 and 0.625)
    arc (90:-90:0.33 and 0.625);
\draw [rotate around={-30:(4,1.25)}] (4,1.25) ++ (130:0.33 and 0.625) arc
    (130:-115:0.33 and 0.625) (4,1.25) ++ (155:0.33 and 0.625) arc (155:215:0.33 and 0.625);
\node at (4.5,2)  {$0$};
\draw [line width=5, white] (4.5625,1.625) -- (6,2.5);
\draw [<-, align=left] (4.5625,1.625) -- (6,2.5) node [right]
    {$0$-framed 3D $1$-surgery\\on $\beta_2$ at $t=\frac{1}{2}$};
\draw [line width=5, white] (4.5625,-1.625) -- (6,-2.5);
\draw [<-] (4.5625,-1.625) -- (6,-2.5) node [right] {$S^2_\text{bot} = S^2\times 0$};
\fill [lightgray] (7,-0.375) arc (0:-180:1.5 and 0.5) -- (7,-0.375);
\draw (7,-0.375) arc (0:-180:1.5 and 0.5);
\end{tikzpicture}
\[\text{(c) the Kirby diagram for the cobordism }(S^2\times I\sharp S^1\times S^2, S^2\times 0\cup S^2\times 1)\]
\caption{4D cobordism}
\label{fig:cobordism4d}
\end{figure*}
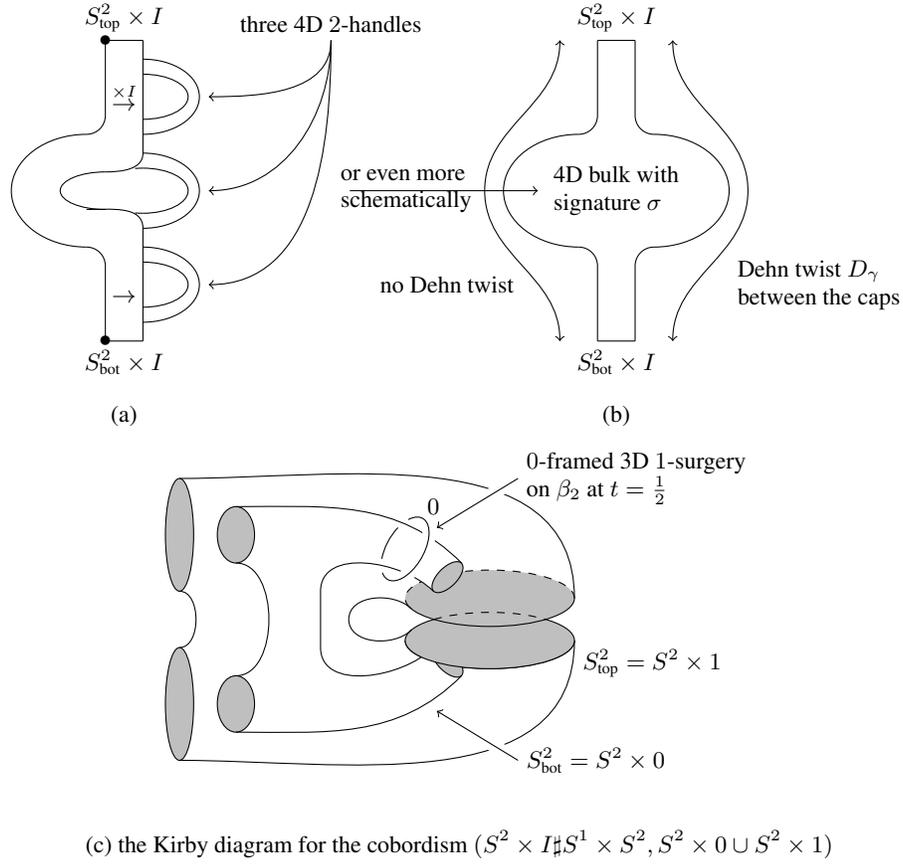

We must draw an \emph{exact} handle diagram and manipulate it according to the rules of the ``Kirby calculus'' to compute the signature of the 4D bulk, $\sigma$, the Dehn twist $D_\gamma$ between the caps, and what $3$-manifold arises on the right edge of Figure \ref{fig:cobordism4d}(b), the other end of the 4D cobordism.

But before drawing the diagram, we should explain how each 3D $1$-surgery effects the same transformation of the $3$-manifold as a pair of 3D (index $2$, index $1$)-handle attachments.  This identity explains why the eight steps of Section \ref{sec:3d} become five in the 4D perspective; each ($2,1$)-handle attachment pair becomes a single 3D $1$-surgery.  A 3D $1$-surgery has two steps: a tubular neighborhood $\mathfrak{N}(S^1) = S^1\times D^2$ is deleted from the $3$-manifold, and then another $D^2\times S^1$ is reglued according to the framing instructions.  In the case that the core circle, $S^1\times 0\subset S^1\times D^2$, lies in a level of a foliation by surfaces, $\mathfrak{N}$ may be arranged to consist of an interval $[-\epsilon,\epsilon]$ of annular strips in surface levels (leaves) with the annuli degenerating to circles (i.e., annuli of zero thickness) at both $-\epsilon$ and $\epsilon$.  Then the replacement operation on levels $[-\epsilon,0]$ is precisely the attachment of a 3D $2$-handle to the circle ($=$ degenerate annulus) at level $=-\epsilon$.  Similarly the replacement operation on levels $[0,\epsilon]$ is the inverse or ``Poincar\'{e} dual'' 3D $1$-handle attachment.  This dual attachment is along the non-attaching (belt) region of the previously attached $2$-handle.

\begin{figure*}
\begin{tikzpicture}
\draw (0.2,0) arc (0:180:0.2 and 0.1);
\draw (0.1,0) arc (0:-180:0.1 and 0.05);
\draw (0,0) ellipse (0.5 and 0.25);
\draw (-1,0.6) to [out=10, in=170] (0,0.75) to [out=0, in=90] (1,0)
    to [out=-90, in=0] (0,-0.75) to [out=170, in=10] (-1,-0.6);
\draw (-1,0.4) ellipse (0.15 and 0.2);
\draw (-1,-0.4) ellipse (0.15 and 0.2);
\draw (-1,0.2) arc (90:-90:0.15 and 0.2);
\node at (-1,0.4) [above left] {$\beta_1^+$};
\node at (-1,-0.4) [below left] {$\beta_1^-$};
\node at (0,0.47) {$\alpha_2$};
\node at (0,-1.125) {$Y$};
\node at (1.1875,0) [right] {$=$};
\begin{scope} [shift={(5.25cm,0cm)}]
\fill [gray] (-1.2165,0.375) ellipse (0.7835 and 0.25);
\fill[lightgray!50!gray] (0,0) ellipse (1 and 1);
\fill[gray] (0.5,0) ellipse (1 and 1);
\begin{scope}
    \clip (0,0) ellipse (1 and 1);
    \fill [white] (0.5,0) ellipse (1 and 1);
\end{scope}
\draw (0,0) ellipse (1 and 1);
\draw (0.5,0) ellipse (1 and 1);
\draw (-2,0.375) arc (180:58:0.7835 and 0.25);
\draw (-2,0) arc (180:68:0.75 and 0.25);
\fill [gray] (-2.53,-0.1875) ellipse (0.75 and 0.75);
\fill [lightgray!50!gray] (-2.53,0.1875) ellipse (0.75 and 0.75);
\begin{scope}
    \clip (0,0) (-2.53,-0.1875) ellipse (0.75 and 0.75);
    \fill [white] (-2.53,0.1875) ellipse (0.75 and 0.75);
\end{scope}
\draw (-3.28,-0.1875) arc (180:335:0.75 and 0.75);
\draw (-3.28,-0.1875) arc (180:27:0.75 and 0.75);
\draw (-3.28,0.1875) arc (180:320:0.75 and 0.75);
\draw (-3.28,0.1875) arc (180:-3:0.75 and 0.75);
\path [fill=lightgray, draw=black] (-2,0.375) -- (-2,-0.375) arc (-180:0:0.7835 and 0.25)
    arc (202:158:1 and 1) arc (0:-180:0.7835 and 0.25);
\draw (-2,0.375) arc (-180:-206:0.7835 and 0.25);
\draw [dashed] (-0.433,0.375) arc (0:58:0.7835 and 0.25);
\draw (-0.5,0) arc (0:-180:0.75 and 0.25);
\draw [dashed] (-0.5,0) arc (0:68:0.75 and 0.25);
\draw (-1.25,0) ++ (131:0.75 and 0.25) arc (131:120:0.75 and 0.25);
\draw [dashed] (-2,0) arc (180:131:0.75 and 0.25);
\draw [dashed] (-0.433,-0.375) arc (0:180:0.7835 and 0.25);
\node at (-3.5,0.4) [above] {$\beta_1^+$};
\node at (-3.5,-0.4) [below] {$\beta_1^-$};
\node at (-0.5,0) [right] {$\alpha_2$};
\node at (-1.25,-1.125) {$Y$};
\draw [->] (1.5,0.75) -- (2,1);
\node at (1.75,0.875) [below right] {time (up in Figure \ref{fig:cobordism3d})};
\end{scope}
\end{tikzpicture}
\caption{}
\label{fig:kirby1}
\end{figure*}

\begin{figure*}
\begin{tikzpicture}
\draw (-1.75,-0.53125) ++ (55:0.5 and 0.46875) arc (55:180:0.5 and 0.46875);
\draw  (-0.625,0.375) ++ (90:0.125 and 0.875) arc (90:270:0.125 and 0.875);
\draw  (-0.625,0.375) ++ (0:0.125 and 0.875) arc (0:-60:0.125 and 0.875);
\draw [white, line width=3] (-0.625,0.375) ++ (0:0.625 and 0.375) arc (0:180:0.625 and 0.375);
\draw (-0.625,0.375) ++ (0:0.625 and 0.375) arc (0:180:0.625 and 0.375);
\draw [white, line width=3] (-0.625,0.375) ++ (0:0.5 and 0.25) arc (0:90:0.5 and 0.25) to
    [out=180, in=0] (-1.375,-0.25) to [out=180, in=-90] (-1.625,0.125);
\draw (-0.625,0.375) ++ (0:0.5 and 0.25) arc (0:90:0.5 and 0.25) to [out=180, in=0] (-1.375,-0.25) to
    [out=180, in=-90] (-1.625,0.125);
\draw [white, line width=3] (0.25,-1) arc (-90:90:1 and 1) to [out=180, in=0] (-0.25,0.875) to
    [out=180, in=0] (-0.625,0.9375) to [out=180, in=0] (-1,0.875) to [out=180, in=0] (-1.5,1) arc (90:270:0.75 and 0.75) to [out=0, in=180] (-0.875,-0.375) to [out=0, in=180] (0.25,-1);
\draw (0.25,-1) arc (-90:90:1 and 1) to [out=180, in=0] (-0.25,0.875) to [out=180, in=0]
    (-0.625,0.9375) to [out=180, in=0] (-1,0.875) to [out=180, in=0] (-1.5,1) arc (90:270:0.75 and 0.75) to [out=0, in=180] (-0.875,-0.375) to [out=0, in=180] (0.25,-1);
\draw [white, line width=3] (0.625,0) ellipse (1 and 1);
\draw (0.625,0) ellipse (1 and 1);
\draw [white, line width=3] (2,0) arc (0:160:1 and 1) to [out=-110, in=0] (-0.625,-0.25) to
    [out=180, in=0] (-1.5,0) arc (90:270:0.5 and 0.5) to [out=0, in=180] (-0.875,-0.875) to [out=0, in=180] (-0.25,-1);
\draw [white, line width=3] (2,0) arc (0:-130:1 and 1) to [out=-140, in=0] (-0.25,-1);
\draw (2,0) arc (0:160:1 and 1) to [out=-110, in=0] (-0.625,-0.25) to [out=180, in=0] (-1.5,0)
    arc (90:270:0.5 and 0.5) to [out=0, in=180] (-0.875,-0.875) to [out=0, in=180] (-0.25,-1);
\draw (2,0) arc (0:-130:1 and 1) to [out=-140, in=0] (-0.25,-1);
\draw [white, line width=3] (-0.625,0.375) ++ (-45:0.625 and 0.375) arc (-45:-180:0.625 and 0.375);
\draw (-0.625,0.375) ++ (0:0.625 and 0.375) arc (0:-180:0.625 and 0.375);
\draw [white, line width=3] (-1.625,0.125) to [out=90, in=180] (-1.375,0.5) to [out=0, in=180]
    (-0.625,0.125) arc (-90:0:0.5 and 0.25);
\draw (-1.625,0.125) to [out=90, in=180] (-1.375,0.5) to [out=0, in=180] (-0.625,0.125)
    arc (-90:0:0.5 and 0.25);
\draw [white, line width=3] (-0.625,0.375) ++ (0:0.125 and 0.875) arc (0:90:0.125 and 0.875);
\draw (-0.625,0.375) ++ (0:0.125 and 0.875) arc (0:90:0.125 and 0.875);
\draw [white, line width=3] (-0.625,0.375) ++ (-60:0.125 and 0.875) arc (-60:-90:0.125 and 0.875);
\draw (-0.625,0.375) ++ (-60:0.125 and 0.875) arc (-60:-90:0.125 and 0.875);
\draw [white, line width=3] (-1.75,-0.53125) ++ (-180:0.5 and 0.46875) arc (-180:55:0.5 and 0.46875);
\draw (-1.75,-0.53125) ++ (-180:0.5 and 0.46875) arc (-180:55:0.5 and 0.46875);
\node (1) at (-0.625,1.25) {$\bullet$};
\node (2) at (-0.625,2.25) {4D $1$-handle (dual representation of $\alpha_2$)};
\path [->] (2) edge (1);
\node at (2.5,1.75) {$0$-framed 4D $2$-handle from $P_{\alpha_2}$};
\path [white, line width=3] (0,1.5) edge (-0.25,0.75);
\path [->] (0,1.5) edge (-0.25,0.75);
\node [align=left] (3) at (4,0.25) {$0$-framed 4D\\$2$-handle from $P_{\beta_2}$};
\path [white, line width=3] (3) edge (1.75,0);
\path [->] (3) edge (1.75,0);
\node [align=left] (4) at (-4.75,0.25) {$-1$-framed 4D $2$-handle\\from $P_{\alpha_2+\gamma}$};
\path [white, line width=3] (4) edge (-1.75,0.125);
\path [->] (4) edge (-1.75,0.125);
\node at (2.125,-0.625) {$\beta_1^+$};
\node at (0.9375,0) {$\beta_1^-$};
\node at (-0.125,-1.75) {Wilson loop of charge $a$ along $\beta_1$};
\path [->] (-1.75,-1.5) edge (-1.75,-1.125);
\node at (-0.125,-2.25) {(a)};
\end{tikzpicture}
\[\text{temporarily drop the Wilson loop and $\beta_1^\pm$ and simplify the link diagram:}\]
\begin{tikzpicture}
\draw (1.75,1) arc (90:-90:0.125 and 0.375);
\draw (4.5,0) arc (0:-180:1 and 1);
\draw (0,0) arc (-180:-90:0.375 and 0.375) to [out=0, in=180] (1.75,0.5);
\draw [white, line width=3] (1.75,0) ellipse (1.25 and 0.75);
\draw (1.75,0) ellipse (1.25 and 0.75);
\draw [white, line width=3] (0,0) arc (180:90:0.375 and 0.375) to [out=0, in=180] (1.75,-0.5)
    arc (-90:90:1 and 0.5);
\draw (0,0) arc (180:90:0.375 and 0.375) to [out=0, in=180] (1.75,-0.5) arc (-90:90:1 and 0.5);
\draw [white, line width=3] (4.5,0) arc (0:180:1 and 1);
\draw (4.5,0) arc (0:180:1 and 1);
\draw [white, line width=3] (1.75,1) arc (90:270:0.125 and 0.375);
\draw (1.75,1) arc (90:270:0.125 and 0.375);
\node at (1.75,1) {$\bullet$};
\node at (2.25,-1.5) {(b)};
\node at (1.125,0.525) {$\vdots$};
\node at (4.5,0.75) {$0$};
\path [->] (5,0) edge node [above] {handle slide along $\vdots$} (8.5,0);
\begin{scope}[shift={(9,0)}]
\draw (1.75,1) arc (90:-90:0.125 and 0.375);
\draw [white, line width=3] (0.5,0) arc (180:0:1.25 and 0.75);
\draw (0.5,0) arc (180:0:1.25 and 0.75);
\draw [white, line width=3] (3.5,0) ellipse (1 and 1);
\draw (3.5,0) ellipse (1 and 1);
\draw (0,0) arc (180:270:0.375 and 0.375) to [out=0, in=-90] (1.25,0.375) to [out=90, in=90]
    (0.75,0);
\draw [white, line width=3] (0,0) arc (180:90:0.375 and 0.375) to [out=0, in=180] (1.75,-0.375) to
    [out=0, in=-90] (2.625,0) arc (180:0:0.125 and 0.125)to [out=-90, in=0] (1.75,-0.625) to [out=180, in=-90] (0.75,0);
\draw (0,0) arc (180:90:0.375 and 0.375) to [out=0, in=180] (1.75,-0.375) to [out=0, in=-90]
    (2.625,0) arc (180:0:0.125 and 0.125)to [out=-90, in=0] (1.75,-0.625) to [out=180, in=-90] (0.75,0);
\draw [white, line width=3] (0.5,0) arc (-180:0:1.25 and 0.75);
\draw (0.5,0) arc (-180:0:1.25 and 0.75);
\draw [white, line width=3] (1.75,1) arc (90:270:0.125 and 0.25);
\draw (1.75,1) arc (90:270:0.125 and 0.375);
\node at (1.75,1) {$\bullet$};
\node at (2.25,-1.5) {(c)};
\node at (4.5,0.75) {$0$};
\end{scope}
\draw [->] (8.5,-1.75) -- (5,-2.25);
\node [rotate=8] at (6.75,-1.75) {Morse cancel};
\begin{scope}[shift={(-0.75,-3)}]
\draw (4.25,0) ellipse (1 and 1);
\draw (0,0) arc (180:270:0.375 and 0.375) to [out=0, in=-90] (1.25,0.375) to [out=90, in=90]
    (0.75,0);
\draw [white, line width=3] (0,0) arc (180:90:0.375 and 0.375) to [out=0, in=180] (1.75,-0.375) to
    [out=0, in=-90] (2.625,0) arc (180:0:0.125 and 0.125)to [out=-90, in=0] (1.75,-0.625) to [out=180, in=-90] (0.75,0);
\draw (0,0) arc (180:90:0.375 and 0.375) to [out=0, in=180] (1.75,-0.375) to [out=0, in=-90]
    (2.625,0) arc (180:0:0.125 and 0.125)to [out=-90, in=0] (1.75,-0.625) to [out=180, in=-90] (0.75,0);
\end{scope}
\node at (2.25,-4.5) {(d)};
\node at (6.75,-3) {$=$};
\node at (7.5,-3.75) {$\overset{\partial}{\cong} S^1\times S^2$};
\begin{scope}[shift={(9,-3)}]
\draw (1.575,0) ellipse (0.5 and 0.5);
\draw (2.925,0) ellipse (0.5 and 0.5);
\node at (2.075,0.25) [above] {$+1$};
\node at (3.425,0.25) [above] {$0$};
\node at (2.25,-1.5) {(e)};
\end{scope}
\end{tikzpicture}
\caption{}
\label{fig:kirby2}
\end{figure*}

\begin{figure*}
\begin{tikzpicture}[scale=1.25]
\draw [->] (0,0) ++ (35:0.5 and 0.5) arc (35:270:0.5 and 0.5);
\draw (0,0) ++ (270:0.5 and 0.5) arc (270:370:0.5 and 0.5);
\draw (0.7,-0.4) ++ (120:0.25 and 0.25) arc (120:-90:0.25 and 0.25);
\draw [->] (0.7,-0.4) ++ (-200:0.25 and 0.25) arc (-200:-90:0.25 and 0.25);
\draw (0.75,0) ++ (90:0.35 and 0.35) arc (90:195:0.35 and 0.35);
\draw (0.75,0) ++ (230:0.35 and 0.35) arc (230:290:0.35 and 0.35);
\draw (1.9,0) ellipse (1 and 0.6);
\draw (1.9,0) ++ (90:0.8 and 0.4) arc (90:-90:0.8 and 0.4);
\draw (1.9,0.4) to [out=180, in=20] (1.1,-0.2625);
\draw [white, line width=5] (1.9,-0.4) to [out=180, in=0] (0.75,0.35);
\draw (1.9,-0.4) to [out=180, in=0] (0.75,0.35);
\draw [->] (1.9,-0.6) ++ (205:0.125 and 0.125) arc (205:270:0.125 and 0.125);
\draw [white, line width=5] (1.9,-0.6) ++ (270:0.125 and 0.125) arc
    (270:375:0.125 and 0.125);
\draw (1.9,-0.6) ++ (270:0.125 and 0.125) arc (270:515:0.125 and 0.125);
\draw [white, line width=5] (3.1,0) ++ (185:0.5 and 0.5) arc (185:-125:0.5 and 0.5);
\draw (3.1,0) ++ (185:0.5 and 0.5) arc (185:-125:0.5 and 0.5);
\draw (3.1,0) ++ (-145:0.5 and 0.5) arc (-145:-152.5:0.5 and 0.5);
\draw [white, line width=5] (1.9,0.5) ++ (30:0.3 and 0.3) arc (30:320:0.3 and 0.3);
\draw (1.9,0.5) ++ (30:0.3 and 0.3) arc (30:320:0.3 and 0.3);
\draw (1.9,0.5) ++ (-10:0.3 and 0.3) arc (-10:2.5:0.3 and 0.3);
\node at (1.9,0.8) {$\bullet$};
\node at (0,-0.5) [below] {$a$};
\node at (0.7,-0.65) [below] {$b_1$};
\node at (1.9,-0.725) [below] {$b_2$};
\node at (3.6,0.25) [above] {$0$};
\node [align=left] at (4.5,0) {$2$-handle\\slide};
\draw [->] (3.875,0) -- (5.125,0);
\begin{scope} [shift={(5.4,0)}]
\draw (0,0) arc (180:390:0.25 and 0.25);
\draw (0,0) arc (180:60:0.25 and 0.25);
\draw (0.75,0) ++ (90:0.375 and 0.375) arc (90:195:0.375 and 0.375);
\draw (0.75,0) ++ (-140:0.375 and 0.375) arc (-140:-100:0.375 and 0.375);
\draw (0.9,-0.375) to [out=0, in=35] (1.375,0.625);
\draw (2.25,0) ellipse (1.5 and 1.125);
\draw [white, line width=5] (3.5,0) arc (0:-180:1.25 and 0.875);
\draw (3.5,0) arc (0:-226:1.25 and 0.875);
\draw [white, line width=5] (0.75,0.375) to [out=0, in=180] (2.25,-0.375);
\draw (0.75,0.375) to [out=0, in=180] (2.25,-0.375) to [out=0, in=-90] (3,0)
    arc (180:0:0.25 and 0.25);
\draw [->] (0.625,-0.34375) ++ (-170:0.125 and 0.125) arc (-170:-90:0.125 and 0.125);
\draw (0.625,-0.34375) ++ (120:0.125 and 0.125) arc (120:-90:0.125 and 0.125);
\draw [white, line width=5] (1.1875,0.625) ++ (-15:0.25 and 0.25)
    arc (-15:170:0.25 and 0.25);
\draw [->] (1.1875,0.625) ++ (-85:0.25 and 0.25) arc (-85:45:0.25 and 0.25);
\draw (1.1875,0.625) ++ (45:0.25 and 0.25) arc (45:170:0.25 and 0.25);
\draw (1.1875,0.625) ++ (210:0.25 and 0.25) arc (210:240:0.25 and 0.25);
\draw (2.25,1.125) ++ (-155:0.25 and 0.25) arc (-155:160:0.25 and 0.25);
\draw [white, line width=5] (4.1,0) ++ (-130:0.5 and 0.5) arc (-130:160:0.5 and 0.5);
\draw (4.1,0) ++ (-130:0.5 and 0.5) arc (-130:200:0.5 and 0.5);
\node at (2.25,1.375) {$\bullet$};
\node at (0.625,-0.46875) [below] {$b_1$};
\node at (1.1875,0.875) [above] {$b_2$};
\node at (4.6,0.25) [above] {$0$};
\end{scope}
\draw [->] (5.125,-2) -- (3.875,-2.5);
\node [rotate=26, align=left] at (4.5,-2.2) {Morse\\cancel};
\begin{scope}[shift={(-2,-4.0)}]
\draw (0.25,0) arc (180:400:0.125 and 0.125);
\draw [->] (0.375,0) ++ (110:0.125 and 0.125) arc (110:180:0.125 and 0.125);
\draw (0.75,0) ++ (90:0.375 and 0.375) arc (90:185:0.375 and 0.375);
\draw (0.75,0) ++ (-150:0.375 and 0.375) arc (-150:-110:0.375 and 0.375);
\draw (0.75,-0.375) to [out=0, in=35] (1.375,0.625);
\draw [white, line width=5] (3.5,0) arc (0:-180:1.25 and 0.875);
\draw (3.5,0) arc (0:-226:1.25 and 0.875);
\draw [white, line width=5] (0.75,0.375) to [out=0, in=180] (2.25,-0.375);
\draw (0.75,0.375) to [out=0, in=180] (2.25,-0.375) to [out=0, in=-90] (3,0)
    arc (180:0:0.25 and 0.25);
\draw (0.5625,-0.3125) ++ (110:0.125 and 0.125) arc (110:-120:0.125 and 0.125);
\draw [->] (0.5625,-0.3125) ++ (-180:0.125 and 0.125) arc (-180:-120:0.125 and 0.125);
\draw [->] (0.9,-0.325) ++ (170:0.125 and 0.125) arc (170:-60:0.125 and 0.125);
\draw (0.9,-0.325) ++ (-120:0.125 and 0.125) arc (-120:-60:0.125 and 0.125);
\draw (4.1,0) ellipse (0.5 and 0.5);
\node at (0.375,-0.125) [left] {$a$};
\node at (0.4425,-0.625) {$b_1$};
\node at (1,-0.75) {$b_2$};
\node at (3.25,0.25) [above] {$+1$};
\node at (4.6,0.25) [above] {$0$};
\draw [->] (4.875,0) -- (6.125,0);
\end{scope}
\begin{scope} [shift={(5,-4.0)}]
\draw (0.5,0) ellipse (0.5 and 0.5);
\draw (1.75,0) ellipse (0.5 and 0.5);
\draw [white, line width=3] (0.125,0.3125) ++ (20:0.125 and 0.125)
    arc (20:-280:0.125 and 0.125);
\draw [->] (0.125,0.3125) ++ (-280:0.125 and 0.125) arc (-280:-180:0.125 and 0.125);
\draw (0.125,0.3125) ++ (-180:0.125 and 0.125) arc (-180:20:0.125 and 0.125);
\draw [white, line width=3] (0,0) ++ (50:0.125 and 0.125) arc (50:-250:0.125 and 0.125);
\draw (0,0) ++ (50:0.125 and 0.125) arc (50:-180:0.125 and 0.125);
\draw [->] (0,0) ++ (-250:0.125 and 0.125) arc (-250:-180:0.125 and 0.125);
\draw [white, line width=3] (0.125,-0.3125) ++ (95:0.125 and 0.125)
    arc (95:-205:0.125 and 0.125);
\draw [->] (0.125,-0.3125) ++ (95:0.125 and 0.125) arc (95:-180:0.125 and 0.125);
\draw (0.125,-0.3125) ++ (-180:0.125 and 0.125) arc (-180:-205:0.125 and 0.125);
\node at (1,0.25) [above] {$+1$};
\node at (2.25,0.25) [above] {$0$};
\node at (-0.4,0.375) {$a$};
\node at (-0.4,0) {$b_1$};
\node at (-0.4,-0.375) {$b_2$};
\end{scope}
\draw[->] (7.5,-4) -- (8.5,-4);
\begin{scope}[shift={(9.0,-2.5)}]
\draw (0,0) arc (180:0:1.125 and 1.125);
\draw (0.5,0) arc (180:0:0.75 and 0.75);
\draw (1,0) arc (180:0:0.375 and 0.375);
\draw (2.25,-2.625) arc (0:-180:1.125 and 1);
\draw (2,-2.625) arc (0:-180:0.75 and 0.625);
\draw (1.75,-2.625) arc (0:-180:0.375 and 0.25);
\draw (2.25,0) -- (2.25,-2.625)
    (2,0) -- (2,-2.625)
    (1.75,0) -- (1.75,-2.625);
\draw (0.375,-0.25) to [out=90, in=90] (0,-0.375);
\draw [white, line width=5] (0,0) -- (0,-0.125) to [out=-90, in=-90] (0.375,-0.25);
\draw (0,0) -- (0,-0.125) to [out=-90, in=-90] (0.375,-0.25);
\draw (0.875,-0.25) to [out=90, in=90] (0.5,-0.375);
\draw [white, line width=5] (0.5,0) -- (0.5,-0.125) to [out=-90, in=-90] (0.875,-0.25);
\draw (0.5,0) -- (0.5,-0.125) to [out=-90, in=-90] (0.875,-0.25);
\draw (1.375,-0.25) to [out=90, in=90] (1,-0.375);
\draw [white, line width=5] (1,0) -- (1,-0.125) to [out=-90, in=-90] (1.375,-0.25);
\draw (1,0) -- (1,-0.125) to [out=-90, in=-90] (1.375,-0.25);
\draw [rounded corners] (0,-0.375) -- (0,-0.5) -- (1,-1.5) -- (0,-2.5) -- (0,-2.625);
\draw [-<] (0,-2.5625) -- (0,-2.625);
\draw [rounded corners] (0.5,-0.375) -- (0.5,-0.5) -- (0,-1) -- (0,-2) -- (0.5,-2.5) -- (0.5,-2.625);
\draw [-<] (0.5,-2.5625) -- (0.5,-2.625);
\draw [rounded corners, ->] (1,-0.375) -- (1,-1) -- (0.5,-1.5) -- (1,-2) -- (1,-2.625);
\draw [white, line width=5] (0.125,-0.625) -- (0.875,-1.375);
\draw (0.125,-0.625) -- (0.875,-1.375);
\draw [white, line width=5] (0.625,-1.625) -- (0.875,-1.875);
\draw (0.625,-1.625) -- (0.875,-1.875);
\draw [white, line width=5] (0.125,-2.125) -- (0.375,-2.375);
\draw (0.125,-2.125) -- (0.375,-2.375);
\draw (3,-2) ellipse (0.5 and 0.5);
\node at (3.5,-1.75) [above] {$0$};
\node at (-0.1875,-2.5) [below] {$a$};
\node at (0.3125,-2.5) [below] {$b_1$};
\node at (0.875,-2.5) [below] {$b_2$};
\node at (1.75,-4.25) {Wilson loops};
\end{scope}
\end{tikzpicture}
\caption{}
\label{fig:loops}
\end{figure*}

To simplify notation, the calculation in Figure \ref{fig:kirby2} is done with $S^2_\text{top}$ and $S^2_\text{bot}$ both capped off by $3$-balls, and the resulting $S^1\times S^2$ itself capped off (on the left) by \tikz[baseline=-0.5ex]{\draw (0,0) ellipse (0.25 and 0.25); \node at (0,0.25) {$\bullet$};}, i.e. $S^1\times D^3$.  This allows us to read off the $3$-manifold and its right boundary.  Specifically the canceling $2$-handle (\ref{fig:kirby2}(c) $\rightarrow$ \ref{fig:kirby2}(d)) yields a 2-handle diagram in $S^3$ rather than in $\partial ( \tikz[baseline=-0.5ex]{\draw (0,0) ellipse (0.25 and 0.25); \node at (0,0.25) {$\bullet$};} )$; then the $+1$-framed handle cobords $S^3$ back to $S^3$ with a $\mathbb{C}P^2\setminus (D^4 \amalg D^4)$ as bulk.  Finally the $0$-framed handle gives a boundary-connected sum with $S^2\times D^2$.  In particular, the signature $\sigma=1$ and the right boundary is (again, up to diffeomorphism) $S^2\times I\sharp \left(S^1\times S^2\right)$.  For the record, the entire closed cobordism is diffeomorphic to $\mathbb{C}P^2 \setminus \left( \mathfrak{N}\left(S^1\right) \amalg \mathfrak{N} \left(S^1 \right) \right)$, complex projective space minus the disjoint union of two open neighborhoods of a circle.
One $\mathfrak{N}\left(S^1\right)$ is the filling $S^1 \times D^3$; the other is dual to the zero framed 2-handle.

Since the Kirby calculus is generally applied to 4-manifolds with connected boundary, in using it to describe a relative cobordism we have done two things. First we closed
the 3-manifolds by adding 3-balls. Second we filled the left boundary with $S^1 \times D^3$ to reduce to the case of connected boundary.

From the computed signature $\sigma=1$, and given the choice of Lagrangian $L_\Sigma \supset\operatorname{span}\left(\beta_1\right)$ (implied by the skein choice for enumerating basis states in Sec. \ref{gsdRev}), one computes the abelian phase $w$ in the formula:
\begin{align}
w D_\gamma^\dagger |\tilde{\Psi} \rangle = \mathcal{D}^3 P_{\beta_2} P_{\alpha_2+\gamma} P_{\alpha_2 } | \tilde{\Psi} \rangle.
\end{align}
The overall normalization, set by $\mathcal{D}^3$ is not computed simply from the topology of the four-manifold, but requires additional
TQFT data; here we obtain the overall normalization from the previous algebraic calculation.
The right side of the 4D cobordism, $M$, is a product up to an extraneous $\sharp S^1\times S^2$.  Using the canonical generator, $x$, \cite{walker1991} of the Hilbert space $\mathcal{V}\left(S^2\right)$, the $S^1\times S^2$ factor can be removed and $M$ induces a unitary from bottom to top exactly as a diffeomorphism would.  By Ref. \onlinecite{walker1991}
the abelian phase factor:
\begin{equation}\label{eq:factor}
w = e^\frac{2\pi ic(3\sigma)}{24}.
\end{equation}
$c$ is the exponentiated central charge of the TQFT, i.e.
\[\sum_{a \in \mathcal{C}}d_a^2 e^{i\theta_a} = e^\frac{2\pi ic}{8}.\]
The basic unit of tangential framing on a $3$-manifold is the Pontryagin number $p_1$ on a bounding $4$-manifold.  The Hirzebruch signature formula states (in dimension $4$):
\[\sigma = \frac{p_1}{3}.\]
This accounts for the factor of three on line (\ref{eq:factor}) above.  For notational convenience we did cap off, ``on the left,'' $S^1\times S^2$ by \tikz[baseline=-0.5ex]{\draw (0,0) ellipse (0.25 and 0.25); \node at (0,0.25) {$\bullet$};} $= S^1\times D^3$ but $\sigma\left(S^1\times D^3\right) = 0$ and so by additivity of signature under gluing along \emph{full} boundary components, this convenience does not affect the calculation of $\sigma$.

A final point is to extract the Dehn twist $D_\gamma$ from Figure \ref{fig:kirby2} and thus provide an independent check on the purely 3D calculation carried out in Figure \ref{fig:neighborhood1}.  With only a little extra work we can calculate the Dehn twist not only in the case where all particle type measurements are of the trivial charge but also in the more general case where we assume the measurements along $\alpha_2$ and $\alpha_2 + \gamma$ yield $b_1$ and $b_2$, respectively. Returning to Figure \ref{fig:kirby2}, we not only restore the deleted $a$-labeled Wilson loop but also add new loops labeled $b_1$ and $b_2$ (respectively) parallel to the cores of the newly glued solid tori.  We get Figure \ref{fig:loops}.

In the case that $a$, $b_1$, and $b_2$ are all Abelian particles, the Wilson loops generate a phase shift the overall phase induced by $\sigma$:
 \[ |\mathcal{N}| P_{\beta_2} P_{\alpha_2+\gamma} P_{\alpha_2} \vert b,0,0\rangle = e^\frac{2\pi ic}{8} e^{-i\theta_{a+b_1-b_2}}\vert b,0,0\rangle,\]
agreeing with the calculation of Sec. \ref{algebraicSecOther} for the overall normalization set by $|\mathcal{N}| = \mathcal{D}^3$. It is straightforward to
also consider the case where the intial and final projections along $\beta_2$ yield $b_0$ and $b_3$ respectively, though this is not explicitly shown in the figures,
where we also obtain agreement with the calculation of Sec. \ref{algebraicSecOther}.

\section{Discussion}
\label{disc}

\begin{figure}
	\centering
	\includegraphics[width=3.6in]{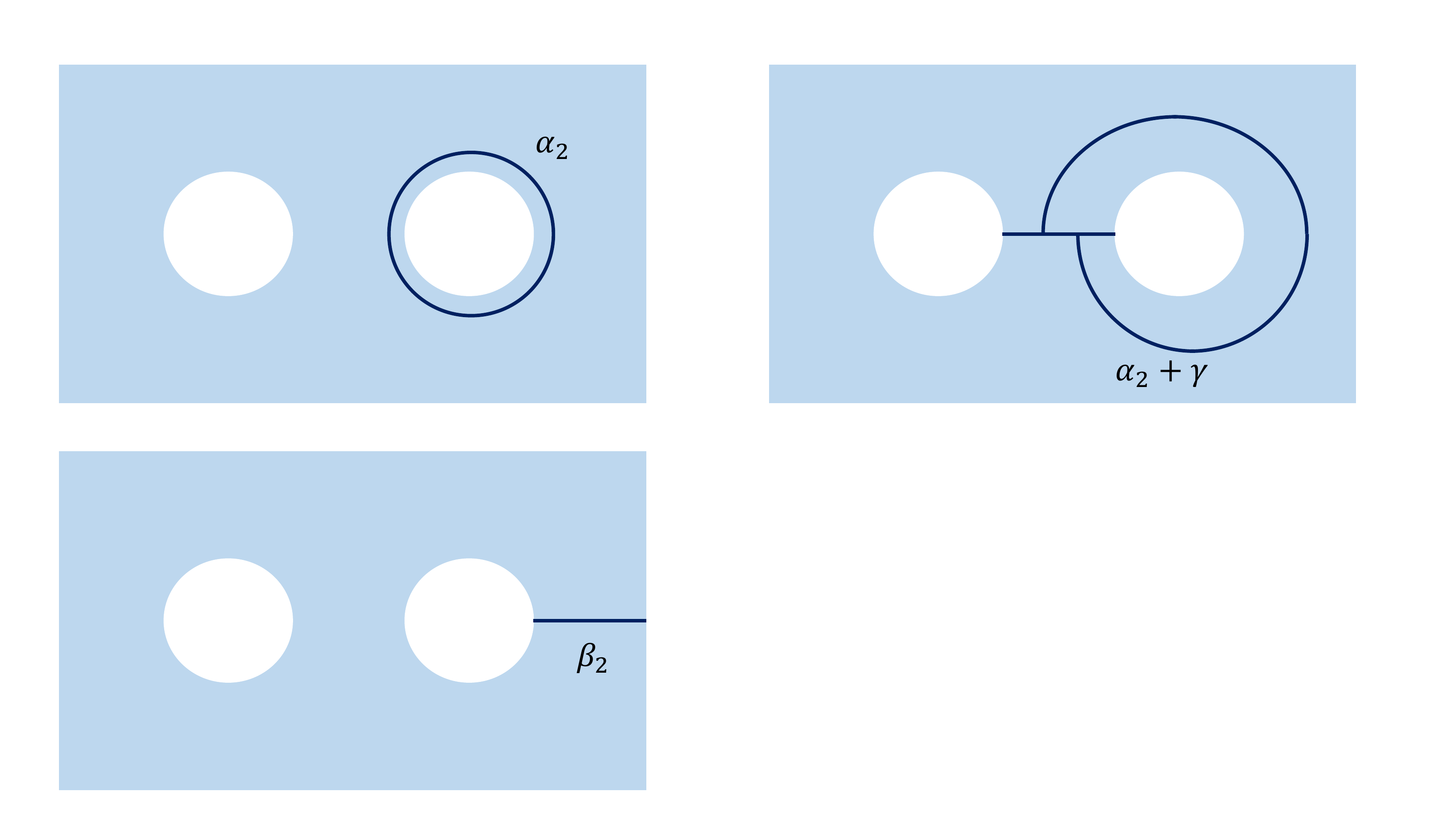}
	\caption{\label{gappedLines} A generic topological phase described by a unitary braided tensor category $\mathcal{B}$, together with
three disconnected gapped boundaries. When $\mathcal{B} = \mathcal{C} \otimes \overline{\mathcal{C}}$, the considerations of this paper are
associated with projections along the loop $\alpha_2$, the graph $\alpha_2 + \gamma$, which connects two
disconnected gapped boundaries, and the line $\beta_2$, which also connects two disconnected gapped boundaries. One could consider these sequences
of projections for more general $\mathcal{B}$ as well. }
\end{figure}

We have shown in this paper that a sequence of topological charge projections (measurements) can generate
Dehn twists on surfaces of non-trivial topology. We have seen that the mathematical theory yields a
well-defined overall phase $e^{2\pi i c /8}$ to the process that we have described. A
natural question is whether it is possible that other sequences of projections could give rise to the same operation, but with
a different overall phase $e^{2\pi i n c/8}$, for some integer $n$. This is indeed possible.
To show this, we use the Lantern relation, which relates a sequence of four Dehn twists to a sequence
of three Dehn twists.  Consider a disk with three punctures,
labelled $1$, $2$, $3$, arranged clockwise around a triangle. The Lantern relation is the identity:
\begin{align}
\label{lantern}
D_1  = D_{12} D_{13} D_{23} (D_2 D_3 D_{123})^{-1},
\end{align}
where $D_i$, $D_{ij}$, and $D_{123}$ are each right-handed Dehn twists around the $i$th puncture, the loop containing
the punctures $i$ and $j$, and the loop containing all three punctures, respectively.
In the preceding sections, we have shown that each Dehn twist can be related to a series of projections, multiplied
by an overall phase $e^{2\pi i c/8}$. From Eq. (\ref{lantern}), we see that $D_1$ can be expressed purely as a series of projections, and the overall
phase will cancel out in the product, realizing the case $n = 0$ above. The case for general $n$ is a straightforward generalization.

While the overall phase is mathematically well-defined, whether it is physically well-defined depends on how the topological charge projections
$P_\omega^{(a)}$ are realized. If they are realized as a non-Abelian Berry phase by adiabatically tuning microscopic parameters, as described in Sec. \ref{tprojSec},
then the overall phase will obtain a non-topological dynamical contribution as the ground state energy of the system changes in time. It is not clear
whether there is a physical procedure for performing the topological charge measurements in such a way that keeps the overall phase of the process
topologically protected and well-defined.

We note that in this paper we have assumed that the topological phases of interest are described by unitary modular tensor categories (UMTC),
which physically means that they arise from systems where the microscopic constituents are bosonic degrees of freedom. For systems where the microscopic
constituents are fermionic degrees of freedom, the corresponding braided tensor category is not modular. While similar results as those shown
here are expected to go through, we have not fully analyzed this case. In the discussions of the topology of the space-time history, one would
need to consider a Spin structure on the manifolds to carry out the analysis. One could, moreover, consider realizing the mapping class group
of non-orientable surfaces through projections, which would require considering a Pin structure on the manifolds of interest. We leave these
analyses for future work.

Furthermore, in this paper we made the connection between a topological phase, described by a UMTC $\mathcal{C}$, on a genus $g$ surface and a topological phase
$\mathcal{C} \otimes \overline{\mathcal{C}}$ on a disk with $n_b = g+1$ disconnected gapped boundaries, with a certain type of topological
boundary condition. The topological charge projections along the non-contractible cycles in the high genus surface are then associated with
topological projections along loops and also lines and graphs that connected the boundaries as shown in Fig. \ref{gappedLines}.

A more general scenario is to consider a generic topological phase, $\mathcal{B}$, in the presence of $n_b$ disconnected boundaries,
where each boundary is associated with some topological boundary condition. Topological phases in the presence of several
disconnected boundaries can have topological ground state degeneracies, and
one can then consider topological charge projections along loops and also lines and graphs that connect the different boundaries,
which give rise to unitary transformations on the topological state space. It may be interesting to revisit the topological quantum computing
schemes with the $Z_2$ surface code with these ideas in mind.\cite{fowler2012} In the case where $\mathcal{B}$ is Abelian,
there have been some general results developed in Ref. \onlinecite{barkeshli2013defect2}. It would be interesting to develop a
more general theory of such topologically protected unitary transformations.

A particularly relevant application of these results to the pursuit of universal topological quantum computation is in the context of
the Ising $\otimes$ $\overline{\text{Ising}}$ topological state. It well-known that the ability to perform topology change and Dehn twists
in the Ising topological state can provide the missing topological $\pi/8$ phase gate and thus enable universal
TQC. \cite{bravyi200universal,freedman2006universal} An adaptation of the ideas of Ref. \onlinecite{bravyi200universal,freedman2006universal}
to the case of the Ising $\otimes$ Ising state with genons has also been developed. \cite{barkeshli2013genon} The considerations of this paper
demonstrate that the Ising $\otimes \overline{\text{Ising}}$ state is also capable of supporting universal TQC, as
the required Dehn twists of the Ising state can be implemented through projections in the Ising $\otimes \overline{\text{Ising}}$ state
to appropriate eigenstates of loop, line, and graph operators in the presence of disconnected gapped boundaries. In the appendix we
provide some additional details of this protocol.

\section{Acknowledgments}

We thank P. Bonderson, K. Walker and Z. Wang for discussions on topology and measurements.

\appendix

\section{Topological charge projections and Wilson operators}

The topological charge projections $P_\omega^{(a)}$ discussed in this paper can be written in terms of Wilson loop operators of quasiparticles
by using the modular $S$ matrix. Specifically, let $W_a(\omega)$ be a Wilson loop operator for a quasiparticle $a$ encircling a loop $\omega$.
Then,
\begin{align}
P_{\omega}^{(a)} = \sum_{x \in \mathcal{C}} S_{0a} S_{xa}^* W_x(\omega) .
\end{align}
To understand this, note that a state with definite topological charge $b$ associated to the loop $\omega$ is an eigenstate of
$W_x(\omega)$, with eigenvalue $S_{xb}/S_{0b}$. Therefore, $P_{\omega}^{(a)}$ acting on this state gives
$\sum_{x\in\mathcal{C}} S_{0a} S_{xa}^* S_{xb}/S_{0b} = \delta_{ab}$.

This can be straightforwardly extended to topological charge projections associated to lines which connect distinct disconnected gapped boundaries
in a planar system. Suppose that we have a doubled topological phase of the form $\mathcal{C} \otimes \overline{\mathcal{C}}$, and gapped
boundaries such that quasiparticles of the form $(a,\overline{a})$, with $a \in \mathcal{C}$ and $\overline{a} \in \overline{\mathcal{C}}$,
can be removed at the boundary by local operators (that is, the quasiparticles
of the form $(a,\overline{a})$ are condensed on the boundary). This implies that there exist Wilson line operators $W_{(a,\overline{a})}(\gamma)$,
where $\gamma$ is a path that ends on the gapped boundaries, which keep the system in the ground state subspace. Consequently, one can define
\begin{align}
P_{\gamma}^{(a)} = \sum_{x \in \mathcal{C}} S_{0a} S_{xa}^* W_{(a,\overline{a})}(\gamma) .
\end{align}
This allows one to define a topological charge $a \in \mathcal{C}$ (or any superposition) to the open
line $\gamma$ which connects different gapped boundaries.

\section{Universal Topological Quantum Computation from the $\text{Ising} \otimes \overline{\text{Ising}}$ state}

Here we will provide some additional details about how to implement a universal set of gates for quantum computation
in the Ising $\otimes \overline{\text{Ising}}$ topological state, assuming the ability to carry out the topological charge
projections discussed in this paper.

The Ising $\otimes \overline{\text{Ising}}$ topological state has 9 topologically distinct types of quasiparticles, which we label as
$(a,\overline{b})$, with $a = \mathbb{I}, \sigma, \psi$ being the anyons of the Ising state, and $\overline{b} = \mathbb{I}, \overline{\sigma},\overline{\psi}$
the anyons of the $\overline{\text{Ising}}$ state. For a review of the topological properties of the Ising state, see Ref. \onlinecite{Bonderson07b}.
The quasiparticles have topological twists:
\begin{align}
e^{i\theta_{(a,\overline{b})}} = e^{2\pi i (h_a - h_b)},
\end{align}
where $h_{\mathbb{I}} = 0$, $h_{\psi} = 1/2$, $h_{\sigma} = 1/16$. The Ising anyons have the fusion rules:
\begin{align}
\sigma \times \sigma &= \mathbb{I} + \psi
\nonumber \\
\sigma \times \psi &= \sigma
\nonumber \\
\psi \times \psi &= \mathbb{I}.
\end{align}
The modular $S$ matrix of the Ising state is given by
\begin{align}
S = \frac{1}{2}\left( \begin{matrix}
1 &\sqrt{2} & 1 \\
\sqrt{2} & 0 & -\sqrt{2} \\
1 & -\sqrt{2} & 1 \\
\end{matrix} \right).
\end{align}

It is well-known that the non-Abelian braiding of quasiparticles in the Ising topological state is not sufficient to realize
a topologically protected universal set of gates for quantum computation. If we allow the possibility of measuring the
fusion channel of any $4$ $\sigma$ quasiparticles, then the only missing gate is the single-qubit $\pi/8$ phase gate.\cite{Bravyi00}
Below we will present a protocol to realize a topologically robust $\pi/8$ phase gate by using the Ising $\otimes \overline{\text{Ising}}$
topological state with gapped boundaries and topological charge projections. The protocol presented below is an adaptation of ideas of
Ref. \onlinecite{barkeshli2013genon} for realizing a robust $\pi/8$ phase gate in the Ising $\otimes$ Ising state with genons, which in turn are based on
ideas of Ref. \onlinecite{bravyi200universal,freedman2006universal} for realizing a topologically protected $\pi/8$ phase
gate in Ising systems when topology changes and Dehn twists are allowed.

\subsection{$\pi/8$ phase gate}

\begin{figure*}
	\centering
	\includegraphics[width=7in]{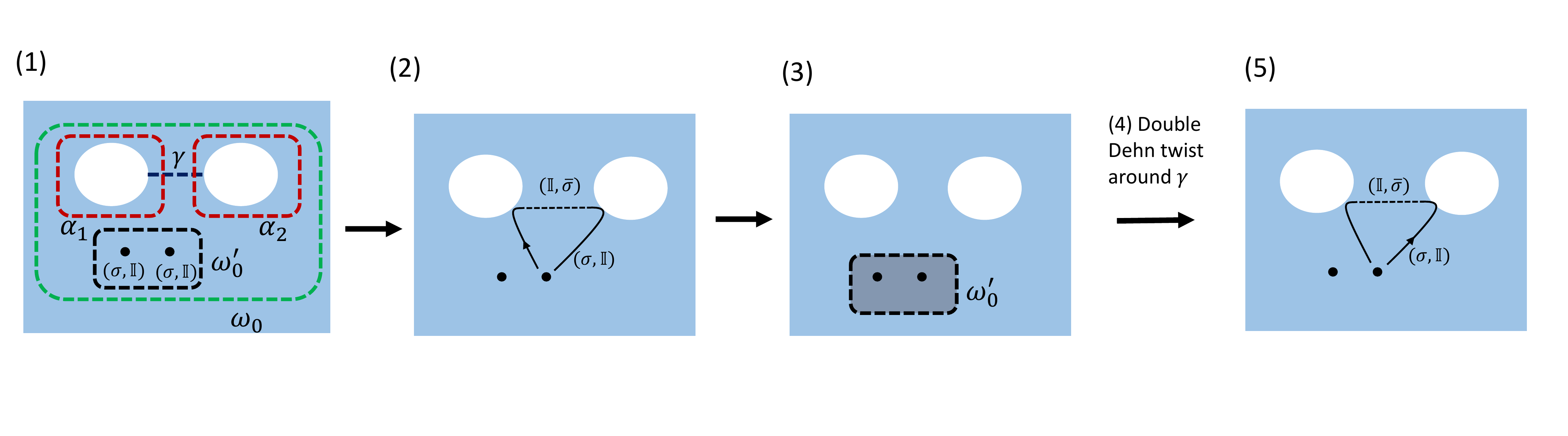}
	\caption{\label{DIsing} Illustration of protocol for $\pi/8$ phase gate. }
\end{figure*}

Let us consider a qubit encoded in the fusion channel of two $(\sigma, \mathbb{I})$ particles
of the Ising $\otimes \overline{\text{Ising}}$ state:
\begin{align}
(\sigma, \mathbb{I}) \times (\sigma, \mathbb{I}) = (\mathbb{I}, \mathbb{I}) + (\psi, \mathbb{I}) .
\end{align}
We consider two disconnected gapped boundaries, shown in Fig. \ref{DIsing}. We then consider the following protocol:
\begin{enumerate}
\item The two disconnected gapped boundaries are initialized in a state which has trivial topological charge, $(\mathbb{I},\mathbb{I})$,
through the loops $\alpha_1$ and $\alpha_2$, as shown. Consequently, the
topological charge through $\omega_0$ is equal to that of $\omega_0'$, and is equivalent to the fusion channel of
the two $(\sigma, \mathbb{I})$ particles of interest.
\item We apply a loop operator which takes one of the $(\sigma, \mathbb{I})$ particles to one of the gapped boundaries,
converts it to a $(\mathbb{I}, \overline{\sigma})$ particle through the action of a local operator on the boundary,
takes it to the other boundary, converts it back to a $(\sigma, \mathbb{I})$ particle, and brings it back to its original location.
\item The topological charge through the loop $\omega_0'$ (see Fig \ref{DIsing}) is projected to the identity, $(\mathbb{I}, \mathbb{I})$.
\item Perform a double ``Dehn twist" around $\gamma$ (see Fig. \ref{DIsing}).
\item Undo step (2) by applying the inverse loop operator.
\end{enumerate}
This protocol applies a relative phase of $e^{i\pi/4}$ to the state of the qubit, depending on whether the fusion channel is
$(\mathbb{I}, \mathbb{I})$ or $(\psi, \mathbb{I})$, thus implementing the single qubit $\pi/8$ phase gate.

To understand this, first observe that after step (1), the topological charge associated to the loop $\alpha_1$ (and also to $\alpha_2$) is $(\mathbb{I}, \mathbb{I})$.
Thus the topological charge associated to the line $\gamma$, which we will denote as $c(\gamma)$, is $c(\gamma) = \frac{1}{2}(\mathbb{I} + \sqrt{2} \sigma + \psi)$.
This is because in this situation, $c(\gamma)$ is related to $c(\alpha_1)$ by the modular $S$ matrix of the Ising state.

\begin{figure}
	\centering
	\includegraphics[width=3.0in]{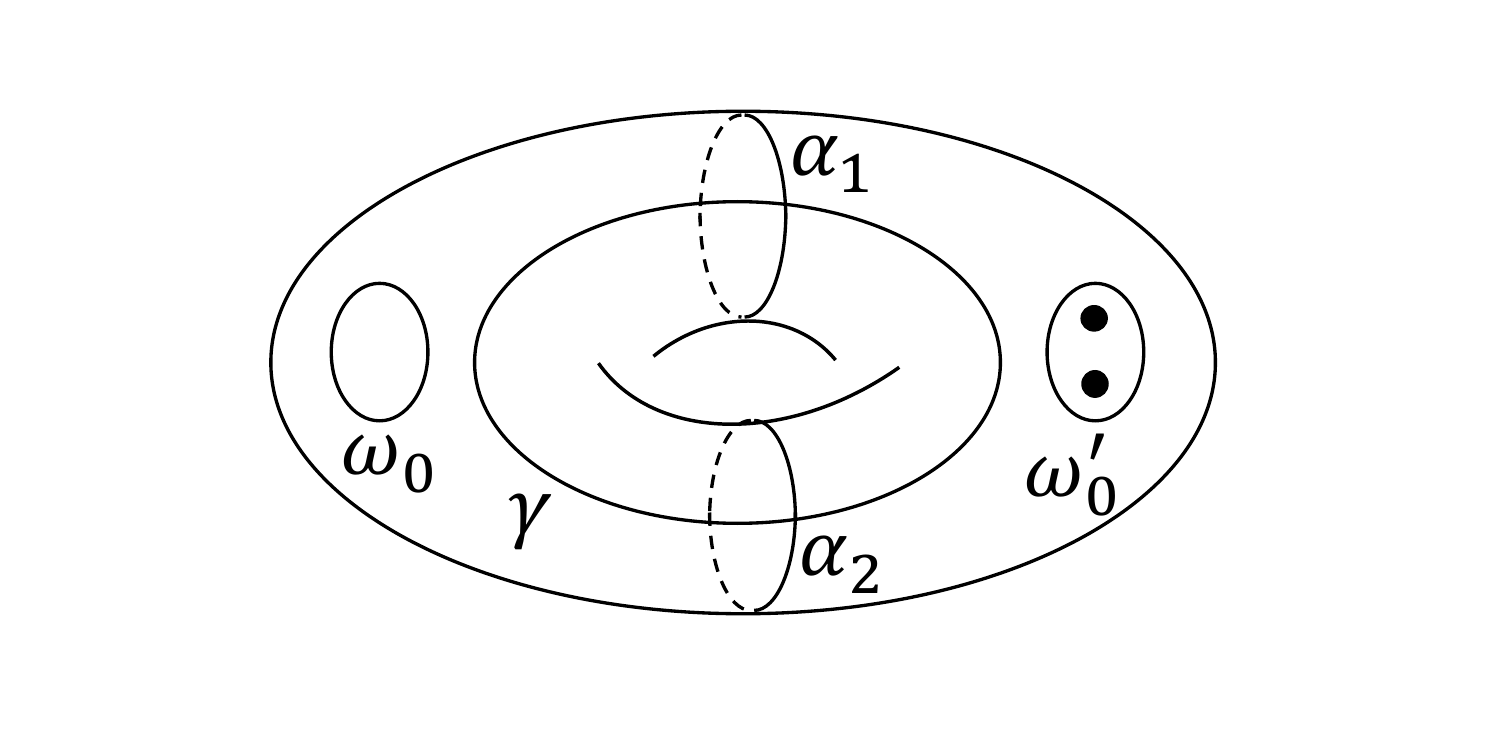}
	\caption{\label{DIsingtorus} The Ising $\times \overline{\text{Ising}}$ state with gapped boundaries maps to a single copy of the Ising state on a torus, with the
loops $\alpha_1$, $\alpha_2$, $\gamma$, $\omega_0$, and $\omega_0'$ as shown.  }
\end{figure}

The topological charge associated to the loop $\omega_0$, which we denote as $c(\omega_0)$, is the state of the qubit, $x$. After step (2), the topological charges become
$c(\omega_0') \otimes c(\gamma) = x \otimes \frac{1}{2}(\mathbb{I} - \psi) + \frac{1}{\sqrt{2}}((\psi,\mathbb{I}) \times x) \otimes \sigma$.
This is because step (2) essentially braids a $\sigma$ particle
around the topological charge $c(\gamma)$. If $c(\gamma) = \mathbb{I}$ or $\psi$, this gives a phase $+1$ or $-1$. However, if $c(\gamma) = \sigma$, then the fusion channel
of the two $(\sigma, \mathbb{I})$ particles encoding the qubit is flipped.
After step (3), the topological charge $c(\gamma)$ is $(\mathbb{I} - \psi)$ if $x = (\mathbb{I},\mathbb{I})$, or $\sigma$ if $x = (\psi,\mathbb{I})$.

Another way to see these results is as follows. The system is effectively equivalent to a single copy of the Ising state on a high genus surface, with
two punctures, one of which has topological charge $c(\omega_0)$, and the other which has topological charge $c(\omega_0')$. The loops $\alpha_1$, $\alpha_2$, and
$\gamma$ map onto the loops shown in Fig. \ref{DIsingtorus}. If the state of the qubit is $x = (\mathbb{I},\mathbb{I})$, then after step (3) both punctures have trivial
topological charge, and $c(\alpha_1) = c(\alpha_2) = \sigma$, which implies that $c(\gamma) = \frac{1}{\sqrt{2}} (1 - \psi)$, by reading off the second row of the $S$-matrix.
However, if instead $x = (\psi,\mathbb{I})$, then after step (3) the system is equivalent to a torus with a single puncture $c(\omega_0)  = \psi$, and
$c(\alpha_1) = c(\alpha_2) = \sigma$. $c(\gamma)$ is then determined by the punctured $S$ matrix of the Ising state, which is $S^\psi_{\sigma,a} = \delta_{a,\sigma}$,
which implies that $c(\gamma) = \sigma$.

The double Dehn twist of step (4) around $\gamma$ then gives a relative phase of $e^{2\pi i 2 h_\sigma} = e^{i \pi/4}$ to the state, depending on whether
$x = (\mathbb{I}, \mathbb{I})$ or $(\psi, \mathbb{I})$.

After step (5), the topological charge $c(\gamma)$ reverts to $\frac{1}{2}(\mathbb{I} + \sqrt{2} \sigma + \psi)$ and
$c(\alpha_1) = c(\alpha_2) = (\mathbb{I},\mathbb{I})$, recovering the initial state, up to the $\pi/8$ phase gate.

In order to implement the double Dehn twist of step (4), we use the protocol described in this paper. We introduce another
disconnected gapped boundary (similar to the extra genus used in the main text), and carry out the relevant sequence of three topological
charge projections.

\subsection{Controlled-Z gate}

It also possible to utilize a similar protocol as described above to implement a controlled-Z gate (CZ) on two qubits.
The controlled-Z gate applies the Pauli $\sigma^z$ operation to the state of the second qubit depending on the state
of the control qubit, and is represented by the matrix:
\begin{align}
CZ = \left(\begin{matrix} 1 & 0 & 0 & 0 \\
0 & 1 & 0 & 0 \\
0 & 0 & 1 & 0 \\
0 & 0 & 0 & - 1\\
\end{matrix} \right)
\end{align}

To implement this, we consider an additional pair of quasiparticles of type $(\sigma, \mathbb{I})$, whose
fusion channel $y = (\mathbb{I},\mathbb{I})$ or $(\psi, \mathbb{I})$ is the state of the second qubit. To implement the CZ gate,
we follow the same steps as described above, except we replace step (4) above with (4'):

(4') Take the additional pair of quasiparticles through a loop involving $\gamma$. This path is similar to the one used in step (2) for a single $(\sigma, \mathbb{I})$
quasiparticle.

Based on the preceding discussion, if $x = (\mathbb{I},\mathbb{I})$, then $c(\gamma) = \frac{1}{\sqrt{2}}(\mathbb{I} - \psi)$, in which case braiding the pair of quasiparticles
in fusion channel $y$ will give a $+1$. On the other hand, if $x = (\psi, \mathbb{I})$, then $c(\gamma) = \sigma$, in which case the braiding of the pair of quasiparticles
in fusion channel $y$ gives $+1$ if $y = (\mathbb{I},\mathbb{I})$ and $-1$ if $y = (\psi, \mathbb{I})$.

\end{document}